\documentclass{aa}  

\usepackage{graphicx}
\usepackage{txfonts}
\usepackage{lipsum}
\usepackage{subcaption}
\usepackage{lscape}
\usepackage{placeins}

\usepackage{xcolor}
\usepackage{ulem}

\begin{document}

\title{Virial-based extraction of structures in numerical simulations: The \textit{vibes} tool}

\author{S. Chevalier\inst{1}\fnmsep\thanks{simon.chevalier3@univ-grenoble-alpes.fr}
        \and F. Louvet\inst{1}\fnmsep\thanks{fabien.louvet@univ-grenoble-alpes.fr}
        \and Y. Bernard\inst{1}
        \and F. Motte\inst{1}
        \and D. J. Price\inst{2}
        \and N. Brucy\inst{3}
        \and M. Valeille-Manet\inst{1, 4}
        \and M. Gonz\'{a}lez-Garcia\inst{5}
        \and E. Moraux\inst{1}
        \and I. Joncour\inst{1}
        \and B. Thomasson\inst{1}
        \and P. Didelon\inst{6}
        }

\institute{Univ. Grenoble Alpes, CNRS, IPAG, 38000 Grenoble, France
           \and School of Physics and Astronomy, Monash University, Clayton, VIC 3800, Australia
           \and ENS de Lyon, CRAL UMR5574, Universite Claude Bernard Lyon 1, CNRS, Lyon 69007, France
           \and Laboratoire d’astrophysique de Bordeaux, Univ. Bordeaux, CNRS, B18N, allée Geoffroy Saint-Hilaire, 33615 Pessac, France
           \and Universidad Internacional de Valencia (VIU), C/Pintor Sorolla 21, E-46002 Valencia, Spain
           \and Laboratoire AIM, CEA/IRFU CNRS/INSU Univ. Paris Diderot, CEA-Saclay, 91191 Gif-sur-Yvette Cedex, France
           }

\date{Received 10 February 2026 / Accepted 30 May 2026}
   
\abstract
{The processes that determine the stellar initial mass function (IMF) and its connection to the core mass function (CMF) are among the major open questions in star formation. The definition of a core remains unclear, yet the way they are extracted from simulations and observations critically shapes the CMF.
Nowadays, cores are mostly detected through their density or intensity only.}
{We aim to explore a new way to define cores in 3D numerical simulations based on a direct application of the virial theorem, and break free from some limitations induced by density-based methods.
We intend to improve the accuracy and the physical meaning of the extracted cores.}
{We developed \textit{vibes}, an innovative method that makes full use of the virial theorem to extract overdensities in simulation snapshots. 
It works by building structures iteratively around density peaks, and applying the virial theorem to the structure at each iteration. 
Then, the structure boundary is set from the evolution of the its energy as it spatially grows.}
{We used STARFORGE simulations to test the sensitivity of the extraction process to the main working parameters (constraints on the structure shape, iteration step, and peak selection criteria).
This sensitivity is observed to be low. 
We compared our extraction with two density-based extraction algorithms, \textit{hop} and \textit{dendrogram}, that are observed to be very sensitive to their input density threshold parameter.
\textit{Vibes} returns structures that are coherent to each other and physically motivated, and it appears much more stable than existing 3D extraction tools.}
{We present \textit{vibes}, a new structure extraction tool based on the virial equilibrium of the dense gas. 
By defining the boundary of the cores on a physical criterion rather than on a user-defined set of density parameters, we expect such extracted cores to be closer to their forsaken definition: gas reservoirs that will form a single star or a close multiple system.
The tool also provides a way to probe the physics of dense structures by following the evolution of their virial energy as they spatially grow.}


\titlerunning{VIrial-Based Extraction of Structures}
\authorrunning{S. Chevalier \& F. Louvet}
\maketitle
\nolinenumbers

\section{Introduction} \label{introduction}

Stars form within dense and compact cores embedded in molecular clouds. 
The nature of these cores and the extent to which their properties are inherited by the resulting stars remain open questions.
In particular, the origin of the initial mass function (IMF) is still debated \cite[see review by][]{Hennebelle&Grudic2024}.

Similarities between the mass distributions of cores and stars have been reported in several star-forming regions, suggesting a direct connection between cores and stars \citep{Motte1998, Konyves2010}.
Comparable results have also been obtained in numerical simulations \citep{Klessen2001, Chabrier&Hennebelle2010}. 
In this framework, the core is a well-isolated mass reservoir whose properties are directly transferred to the protostar(s) it hosts. 
If the properties of stars are indeed inherited from their parent cores, the stellar IMF should be directly derived from the original core mass function (CMF), after considering efficiency and fragmentation effects.
However, this model has difficulties explaining high-mass (M > 8 M\(_\odot\)) star formation. 
Indeed, recent observations challenge this CMF-to-IMF direct relationship, by observing significantly different mass distributions for cores in highly dynamical regions \citep{Motte2018a, Liu2018, Louvet2024}. 
Different models of highly dynamical star formation processes support the idea that the resulting stars are not completely set by the cores where they form but rather by accretion flows at larger scales \citep{Smith2009, Vazquez-Semadeni2019}.
The nature and the very physical reality of these cores is a matter of debate.
Through numerical simulations, cores have been observed to be highly transient and hard to follow in time \citep{Ballesteros-Paredes2018, Smullen2020, Offner2025}. 
Through observations, they appear to be highly resolution-dependent \citep{Louvet2021}.

The objects themselves and their follow-up in time might be affected by the way they are extracted. 
The high time variability of the cores in simulations can reflect their transient nature but also artifacts induced by the extraction method itself, or both \citep{Dib2008, Smullen2020, Offner2025}.
Plenty of different techniques and algorithms have been developed.
Nowadays, extraction methods are mostly limited to density or intensity only. 
Some are based on isocontours, such as \textit{clumpfind} \citep{Williams1994} and \textit{dendrogram} \citep{Rosolowsky2008}.
They are widely used by the community, and applied both to simulations \citep{Smullen2020, Offner2025, Cusack2025} and observations \citep{Sanhueza2019, Yoo2025, Zhao2025}. 
Others use grouping algorithms, such as \textit{hop} \citep{Eisenstein&Hut1998} or \textit{fellwalker} \citep{Berry2015}.
Two-dimensional-only approaches have been specifically designed for observational data.
Among the most regularly used, \textit{getsf} \citep{Menshchikov2021} uses the spatial decomposition of multiwavelength images to extract sources and filaments from the background.
\textit{CuTEx} \citep{Molinari2011} uses second-order differentiation maps to identify peaks.
Different ways to extract structures in simulations have been proposed such as using isocontours of gravitational potential rather than density, and even taking into account the local thermal energy when assigning elements to a structure \citep{Gong&Ostriker2011}. 

As discussed in \cite{Menshchikov2021}, most of those techniques focus on the detection of dense structures and not necessarily on the extraction accuracy. 
Even though those techniques are consistent with the definition of cores as over-densities in the interstellar medium, they might show high variability and a lack of physical considerations. 
To compensate for this second point, they are often combined with post-extraction analysis and selection techniques.
The boundedness of the extracted objects is checked using quantities derived from the virial theorem (for example, through the comparison of the object mass derived from the flux to the mass of the critical Bonnor-Ebert sphere matching the object size \citep{Louvet2024}).
Observational studies can not estimate the collapsing behavior of structures --- except obtaining optically thin and thick tracers that can, in certain conditions, reveal signatures of collapse in the line of sight. 
Instead, observers often use virial parameters as proxies of the virial theorem to address the collapse of a structure. 
The effects of the turbulence and magnetic field can be ignored, and the surface terms of the virial theorem are almost always neglected, an assumption that is open to discussion \citep{Ballesteros-Paredes2006, Dib2007}. 
These terms are often omitted in practice because they are difficult to evaluate reliably.
Those virial parameters compare the different volume terms, generally thermal energy or turbulence to gravitational energy.
This post-extraction analysis can also be used for simulated data \citep[e.g.,][]{Ntormousi&Hennebelle2019}.

In this work, we present a method to extract structures in numerical simulations based on the virial theorem. 
Our algorithm is built to extract over-densities from a background and sets the boundaries of the structures on an energy rather than density criterion.
The algorithm itself is described in detail in Sect. \ref{section-methods}.
In Sect. \ref{section-benchmark}, we test the robustness of the method to its internal parameters.
In Sect. \ref{section-discussion}, we compare our extraction with two other extraction techniques (\textit{hop} and \textit{dendrogram}) and analyze their own parameter sensitivity.
The conclusion of this work is given in Sect. \ref{conclusion}.

\section{Methods} \label{section-methods}

The extraction process has been built to take as input lists of particles and can work on any type of simulation, either Eulerian or Lagrangian. It has been tested on the Lagrangian code GIZMO \citep{Hopkins2015} and the Eulerian code RAMSES \citep{Teyssier2002}.
In the following, we call "cell" a single fluid element, either an Eulerian cell or a Lagrangian particle depending on the nature of the simulation.
The example profiles and results presented in this paper come from an extraction performed on a STARFORGE simulation snapshot presented in Sect. \ref{benchmark-subsection-simulation-overview}.

\subsection{Virial theorem application} \label{methods-subsection-virial-theorem}

The virial theorem describes how the mass distribution within a given volume evolves over time by accounting for all the physical contributions acting on the system. It does so by looking at the second time derivative of the momentum of inertia, \( I = \int_\mathcal{V} \rho r^2 ~dV \), where \(\rho\) is the mass density, \(r\) the distance to the center, and \(V\) the volume of the structure.

The virial theorem was first formulated in its Lagrangian form by \cite{Chandrasekhar&Fermi1953}. 
We use the Eulerian formulation that makes the assumption of a time-independent volume \citep{McKee&Zweibel1992}, given by:
\begin{equation} \label{eq-methods-eulerian-virial-theorem}
    \frac{1}{2} \ddot{I} = W + T + K + M + \dot{\Phi}
\end{equation}
, where \(W\), \(T\), \(K\), and \(M\) are the gravitational, thermal, kinetic and magnetic terms, respectively. 
Furthermore, \(\dot{\Phi}\) corresponds to the rate of change of the flux of the momentum of inertia across the surface. 
These five terms are given by:
\begin{equation} \label{eq-methods-virial-gravity-term}
    W = - \int_\mathcal{V} \rho \nabla \phi \cdot \mathbf{r} ~dV
\end{equation}
\begin{equation} \label{eq-methods-virial-thermal-term}
    T = \underbrace{ \left( \int_\mathcal{V} 3 P ~dV \right) }_{T_V} + \underbrace{ \left( - \int_\mathcal{S} P ~\mathbf{r} \cdot \mathbf{dS} \right) }_{T_S}
\end{equation}
\begin{equation} \label{eq-methods-virial-kinetic-term}
    K = \underbrace{ \left( \int_\mathcal{V} \rho v'^2 ~dV \right) }_{K_V} + \underbrace{ \left( - \int_\mathcal{S} \left( \rho \mathbf{v} \otimes \mathbf{v} \cdot \mathbf{r} \right) \cdot \mathbf{dS} \right) }_{K_S} 
\end{equation}
\begin{equation} \label{eq-methods-virial-magnetic-term}
    M = \underbrace{ \left( \int_\mathcal{V} \frac{B^2}{2 \mu_0} ~dV \right) }_{M_V} + \underbrace{ \left( \int_\mathcal{S} \left( \mathbf{T_M} \cdot \mathbf{r} \right) \cdot \mathbf{dS} \right) }_{M_S}
\end{equation}
\begin{equation} \label{eq-methods-virial-add-dflux-term}
    \dot{\Phi} = - \frac{1}{2} \frac{d}{dt} \int_\mathcal{S} r^2 \rho \mathbf{v} \cdot \mathbf{dS} 
\end{equation}
, where \( (v')^2 \) is the modified square velocity detailed below.
The volume and surface contributions are denoted as \(X_V\) and \(X_S\) (\(X\) being \(T\), \(K\), or \(M\)), and are computed separately.
The quantities are integrated in the structure frame: the position and velocity origins are set to the center of mass position and velocity, respectively, given by 
\( \mathbf{r}_{\rm com} = \frac{\sum_{i \in \mathcal{V}} m_i \mathbf{r}_i }{\sum_{i \in \mathcal{V}} m_i} \)
and
\( \mathbf{v}_{\rm com} = \frac{\sum_{i \in \mathcal{V}} m_i \mathbf{v}_i }{\sum_{i \in \mathcal{V}} m_i} \), where \( \mathbf{r} \) and \( \mathbf{v} \) are the position and velocity relative to the structure center-of-mass.
The full derivation of the theorem with the description of the variables is given in Appendix \ref{appendix-virial-theorem-derivation}.

The gravitational potential is denoted as \( \phi \).
The gravitational term W takes into account the surrounding potential contributions and is only equal to the gravitational energy of the structure if there is no external potential. 
The thermal, kinetic, and magnetic terms (T, K, and M, respectively) include both volume and surface contributions.
The thermal surface contributions \(T_S\) corresponds to the effect of the external pressure on the volume. 
The kinetic surface term \(K_S\) corresponds to the flux of momentum across the surface, and is always negative: if the velocity field is oriented outward, the volume loses positive momentum, while if it is oriented inward, the volumes earns negative momentum. 
Effects of mass transfers across the surface are taken into account in the last term \(\dot{\Phi}\). 
The surface contributions are often neglected, while they have been found to be in the same order of magnitude than the volume contributions \citep{Ballesteros-Paredes2006, Dib2007}.
A more detailed analysis of the relative importance of each component will be the scope of a future work.
The volume integration of the kinetic term \( K \) accounts for all contributions.
However, as discussed in previous studies \citep{Ballesteros-Paredes2006}, the kinetic term does not always provide support against the collapse but can, in case of collapse, contribute to it. 
As a consequence, we choose to remove the velocity contributions oriented toward the center from the volume kinetic term \(K_V\).
Keeping them would lead to an overestimation of the kinetic term support in regions with strong infall motions.
The kinetic energy is then computed using a modified squared velocity given by:
\begin{equation*}
    v'^2 = v^2 - v_{r ~[v_r<0]}^2
\end{equation*}
, where \( (v')^2 \) is the modified square velocity used for the calculation of the volume kinetic contribution \(K_V\) (Eq. \ref{eq-methods-virial-kinetic-term}), \(v\) is the velocity, and \(v_r\) its radial component.
The radial component is removed if and only if it is negative: components oriented outward are kept, as well as rotation components.
Doing so, we use an adapted version of the virial theorem, that is consistent with our goal of extracting collapsing structures.
Regarding \(\dot{\Phi}\) (Eq. \ref{eq-methods-virial-add-dflux-term}), we have no information about the time derivative since we are extracting structures from single simulation snapshots. 
We approximate it by injecting the momentum conservation equation in the surface integral (see Appendix \ref{appendix-virial-theorem-derivation}).

\begin{figure}[!ht]
    \centering
    \includegraphics[width=9cm]{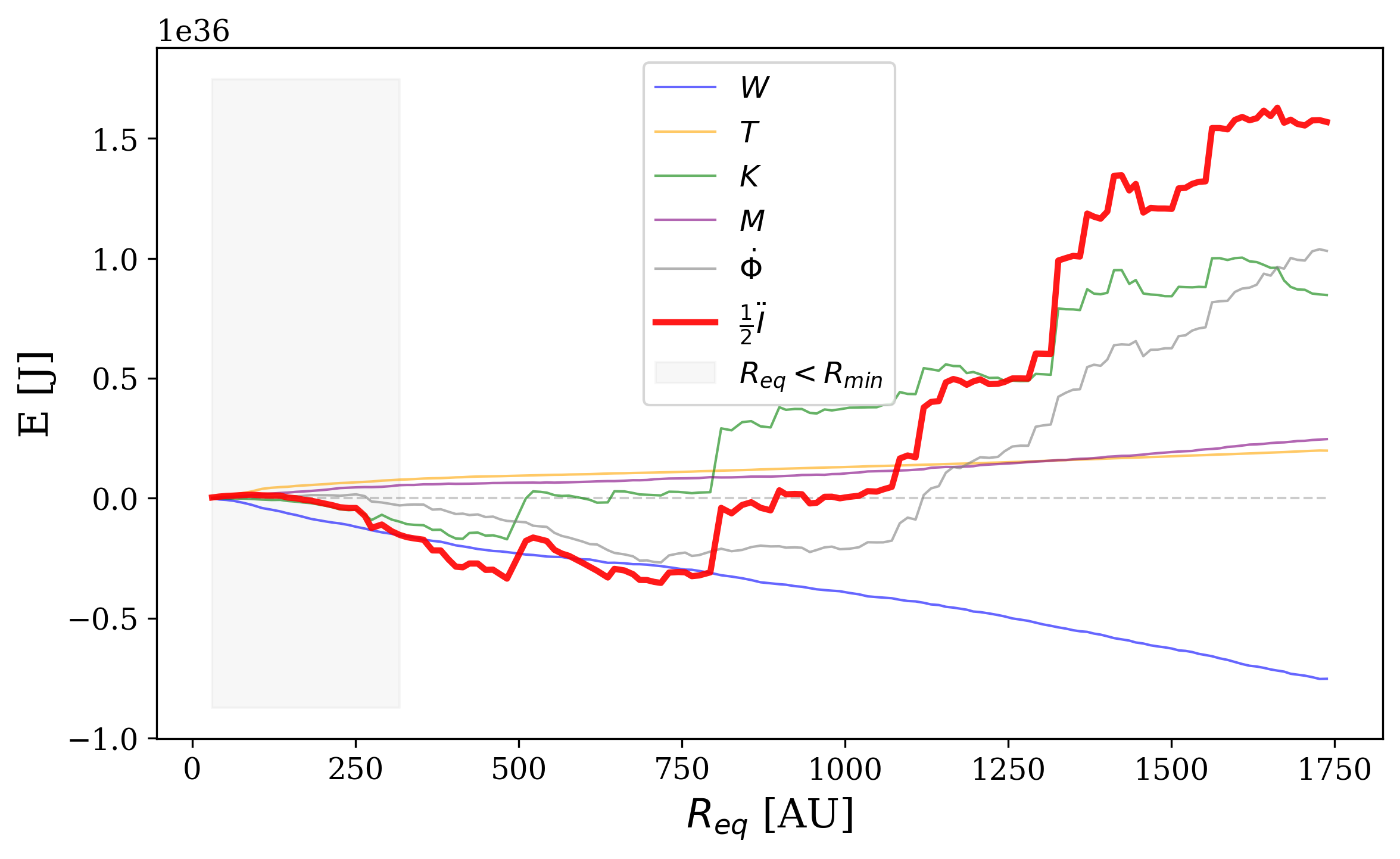}
    \caption{Structure energy components as a function of the equivalent radius \( R_{\rm eq} \) in au (Eq. \ref{eq-methods-equivalent-radius}).
    The contributions of the virial theorem are plotted separately: gravity \( W \) in blue, thermal \( T \) in orange, kinetic \( K \) in green, magnetic \( M \) in purple, and additional surface term \( \dot{\Phi} \) in gray.
    The sum of all the contributions \( \frac{1}{2} \ddot{I} \) is given in red. 
    The light gray region corresponds to \( R_{\rm eq} < R_{\rm min} \).
    The dashed horizontal line gives E=0.}
    \label{fig-method-virial-theorem-profile}
\end{figure}

At each structure building iteration, all the virial contributions detailed above are estimated. 
An example of the structure energy evolution as it is built iteratively is given in Fig. \ref{fig-method-virial-theorem-profile}. 
The energy is plotted as a function of the equivalent radius given by:
\begin{equation} \label{eq-methods-equivalent-radius}
    R_{\rm eq} = \left( \frac{3}{4 \pi} \mathcal{V} \right)^{1/3}
\end{equation}
, where \(\mathcal{V}\) is the physical volume.

The acceleration term \( \Ddot{I} \) gives the rate of change of the structure mass distribution. 
If \( \Ddot{I} < 0 \), the structure should be contracting, while if \( \Ddot{I} > 0 \) the structure should be expanding. 
However, \( \Ddot{I} \) is an acceleration term, thus not a direct indicator of the actual collapse of the structure. 
In other words, \( \Ddot{I} \) can be negative for a structure collapsing faster and faster, but also for a structure expanding slower and slower. 
In that sense, the first derivative of the momentum of inertia \( \dot{I} \) is a much more immediate and reliable indicator of actual collapse \citep[see discussion in][]{Ballesteros-Paredes2006}.
Nevertheless, the first derivative of the momentum of inertia conveys much less physical interpretability than its second derivative, which provides information about the boundedness of the structure and the balance of all the energy contributions.

Since we aim to focus on the evolution of the gas mass distribution, the sink particles are not directly taken into account in the calculations.
If included here, sinks could strongly affect the calculations of the center of mass position and velocity, and mislead the study of the motion of the gas envelop itself.
However, the gravitational potential (\( \phi \) in Eq. \ref{eq-methods-virial-gravity-term}) takes into account the presence of the sink masses.
As a consequence, the sink particles participate in the gravitational term through the potential.

As in \cite{Dib2007}, we aim to lower numerical noise by transforming the surface terms into volume integrals using the divergence theorem. 
The volume integrals are estimated through summations over all the cells belonging to the volume of the structure. 
The gradient estimation follows the methods from \cite{Hopkins2015} (see Appendix \ref{appendix-gradient-calculation}).
When applying the virial theorem, we make an important distinction between Eulerian and Lagrangian simulations. 
We estimate the energy contributions by integrating over cell volumes.
For Lagrangian simulations, the particle physical volume is not well defined and summations over particle volumes may lead to inconsistencies \citep{Lucy1977,Gingold&Monaghan1977}. 
To circumvent this problem, we build local AMR grids around the density peaks, so that each AMR cell contains two Lagrangian particles or less. 
Then, physical quantities and gradients in the AMR cells are interpolated from the surrounding Lagrangian particles \citep{Price&Federrath&Brunt2011}.
For Eulerian simulations, we apply the same gradient estimation method and thus compute for each grid cell an artificial smoothing length. Additional details are given in Appendix \ref{appendix-gradient-calculation}.

We then use this formulation of the virial theorem to extract collapsing structures.
We do so by spatially building structures around density peaks.
The building is performed iteratively over density, adding the densest parts of the structure neighborhood.
At each iteration, we apply the virial theorem to the structure, and use the evolution of its energy as it grows in space to set its boundary.
We choose to iterate over density rather than using a virial-based criterion. 
Indeed, we want to extract overdensities, while an iteration based on the energy of the structure may preferentially follow low-density velocity gradients rather than dense material. 
Moreover, iterating on the energy is significantly more computationally expensive.

\subsection{Peak detection and sorting} \label{methods-subsection-peak-sorting}

We start by extracting all the density maxima and sort them to keep only the significant peaks. 
The detection of the local density maxima is performed over the whole density cube. 
We call local density maximum a cell that is denser than its \(N_{\rm n}\) closest neighbors. 
We introduce here \(N_{\rm n}\), the number of closest neighbors than define the neighborhood of each cell.
We set a peak minimum density threshold at \( \rho_{\rm threshold} = 10^{-15} \) kg.m\(^{-3} \), equivalent to \( n_{\rm threshold} \sim 2 \times 10^{5} \) cm\(^{-3} \), which is the typical order of magnitude for prestellar cores \citep{Motte2018b}.

The detected local density maxima are then sorted. 
We aim to remove small density fluctuations that would not be significant enough compared to their local environment. 
We discriminate density maxima on their density contrast relative to their surrounding. 
The threshold in density contrast is set by the peak-to-saddle ratio (psr), defined as the ratio of the peak density to the density of the closest saddle point in 3D.
To find the closest saddle to the peak, we follow the density ridge building a cell group iteratively, starting from the local density maxima of density \( \rho_0 \). 
At each iteration, we add to the group the densest cell of the neighborhood, where the neighborhood is defined as the union of the \(N_{\rm n}\) closest neighbors of all cells belonging to the group at previous iteration.
The iteration stops and we store the peak if the density contrast is sufficient, meaning \( \rho_{\rm n} \) of the last-added cell is low enough compared to the peak density, following \( \rho_0 > \text{psr} \times \rho_{\rm n} \). 
The psr default value is set to 1.75 (see discussion in Sect. \ref{benchmark-subsection-peak-sorting}).
If the density contrast to the saddle is not high enough, the iteration also stops but the peak is discarded. 
This happens if a cell denser than the local maximum is met, \( \rho_{\rm n} > \rho_0 \), before meeting a low-enough saddle. 
Finally, if the last added cell falls beyond 5000 au from the peak, the peak is discarded.

A discussion on those parameters is given in Appendix \ref{appendix-benchmark-peak-sorting}.
The selected peaks will set the regions where we will look for structures to extract. 
As we intend to extract 3D over-densities, we use the peaks considered significant enough as seeds for the structure building process.
We sort all the selected peaks by decreasing density, such that the densest peaks will be explored first.

\subsection{Structure building} \label{methods-subsection-structure-building}

\begin{figure}[!ht]
    \centering
    \begin{subfigure}[t]{4.2cm}
        \centering
        \includegraphics[width=3.5cm]{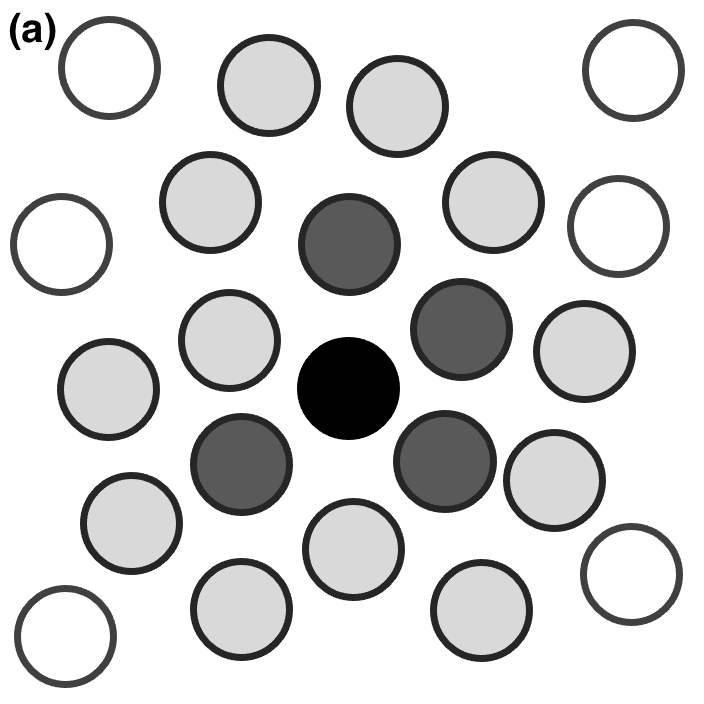}
    \end{subfigure}
    \hfill
    \begin{subfigure}[t]{4.2cm}
        \centering
        \includegraphics[width=3.5cm]{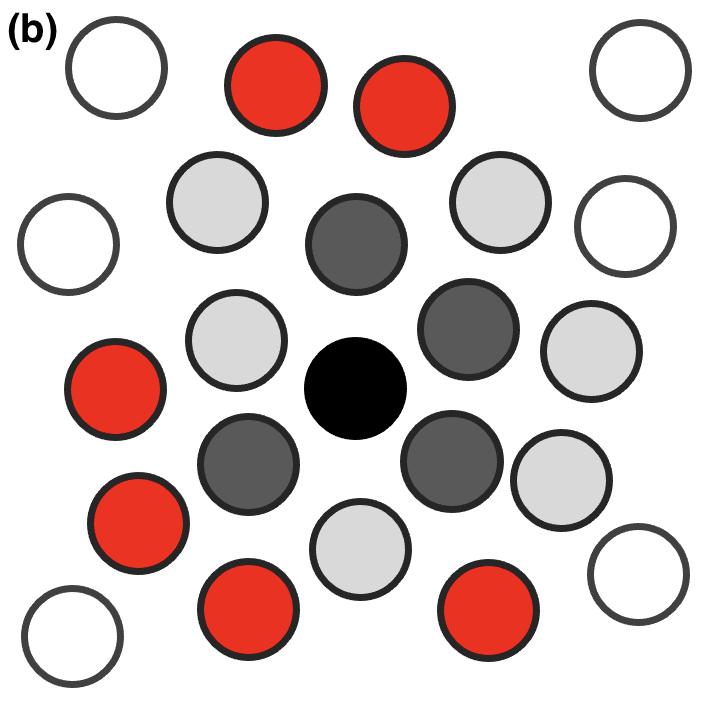}
    \end{subfigure}
    \hfill
    \begin{subfigure}[t]{4.2cm}
        \centering
        \includegraphics[width=3.5cm]{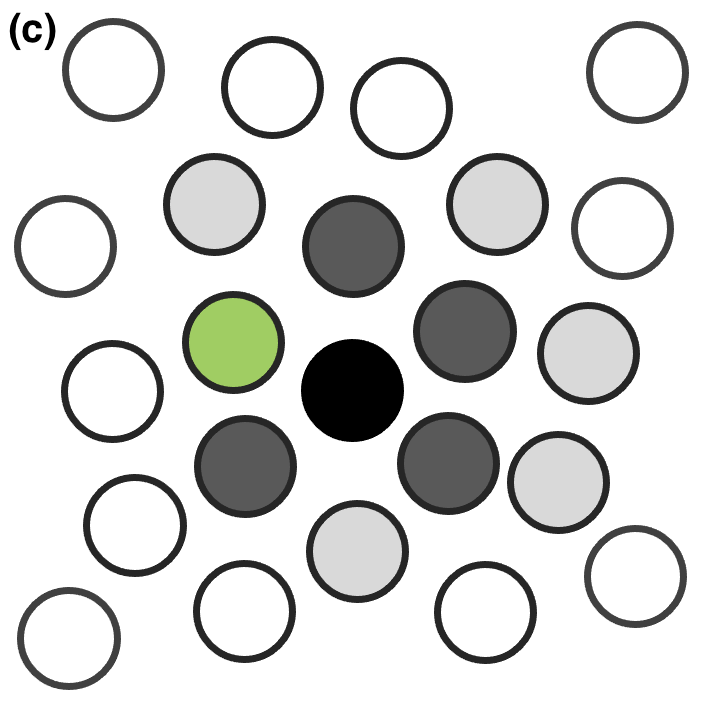}
    \end{subfigure}
    \hfill
    \begin{subfigure}[t]{4.2cm}
        \centering
        \includegraphics[width=3.5cm]{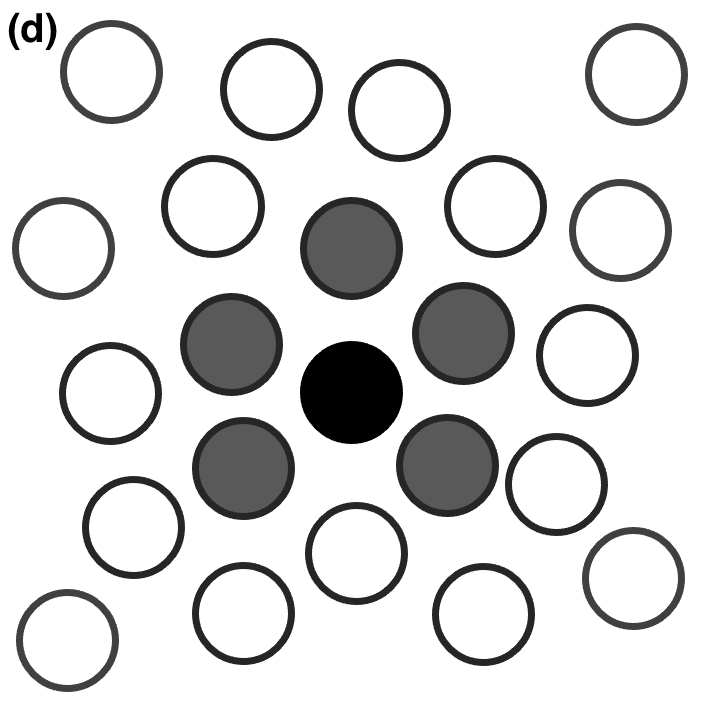}
    \end{subfigure}
    \caption{Illustration of a single cell iteration. 
    Structures are built around density peaks by repeating this process iteratively. 
    The initial peak is given in black. 
    The other cells belonging to the structure are given in dark gray. 
    The cells identified as part of the neighborhood are given in light gray. 
    The neighbors discarded because of the shape constraints are given in red. 
    The cell selected to be added to the structure is given in green.
    \textit{(a)} Neighborhood definition.
    \textit{(b)} Shape-constraint discarding.
    \textit{(c)} Density-based selection.
    \textit{(d)} Cell merging.
    }
    \label{fig-method-cell-by-cell-iteration}
\end{figure}

Once the significant density peaks have been identified, we build structures around them using another independent iterative process.
A structure is built adding the densest neighbors in its vicinity at each iteration.
Each iteration is divided in several steps, whose principle is illustrated in Fig. \ref{fig-method-cell-by-cell-iteration}. 
The process actually works in 3D even though the illustration is shown in 2D for the sake of clarity.
\begin{enumerate}[(i)]
    \item The initial peak is given by the black cell in Fig. \ref{fig-method-cell-by-cell-iteration}a.
    The cells belonging to the initial structure for the iteration are given in dark gray.
    We start by determining the structure neighborhood (light gray cells).
    It is defined as the set of cells composed by the \(N_{\rm n}\) closest neighbors of each cells in the structure. 
    \(N_{\rm n}\) is set to 16 (see discussion in Sect. \ref{methods-subsection-acceleration}).
    \item A constraint on the shape of the structure is applied. 
    The neighbors that would deform too much the structure (red cells in Fig. \ref{fig-method-cell-by-cell-iteration}b) are discarded from this iteration. 
    The shape conditions are detailed below. 
    \item Among the remaining neighbors, the densest cell (green cell in Fig. \ref{fig-method-cell-by-cell-iteration}c) is selected to be added to the structure.
    \item The selected neighbor is added to the structure, increasing its size (see Fig. \ref{fig-method-cell-by-cell-iteration}d).
    The new structure obtained is the initial structure for the next iteration. 
\end{enumerate}

The constraints on the shape are set both on the convexity and the elongation of the structure. 
The convexity is constrained through the ratio of the structure volume, \(\mathcal{V}\), over the convex hull volume:
\begin{equation} \label{eq-methods-convexity-constraint}
    \frac{\mathcal{V}}{\mathcal{V}_{\rm convex-hull}} > C_{\rm convexity}
\end{equation}
The convex hull of a volume is the smallest shape that contains all the segments connecting the volume cells: it is necessarily larger than the structure volume it encompasses, and equal if the structure shape is perfectly convex. 
The elongation of the structure is constrained through the ratio of the distance \( d_{n0} \) from the peak to the last-added cell \( n \) over the equivalent radius of the structure:
\begin{equation} \label{eq-methods-elongation-constraint}
    \frac{d_{n0}}{R_{\rm eq}} < C_{\rm elongation}
\end{equation}
The convexity constraint ensures that the center-of-mass remains inside the structure and prevents the formation of holes.
The segregation on the elongation prevents overly thin structures. 
Those constraints aim to let significant freedom for the structure shape to capture a large panel of possible collapsing objects while ensuring the validity of the virial theorem application (Sect. \ref{methods-subsection-virial-theorem}).
These two parameters \( C_{\rm convexity} \) and \( C_{\rm elongation} \) can be adjusted by the user and take 0.75 and 5 as default values, respectively.
The effect of these parameters is discussed in Sect. \ref{benchmark-subsection-structure-building}.

\subsection{Boundary selection and stopping criteria} \label{methods-subsection-boundary-selection}

\begin{figure}[!ht]
    \centering
    \begin{subfigure}[t]{9cm}
        \centering
        \includegraphics[width=8cm]{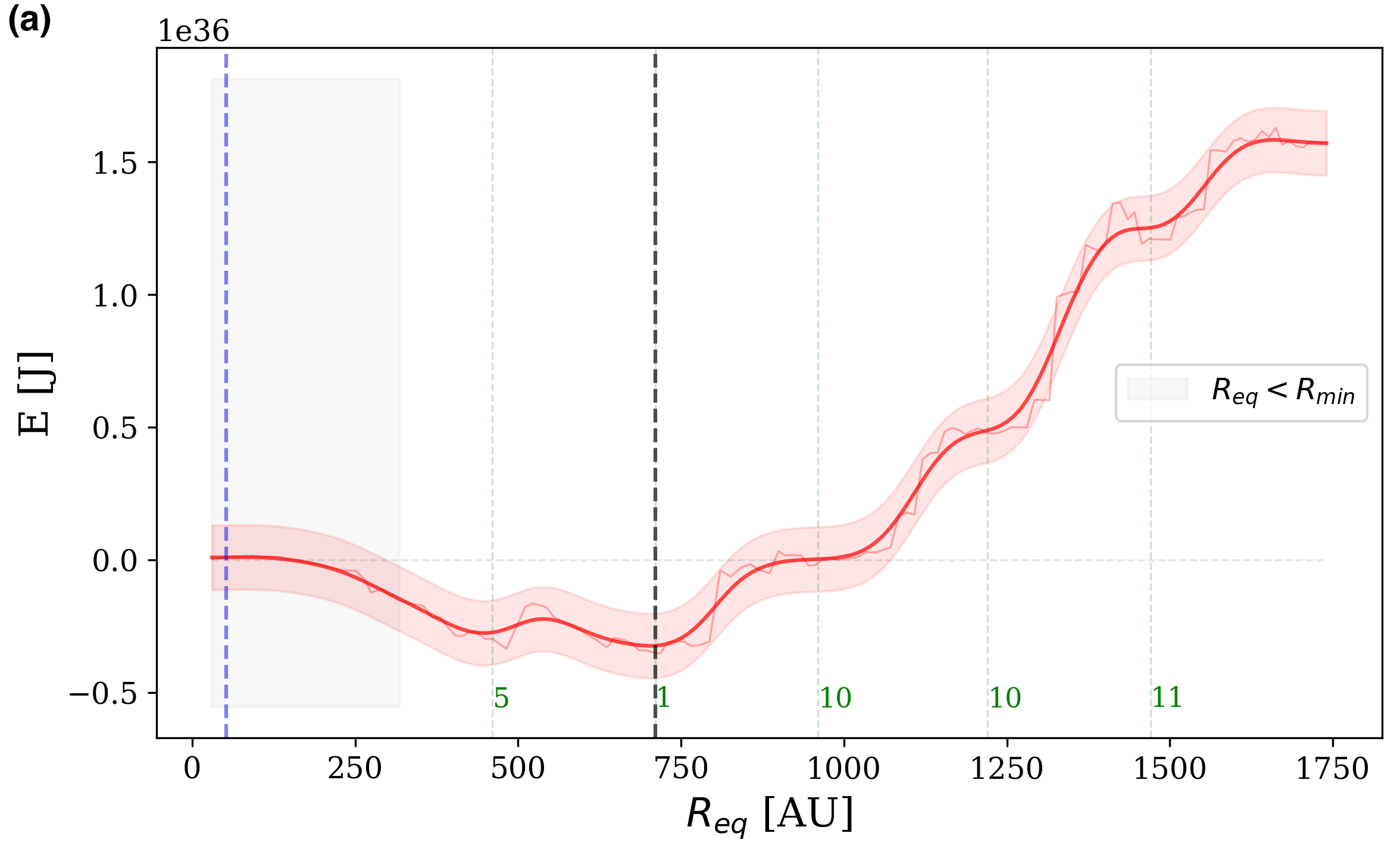}
    \end{subfigure}
    \begin{subfigure}[t]{9cm}
        \centering
        \includegraphics[width=8cm]{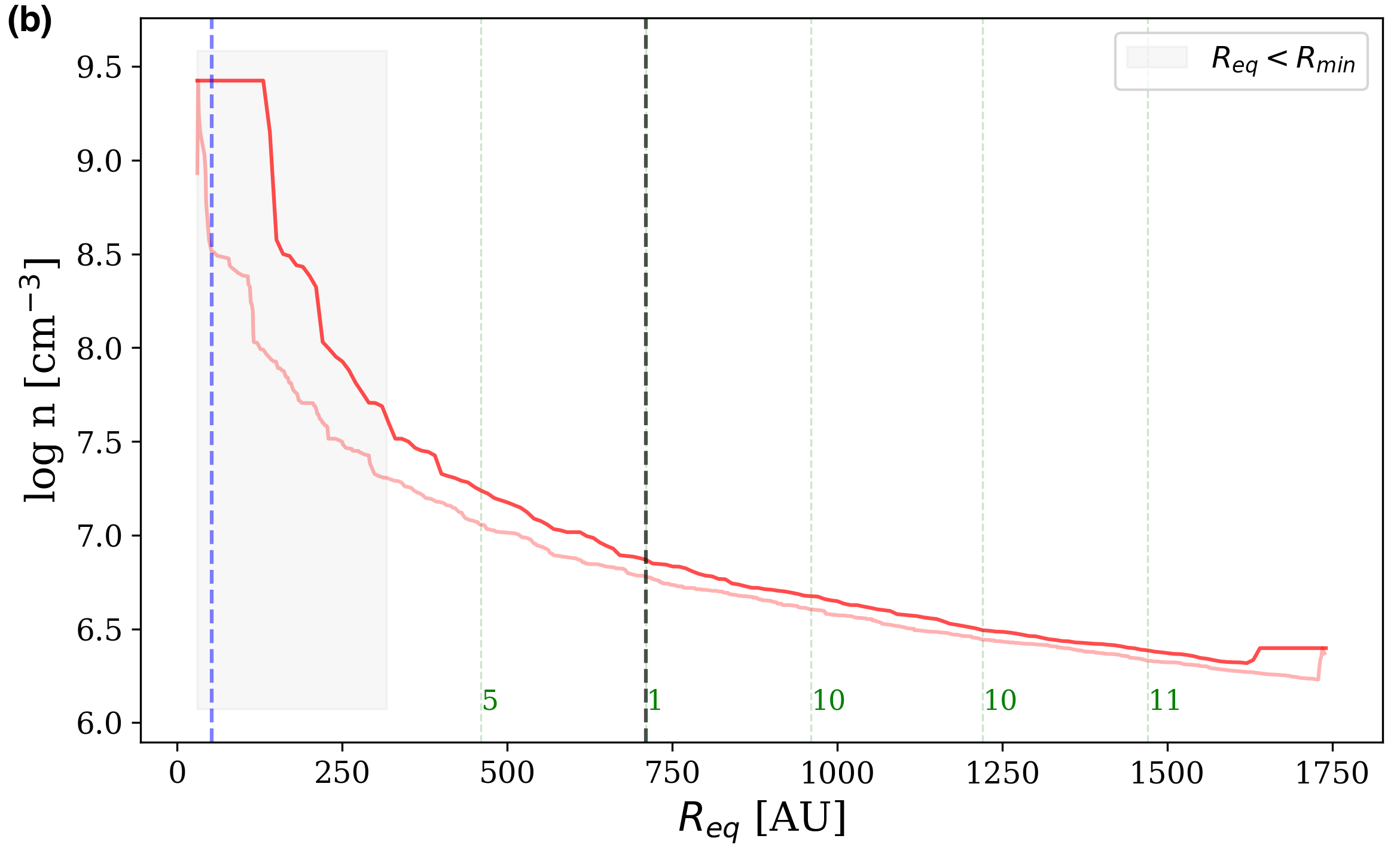}
    \end{subfigure}
    \begin{subfigure}[t]{9cm}
        \centering
        \includegraphics[width=7cm]{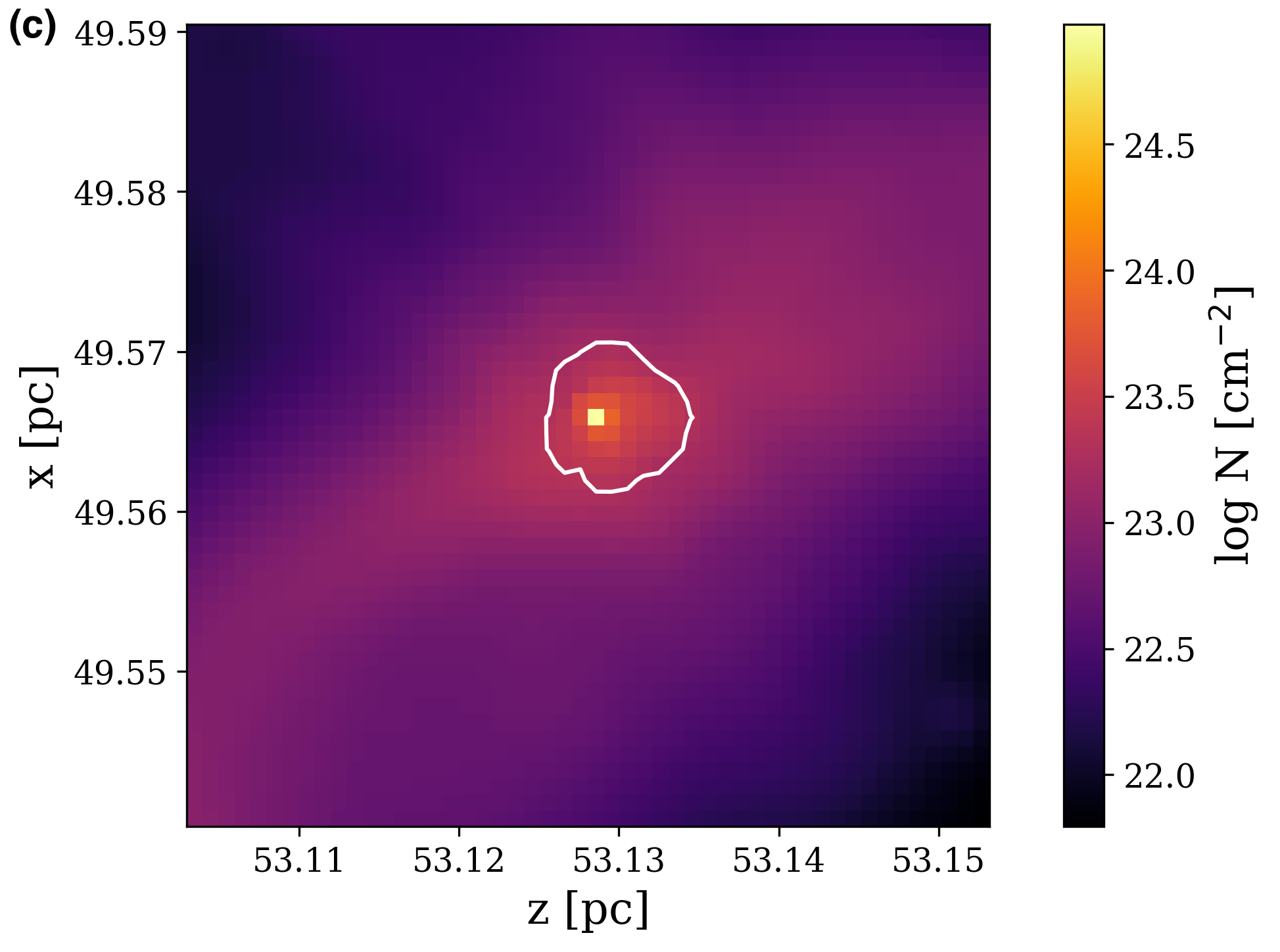}
    \end{subfigure}
    \caption{Illustration of the boundary selection process.
    The dashed vertical green lines correspond to the detected characteristic points, and the small bottom numbers are the associated flag.
    The dashed black line corresponds to the selected boundary of the structure \( R_{\rm select} \).
    The dashed blue line correspond to the position of the sink in the structure. 
    The gray region corresponds to \( R_{\rm eq} < R_{\rm min} \).
    \textit{(a)} Energy of the structure as a function of the structure size. The light red curve is the sum of all the energy contributions (see Sect. \ref{methods-subsection-virial-theorem}). The thick red curve is the smoothed profile used for the boundary selection process. The light red area is the tolerance interval.
    \textit{(b)} Density of the last-added cell as a function of the structure size. The light red curve corresponds to the actual density of the last-added cell. The thick red curve corresponds to the upper-limit profile.
    \textit{(c)} Projected density map. The white contour corresponds to the projected structure boundary \( R_{\rm select} \).
    }
    \label{fig-method-boundary-selection-process}
\end{figure}

The previous step provides an energy profile, giving the virial energy of the structure with respect to its size.
Since we build our structures iteratively, the last part of the energy profile is a direct signature of the contribution of the last-added layers to the integrated acceleration term \( \ddot{I} \). 
If the energy profile increases, it means the outer layer brought a positive contribution, with an acceleration oriented outward. 
Thus we aim to set the structure boundary at energy minima. 
To define the energy minima we use the energy profile as defined by Eq. \ref{eq-methods-eulerian-virial-theorem} and its first derivative.
The boundary selection process is illustrated in the Fig. \ref{fig-method-boundary-selection-process}.

Some noise appears in the energy profile due to the structure building process. Because of the shape constraints, cells may be added from wherever within the neighborhood: two consecutively added cells are not necessarily spatially close. This may introduce local variations in the resulting energy. To limit the effect of this numerical noise in the boundary selection process, the energy profile and its first derivative are slightly smoothed.
The energy profile is given in Fig. \ref{fig-method-boundary-selection-process}a: the light red curve is the original profile, and the opaque red line gives the smoothed profile.

From the smoothed profile, we select the points that are local minima of the energy profile considering a set of abscissa interval sizes. 
We choose the energy minimum within a largest interval, with smallest \(R_{\rm eq}\) in case of a tie.   
If there is not minimum, we look for inflection points as extrema of the derivative: we apply the same interval scheme on the derivative profile.
To each minimum or inflection point, we assign a flag that is directly linked to the minimum interval size.
Such flags provide a confidence level for the boundary selection: the larger is the minimum, the more pronounced is the change in the structure energy with respect to its size.  
We define an ad hoc length scale \(\Delta R_{\rm base}\) that we use to index all the abscissa intervals.
We set \(\Delta R_{\rm base}\) to the maximum value between the numerical resolution and \( 200 \) au:
\begin{equation} \label{eq-methods-rbase-parameter}
    \Delta R_{\rm base} = \max( 200~\text{au}, ~\Delta x_{\rm au} )
\end{equation}
, where \(\Delta x_{\rm au}\) is the highest spatial resolution of the simulation. 
In the example of a STARFORGE simulation, it is taken as the smallest smoothing length.
We scope each point of the energy profile and determine if it corresponds to a minimum in a set of 5 intervals centered on the point itself: 10, 5, 3, 2 and 1 times \(\Delta R_{\rm base}\) (\(\simeq\)200 au, Eq. \ref{eq-methods-rbase-parameter}). 
If a point is a local minimum, it gets attributed the flag 1, 2, 3, 4 or 5 respectively depending on the interval on which it is a local minimum, with flag 1 corresponding to an interval of \(10 \times \Delta R_{\rm base}\). 
To search for local minima on scales lower than \(1\times\Delta R_{\rm base}\), we look at null values from negative to positive on the first derivative of the energy profile ; these points get attributed the flag number 6.
In case no minimum is detected, the boundary is set at inflection points, that can be signatures of a change of collapse regime, for example a change in the dominant term among the various virial contributions.
The inflection points are detected as extrema on the first derivative of the energy profile: the same procedure is applied to the derivative profile, with flags going from 7 to 12.
However, since we want to keep only slope breaks, we do not scope all the minima of the derivative profile, but rather maxima when the derivative is negative and minima when the derivative is positive.
The characteristic points detected on the example profile are given by the green vertical dashed lines in Fig. \ref{fig-method-boundary-selection-process}, with the associated flags being the bottom green numbers.
A summary of the flags is given in Appendix \ref{appendix-flags}.

The core limit is set to the first occurrence of the lowest flag from the previous list.
If several points are detected with the same flag (e.g., two flags number 1), we extract the structure with the smallest \(R_{\rm eq}\). 
The selected equivalent radius \( R_{\rm select} \) that sets the boundary of the structure (black dashed vertical line in Fig. \ref{fig-method-boundary-selection-process}) still has to respect additional criteria.
First, the structure has to be large enough, filling the condition \( R_{\rm select} > R_{\rm min} \), where \( R_{\rm min} \) corresponds to a limit of 50 cells, given by the light gray region in Fig. \ref{fig-method-boundary-selection-process}.
Second, the structure energy has to be negative, with a tolerance: \( \frac{1}{2} \ddot{I} (R_{\rm select}) < \delta_{\ddot{I}}(R_{\rm select}) \). 
The tolerance is set as a fraction of the absolute profile mean: \( \delta_{\ddot{I}} = \frac{1}{4} \overline{ \left| \frac{1}{2} \ddot{I} \right| } \), given by the red interval around the opaque line in Fig. \ref{fig-method-boundary-selection-process}a.
Since no lower flag can be detected afterward, we stop the iteration if a global energy minimum is reached (flag equal to 1).
To ensure such minimum is reached, we require to have iterated further enough, the last iteration radius must be higher than the selected radius plus half of the maximum interval size: \( R_{\rm last} > R_{\rm select} + 5 ~\Delta R_{\rm base} \).

We additionally use the density profile to check if a distinct over-density is reached.
From the structure building step, we obtain a density profile giving the density of the last-added cell.
The density profile is given by the light red line in Fig. \ref{fig-method-boundary-selection-process}b.
In some regions, the profile appears noisy. 
This is mostly due to the constraint on the shape, which makes the iteration follow only partially the density ridges. 
It forces the structure to spatially grow in all directions almost simultaneously to keep a reasonably deformed shape.
As a consequence, cells of lower density can be added in between cells part of the density ridge, creating noise on this profile. 
However, we try to extract the main information considering the highest density values of the profile. 
In the example of a filament, the lower limit of the profile may highly decrease following iteration in directions orthogonal to the filament axis. 
Instead, the upper limit should follow the density ridge values. 
Thus, the density profile used afterward is the upper-limit of the actual profile, given by the opaque line in Fig. \ref{fig-method-boundary-selection-process}b.
It constitutes a step function, \( \rho_{up} \), as a function of \(R_{\rm eq}\) where the value at each point is the local maxima in an arbitrary abscissa interval set to \( \Delta R_{\rm base} \).

We use this upper-density profile to detect neighbor over-densities we consider as distinct. 
We consider reaching a distinct neighbor object either if the density gets higher than the local maximum, or another peak with a sufficient density contrast is detected after a saddle.
In the first case, the condition of reaching a denser cell than the peak is simply given by: \( \rho_{\rm up}(R_{\rm eq}) > \rho_0 \).
In that case, the neighbor detected is critical: since peaks are explored by decreasing density, reaching denser cells than the starting peak means the region has already been explored previously.
In that case, the iteration stops, but a structure might still be extracted at smaller scale if it fills the conditions on the energy profile detailed above.
In the second case, the condition of reaching a peak with a sufficient density contrast is given by: \( \log{\rho_{up}(R_{\rm eq})} - \log{ \min_{R < R_{\rm eq}} \rho_{up}(R)} > \log nsr \), with nsr the neighbor-to-saddle-ratio set to 3.
In that case, we let the iteration continue on a predefined distance of \( 5 ~ \Delta R_{\rm base} \), to allow the detection of a global minimum (flag 1) at the saddle point.
The structure boundary \( R_{\rm select} \) still has to be lower than the point where the neighbor detection criterion is met.

We build a global mask, updated after each structure extraction, identifying the cells belonging to extracted objects. 
Once a structure is defined by the boundary selection process, we check if at least 75 \% of the structure cells are free in the mask (meaning most of the cells have not already been assigned to another structure before) and if the number of such free cells exceeds 50.
In that case, the global mask is updated and the cells of the new extracted object are labeled accordingly.
If more than 25 \% of the cells are already occupied by other structures, the new object is not added to the mask.
The projected mask for the example structure is given by the white contour in Fig. \ref{fig-method-boundary-selection-process}c.

\subsection{Acceleration strategies and approximations} \label{methods-subsection-acceleration}

Several approximations and simplifications are made to accelerate the extraction process.
Only a subdomain of the map (a box of size 10 \(R_{\rm max} \)) is considered around each peak. 
If a neighbor of the structure (in the sense of the \(N_{\rm n}\) closest cells as defined in Sect. \ref{methods-subsection-structure-building}) is out of the subdomain, the iteration stops immediately.
As we aim to study prestellar and protostellar objects, we fix a maximum physical size of \( R_{\rm max} = 5000\) au in equivalent radius. 
Calling \(R_{\rm last}\) the last iteration radius, iterations stops if \( R_{\rm last} > R_{\rm max} \).
To ensure physical consistency in the calculations presented in Sect. \ref{methods-subsection-virial-theorem}, we set the minimum size of the structures to 50 cells. 
Since the local resolution changes from one place to another inside the simulation, the equivalent radius of the structure \( R_{\rm min} \) corresponding to this lower-limit size of 50 cells has to be lower than a critical physical value of \( 2500 \) au, set to half of the maximum physical size. 
If not, the local resolution is considered as not sufficient and the peak is discarded.
If the global mask is not free on the peak cell, meaning the density peak is already part of a previously extracted structure, the peak is skipped.

To reduce the execution time of the extraction process, the three previous steps, namely structure building (Sect. \ref{methods-subsection-structure-building}), virial theorem application (Sect. \ref{methods-subsection-virial-theorem}), and boundary selection (Sect. \ref{methods-subsection-boundary-selection}) are performed almost simultaneously.
Every 10 building iterations, we check if the iteration can be stopped. 
This enables to speed up the global extraction, especially for small structures that are detected very fast and do not require a lot of iterations.

In the structure building process (Sect. \ref{methods-subsection-structure-building}), the number \(N_{\rm n}\) of nearest neighbors defining the local neighborhood of a cell is set to 16. 
This value corresponds to a short range neighborhood that enables a smooth growth of the structures while preventing the formation of internal holes. 
For higher values, the iteration might add neighbors that are not directly connected to the structure: the constraints on the shape will then force the structure to reconnect to those distant neighbors in the following iterations, leading to a discontinuous growth. 
In the peak sorting process (Sect. \ref{methods-subsection-peak-sorting}), the growth smoothness does not matter. 
When looking for local density maxima, we set the neighborhood size to 4 \(N_{\rm n}\) (64 neighbors). 
Increasing the neighborhood size enables to reduce the number of density maxima detected and accelerate this step. Closer maxima would be removed anyway by the peak sorting.
Then, when sorting the peaks, we build a structure to follow the density ridge and see if the contrast to the background in density is sufficient. 
Since we only aim to follow density ridge without any constraint on the structure shape for this step, we accelerate the iteration by increasing the neighborhood size to 2 \(N_{\rm n}\) (32 neighbors).

Moreover, the structures are actually built layer-by-layer rather than cell-by-cell in order to improve the speed of the algorithm. 
It means that, at each iteration, we do not add a single neighbor to the structure, but rather a bunch of neighbor cells.
As an example, all the neighbors in Fig. \ref{fig-method-cell-by-cell-iteration}c are added by decreasing density until the equivalent radius has increased by a layer size \( \Delta R_{\rm layer} \), corresponding to an iteration, that is set to:
\begin{equation} \label{eq-methods-rlayer-parameter}
    \Delta R_{\rm layer} = \frac{\Delta R_{\rm base}}{N_{\rm layer}}
\end{equation}
, where \(N_{\rm layer}\) sets the number of layers per \(\Delta R_{\rm base}\) in equivalent radius. 
Default value is set to \(N_{\rm layer} = 20\), corresponding to \( \Delta R_{\rm layer} \sim 10\) au.
We note that building the structure layer-by-layer induces additional noise in the energy and density profiles.
The structure energy is estimated once every given abscissa interval \( \Delta R_{\rm layer} \) (i.e., once per layer instead of once per cell) to accelerate the process.
The convex-hull used for the constraint on the shape (Sect. \ref{methods-subsection-structure-building}) is updated ten times per layer rather than for each added cell. 
Building structures layer-by-layer rather than cell-by-cell reduces significantly the number of calculations to perform, especially for large structures for which layers may include high number of cells.
We observe the computation time to be linear with the number of layers \(N_{\rm layer}\).
This parameter gives a resolution-equivalent parameter for the structure building and energy calculation.
However, a balance between accuracy and acceleration has to be found here.
If the number of layers is too high, the computation time increases linearly whereas accuracy may not change significantly. 
On the other hand, having a low \( N_{\rm layer} \) value reduces the execution time but affects the accuracy of the resulting energy profile.
The effect of this parameter on the extraction is discussed in Sect. \ref{benchmark-subsection-structure-building}.

\section{Benchmark} \label{section-benchmark}

We tested the effects of the main parameters used in the extraction process detailed in Sect. \ref{section-methods}, namely the peak-density threshold and the peak-to-saddle ratio for the peak sorting step, and the shape constraints and the layer size parameters for the structure building step.

\subsection{Simulation overview} \label{benchmark-subsection-simulation-overview}

We tested our extraction process on a magnetohydrodynamic simulation of star-forming cloud from the STARFORGE project \citep{Grudic2021} using the Lagrangian meshless finite-mass code GIZMO \citep{Hopkins2015}.
The simulation setup is detailed in \cite{Guszejnov2022}.
We use the simulation of a collapsing cloud of mass \( M = 2 \times 10^4 ~M_{\odot} \) with mass resolution \( \Delta m = 10^{-3} ~M_{\odot} \), and initial radius \( R_{\rm cloud} = 10 \) pc.
The cloud was initialized as a sphere with uniform density, surrounded by diffuse gas with a density contrast of 1000, inside a \( 10 ~R_{\rm cloud} \) box.
The temperature was initialized at equilibrium with the interstellar radiation field, for which Solar neighborhood conditions were assumed.
The metallicity was set to solar conditions.
The turbulence was initialized by applying a random velocity field with power spectrum \( E_k \propto k^{-2} \) and scaled to the kinetic virial parameter \( \alpha_{\rm vir} = 2 \).
The magnetic field was initialized as a uniform field set by the energy ratio \( \frac{E_{\rm B}}{|E_{\rm grav}|} = 0.1 \). 
The simulation ran in the ideal MHD framework.
The divergence cleaning for the magnetic field was ensured by adding the source terms from \cite{Powell1999} and \cite{Dedner2002} in conservation equations, and following the constrained gradient method described in \cite{Hopkins2016}.
Sink particles were created when a gas particle reached a density exceeding a given threshold set to \( 3 \times 10^{-14}\) g.cm\(^{-3}\), and satisfied a virial and a tidal criterion \citep{Grudic2021}.
Feedback was taken into account, with stellar radiations, winds, protostellar jets and supernovae included.
We performed our extractions on three snapshots at t = 4.93 Myr (153 sinks), t = 6.41 Myr (373 sinks) and t = 7.88 Myr (498 sinks).

\subsection{Effect of the peak sorting parameters} \label{benchmark-subsection-peak-sorting}

\begin{figure}[!ht]
    \centering
    \includegraphics[width=9cm]{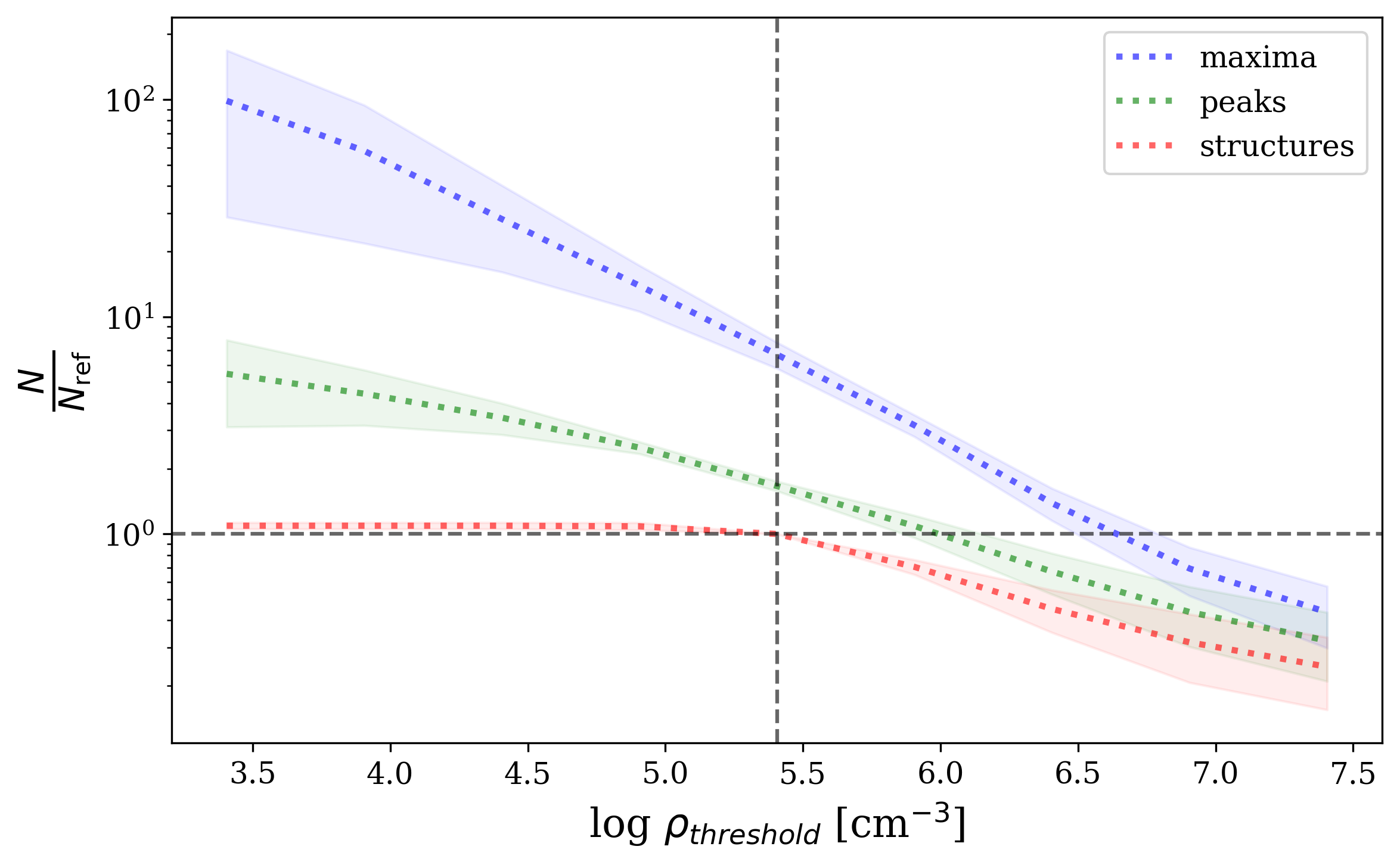}
    \includegraphics[width=9cm]{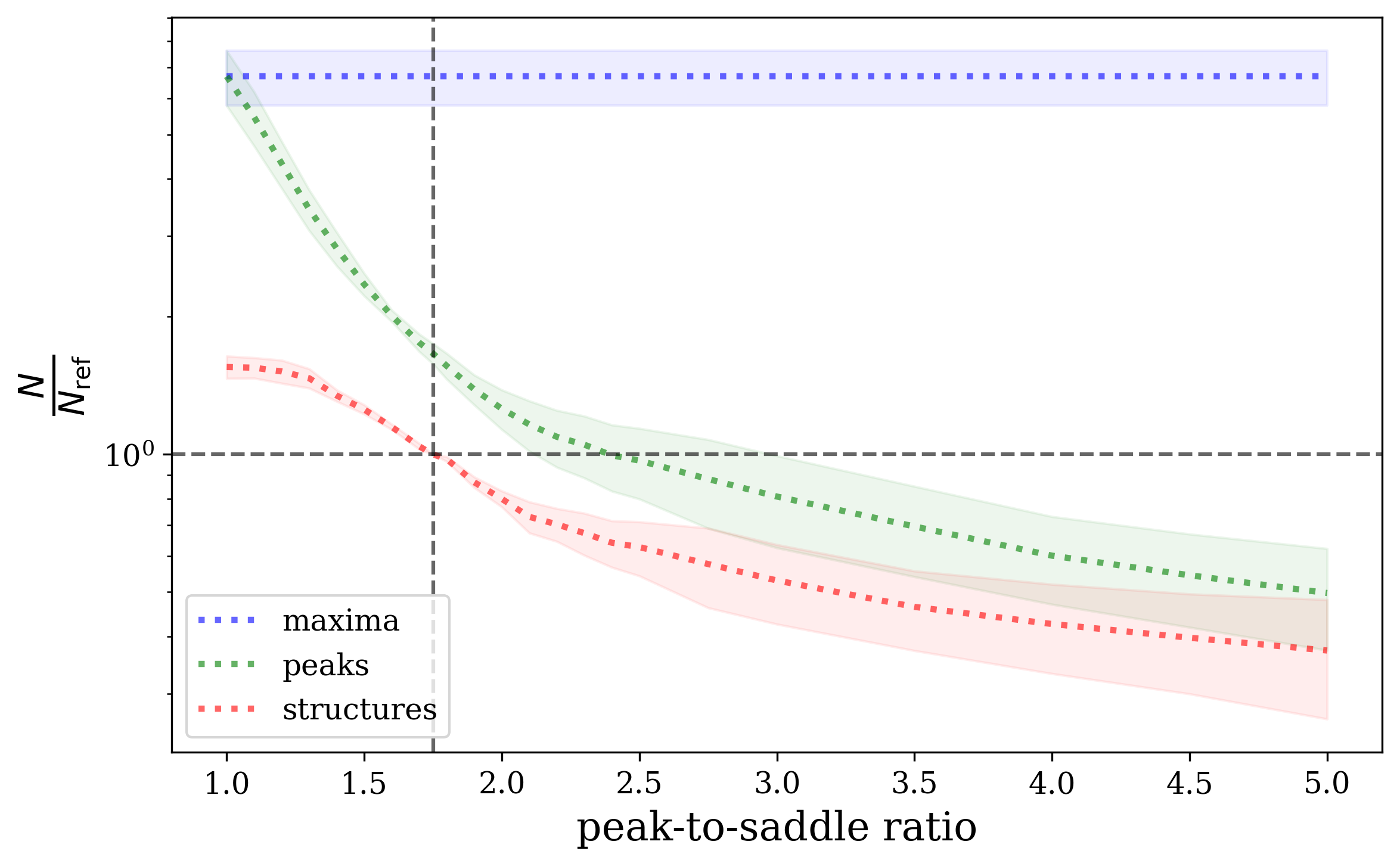}
    \caption{Number of objects relative to the reference number of structures with respect to the peak threshold (top) and the peak-to-saddle ratio (bottom). 
    The reference values are given by the dashed vertical black lines. 
    The dotted red line gives the number of structures extracted at a given parameter value relative to the number of objects from the reference extraction. 
    The dashed horizontal gray line shows \(N=N_{\rm ref}\). 
    The dotted blue and green lines show the number of local density maxima (before sorting) and density peaks (after sorting), respectively, relative to the reference number of extracted structures. The colored lines and intervals correspond to the mean and standard deviation over three snapshots.}
    \label{fig-appendix-benchmark-struc-nb-wrt-peak-sorting}
\end{figure}

Only density maxima above a given density threshold can be used as structure seeds in the extraction process.
However, all the seeds do not necessarily lead to an extracted object. 
The number of structures extracted with respect to the values of the peak sorting parameters is given in Fig. \ref{fig-appendix-benchmark-struc-nb-wrt-peak-sorting}.
Since those parameters only affect the sorting of the peaks and not the iterative construction of the structures, the extracted objects are identical.
The only thing that changes is the number of such extracted objects.
The total number of extracted structures (red line on Fig. \ref{fig-appendix-benchmark-struc-nb-wrt-peak-sorting}) remains almost constant below the reference value \(\rho_{\rm threshold} \sim 2 \times 10^{5}\) cm\(^{-3}\) while the numbers of density maxima and density peaks (blue and green lines, respectively) increase significantly (more than a factor 10 for the number of density maxima).
This confirms this parameter value to be a reasonable choice which removes most of the low-density regions where no additional structures are extracted when explored.
The number of extracted structures drops down for higher threshold values, following the trend of the number of detected density maxima and peaks.

\begin{figure}[!ht]
    \centering
    \includegraphics[width=9cm]{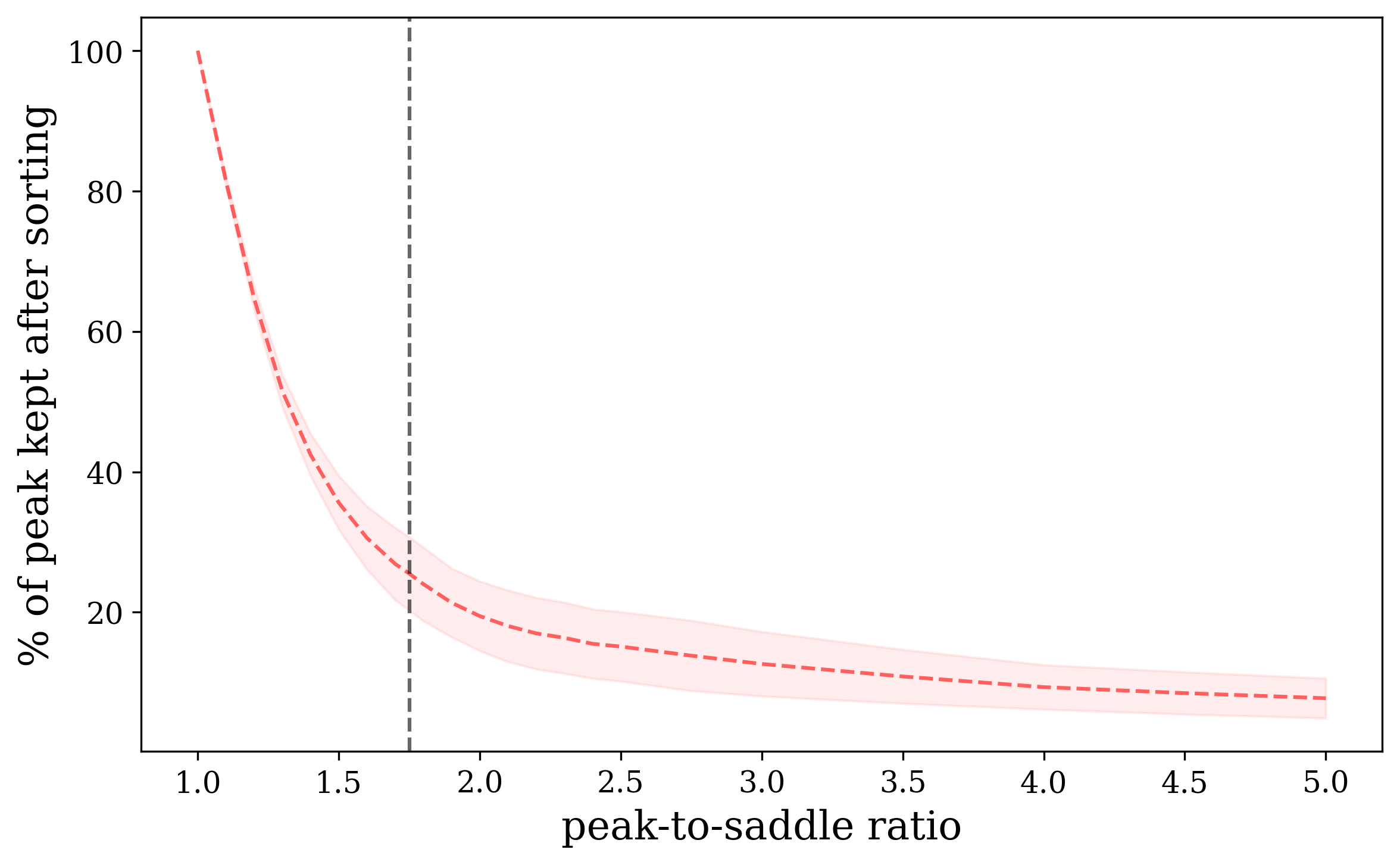}
    \caption{Number of peaks kept after sorting relative to the total number of detected density maxima as a function of the peak-to-saddle ratio. 
    The red line and interval are the mean and the standard deviation, respectively, over three snapshots. 
    The vertical dashed gray line is the default parameter value.}
    \label{fig-benchmark-peak-ratio-wrt-psr}
\end{figure}

The sorting of the local density maxima based on their density contrast is controlled by the peak-to-saddle ratio (see Sect. \ref{methods-subsection-peak-sorting}).
When it is too low, the density fluctuations of the cube are poorly filtered, resulting in keeping a high number of peaks relative to all the density maxima. 
In that case, noise spikes may be considered as significant peaks. 
When this parameter gets too high, the number of significant objects slowly tends toward zero and peaks of interest for structure extraction may be missed.
The total number of detected density maxima (blue line on Fig. \ref{fig-appendix-benchmark-struc-nb-wrt-peak-sorting}) does not change since this parameter only affects the sorting of those maxima. 
The total number of peaks kept after sorting (green line) increases abruptly when the peak-to-saddle ratio decreases, below the reference value psr = 1.75, while the number of extracted structures (red line) seems to converge close to psr = 1.
For psr > 1.75, the number of peaks kept after sorting decreases smoothly, and the resulting number of extracted objects follows the same trend.
This shows only a limited number of structures, with low density contrast, are actually removed by setting psr=1.75.
The proportion of density maxima kept after sorting as a function of the peak-to-saddle ratio is given in Fig. \ref{fig-benchmark-peak-ratio-wrt-psr}.
We observe a significant change of slope close to a value of 2 for the peak-to-saddle ratio, which is commonly taken \citep[see for example][]{Bleuler&Teyssier2014}. 
Below this value, many small density fluctuations, detected as local maxima, are kept as peaks. 
Above psr = 2, most of those fluctuations have been filtered and only a limited number of significant peaks remain.
To avoid prematurely discarding intermediate structures that could be meaningful, we choose a value psr = 1.75 slightly lower than 2. 
On the three studied snapshots, this parameter results in keeping as significant enough about 25 \% of all the density maxima.
However, not all the peaks will necessarily lead to extractions due the various criteria detailed in Sect. \ref{section-methods}. 
With the peak-to-saddle ratio set to 1.75, we observe that about 70 \% of the peaks considered as significant are ultimately associated with extracted structures.
Complements on the effect of the peak sorting parameters on the extraction are provided in Appendix \ref{appendix-benchmark-peak-sorting}.

\subsection{Effect of the structure building parameters} \label{benchmark-subsection-structure-building}

The convexity parameter (Eq. \ref{eq-methods-convexity-constraint}) is defined as the ratio of the structure volume over the volume of its convex hull.
It belongs to the [0, 1] interval, 0 corresponding to no constraint on the convexity, and 1 to a perfectly convex volume. 
The convexity parameters has default value of \( C_{\rm convexity} = 0.75 \).
The elongation parameter (Eq. \ref{eq-methods-elongation-constraint}) is defined as the maximum ratio of the distance from the last-added cell to the local density peak over the equivalent radius. 
It belongs to the [1, \( +\infty \)[ interval, 1 corresponding to a perfectly circular object and \( +\infty \) to no constraint on the elongation. 
We limit the study to the interval [1.5, 15].
We do not test the parameter equal to 1 since it corresponds to the limit case of perfect ball objects.
The elongation parameter has default value of \( C_{\rm elongation} = 5 \).
In the assumption of a cylindrical filament, the elongation limit can be translated to a constraint on the aspect ratio:
\begin{equation} \label{eq-benchmark-aspect-ratio-constrain}
    \frac{L_f}{d_f} \lesssim \sqrt{\frac{3 \left(\left(C_{\rm elongation}\right)^3 - 1\right)}{2}}
\end{equation}
, where \(L_f\) is the filament length and \(d_f\) the diameter.
The calculation is detailed in Appendix \ref{appendix-benchmark-elongation-aspect-ratio}.
For the default value \( C_{\rm elongation} = 5 \), the upper limit for the aspect ratio is \(\sim 14\). 
This is intentionally very permissive, in order to capture the wide panel of shapes collapsing structures can have.
However, the different criteria on the density profile detailed in Sect. \ref{methods-subsection-boundary-selection} prevent the merging of distinct overdensities within filamentary segments: those criteria enforce the iteration to stop before reaching a neighbor peak with a high density contrast.

Those two parameters define the constraint set on the shape of the structure. 
They affect how iteration is performed, especially the way it follows or not the density ridges. 
If the constraint is strong either in elongation or convexity, the structure has to spatially grow almost simultaneously in all directions, including low-density regions. 
This obviously affects the energy estimation as well, and the boundary selection process. 
The peak detection step is not affected by those parameters: iterations start around the same peaks. 
We can perform one-to-one comparisons of the structures from the so-called reference extraction (with default values) and the test extractions varying the parameters.

The layer thickness (Eq. \ref{eq-methods-rlayer-parameter}) corresponds to the resolution of the structure building process.
It is set by the number of layers \( N_{\rm layer} \) that belongs to the [1, \( +\infty \)[ interval, 1 corresponding to a layer size set to \(\Delta R_{\rm base} \sim 200\) au, and \( +\infty \) to a layer thickness that tends toward zero (i.e., a single cell). 
We limit the study to the interval [4, 40] in \( N_{\rm layer} \), corresponding to an interval [5, 50] au in \( \Delta R_{\rm layer} \). 
The default value is set to \( N_{\rm layer} = 20 \), corresponding to a layer thickness of \( \Delta R_{\rm layer} = 10 \) au.

\begin{figure}[!ht]
    \centering
    \includegraphics[width=9cm]{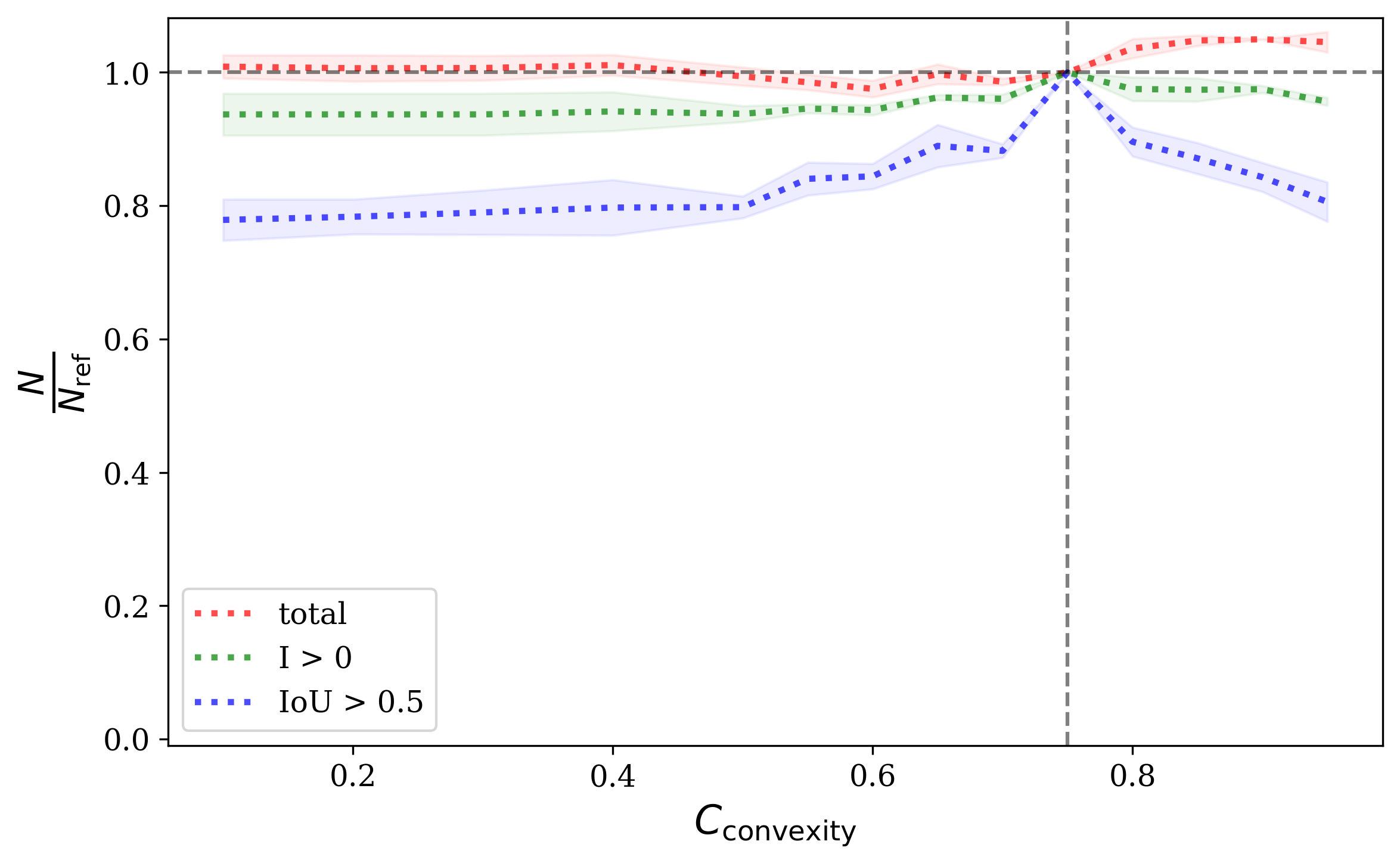}
    \includegraphics[width=9cm]{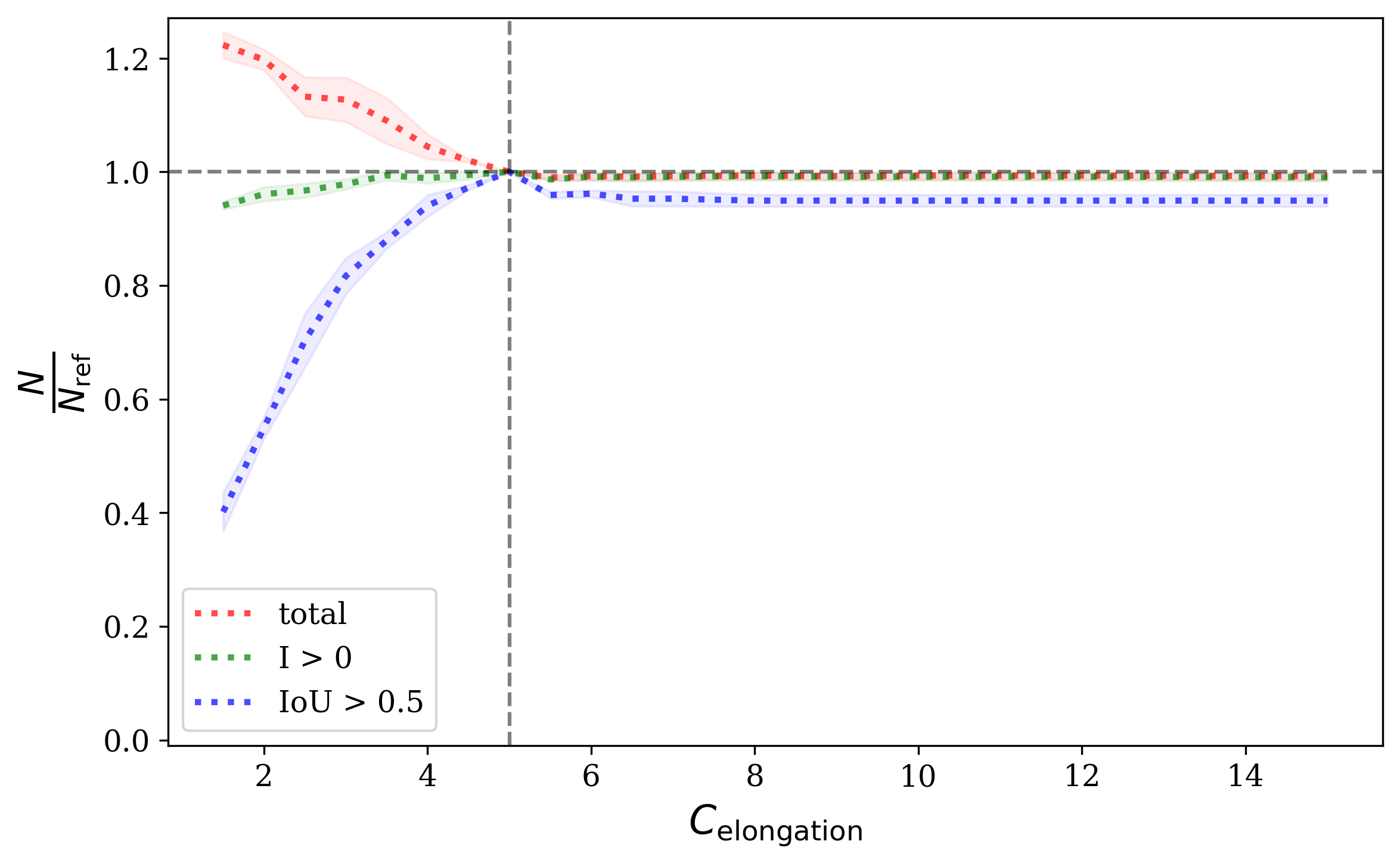}
    \includegraphics[width=9cm]{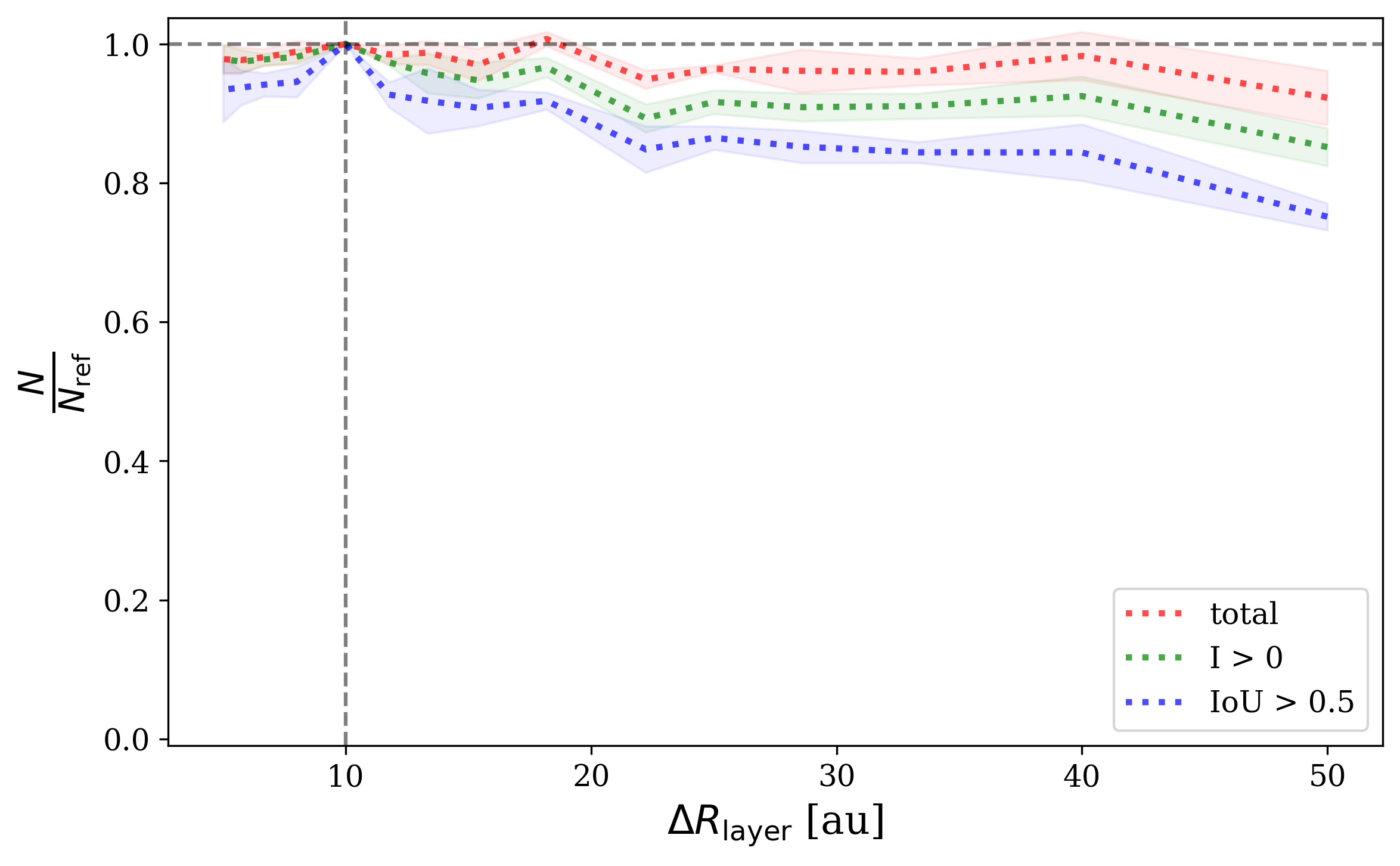}
    \caption{Relative number of extracted objects with respect to the convexity (top), elongation (middle), and layer thickness (bottom) parameters. 
    The dotted red line gives the number of structures extracted at a given parameter value relative to the number of objects from the reference extraction. 
    The black vertical dashed lines give the reference values for each parameter.
    The dotted green and blue lines show the relative number of objects having respectively a nonzero intersection and intersection-over-union higher than 50\% with an object from the reference extraction associated to the same density peak.
    The colored lines and intervals give the mean and deviation over three snapshots.
    The dashed horizontal gray line shows \(N=N_{\rm ref}\).}
    \label{fig-benchmark-struc-nb-wrt-struc-building}
\end{figure}

The number of extracted structures with respect to the values of the parameters is given in Fig. \ref{fig-benchmark-struc-nb-wrt-struc-building}.
When varying the convexity parameter, the number of extracted structures does not vary significantly, with a total number of structures (red points) remaining between 1 and 1.1 relative to the reference extraction, and a number of common structures with intersection-over-union IoU > 0.5 (blue points) remaining above 75 \% of the reference structure number. 
As expected, the accuracy is even better close to the reference value at \( C_{\rm convexity} = 0.75 \).
Varying the elongation parameter has a strong effect at low values, but has a very limited impact for values close to the reference, set to 5, and higher. 
For \( C_{\rm elongation} < 4 \), the relative number of objects increases up to 1.2 relative to the reference, while common structures rarefy with structures having IoU > 0.5 going down to 40\%.
However, the number of structures with IoU > 0.5 with reference remains above 90 \% for \( C_{\rm elongation} > 4 \), and the total number of extracted structures is almost at 1 for such parameter values.
When varying the layer thickness parameter, the number of extracted structures remains almost constant and the number of structures with IoU > 0.5 remaining above 90 \% of the reference number of objects for \(\Delta R_{\rm layer} < 15\) au. 
For layer thickness above 15 au, common structures slightly rarefy but the effect is really weak, having still more 75 \% of objects with IoU > 0.5. 

\begin{figure}[!ht]
    \centering
    \includegraphics[width=9cm]{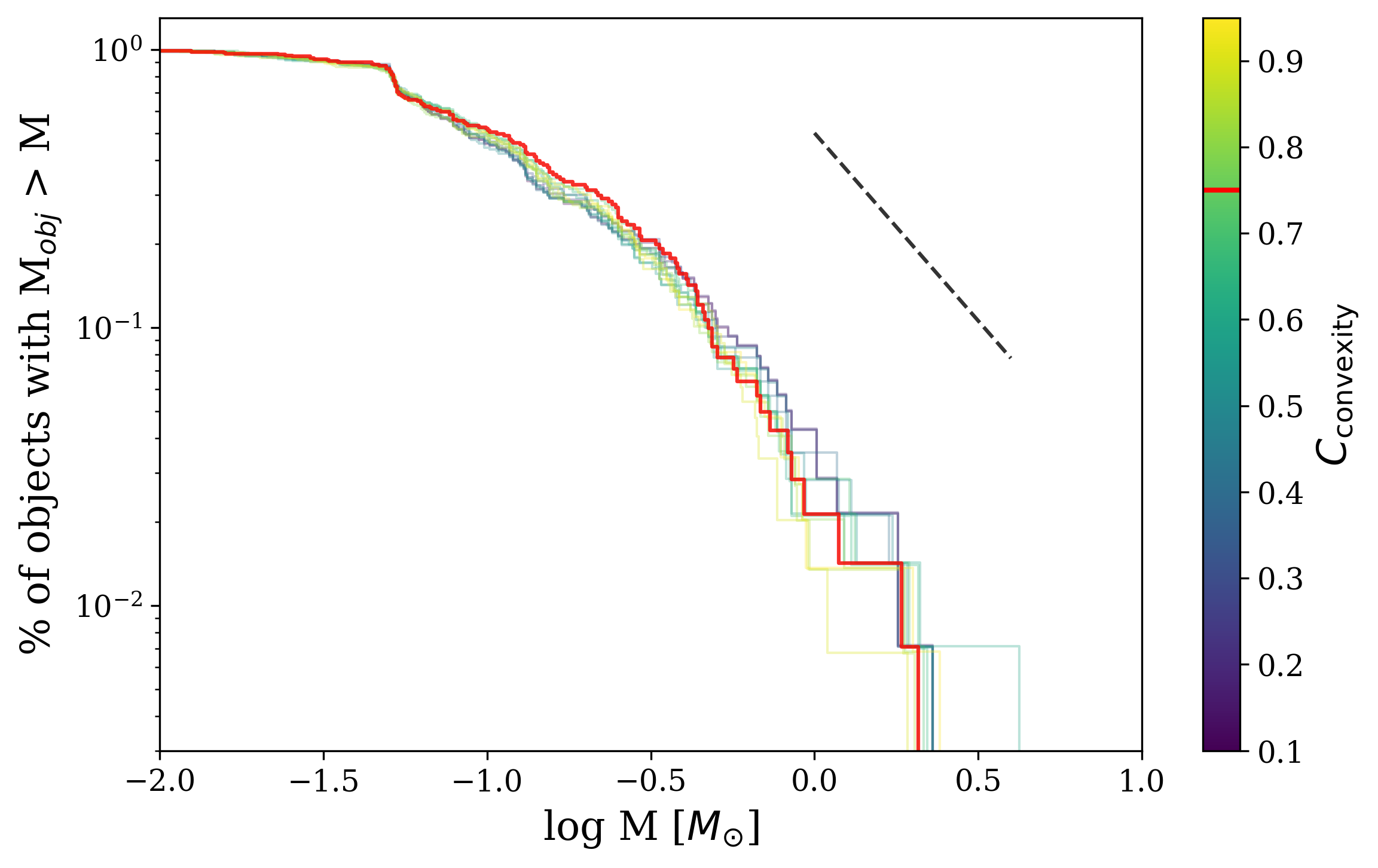}
    \includegraphics[width=9cm]{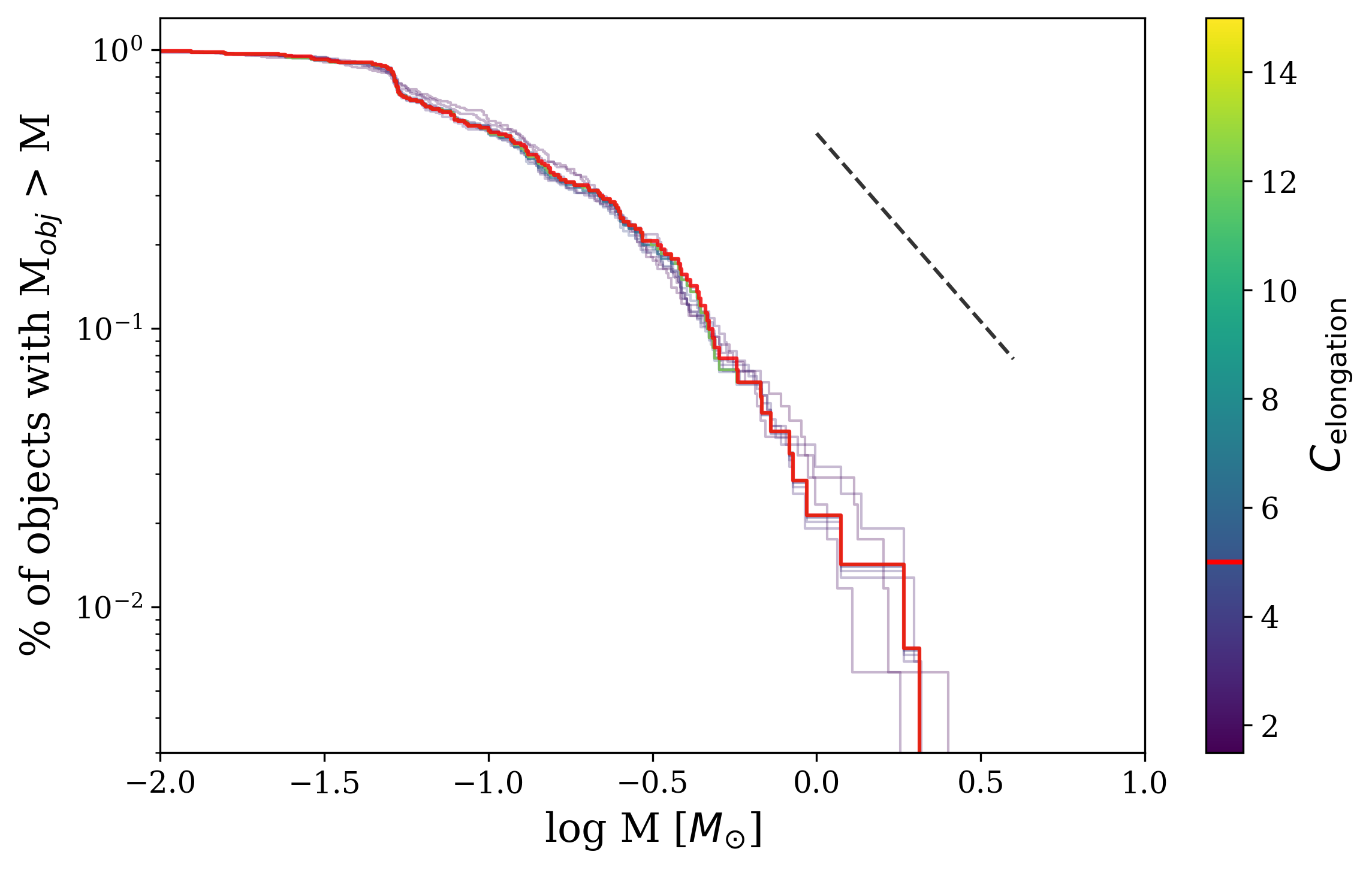}
    \includegraphics[width=9cm]{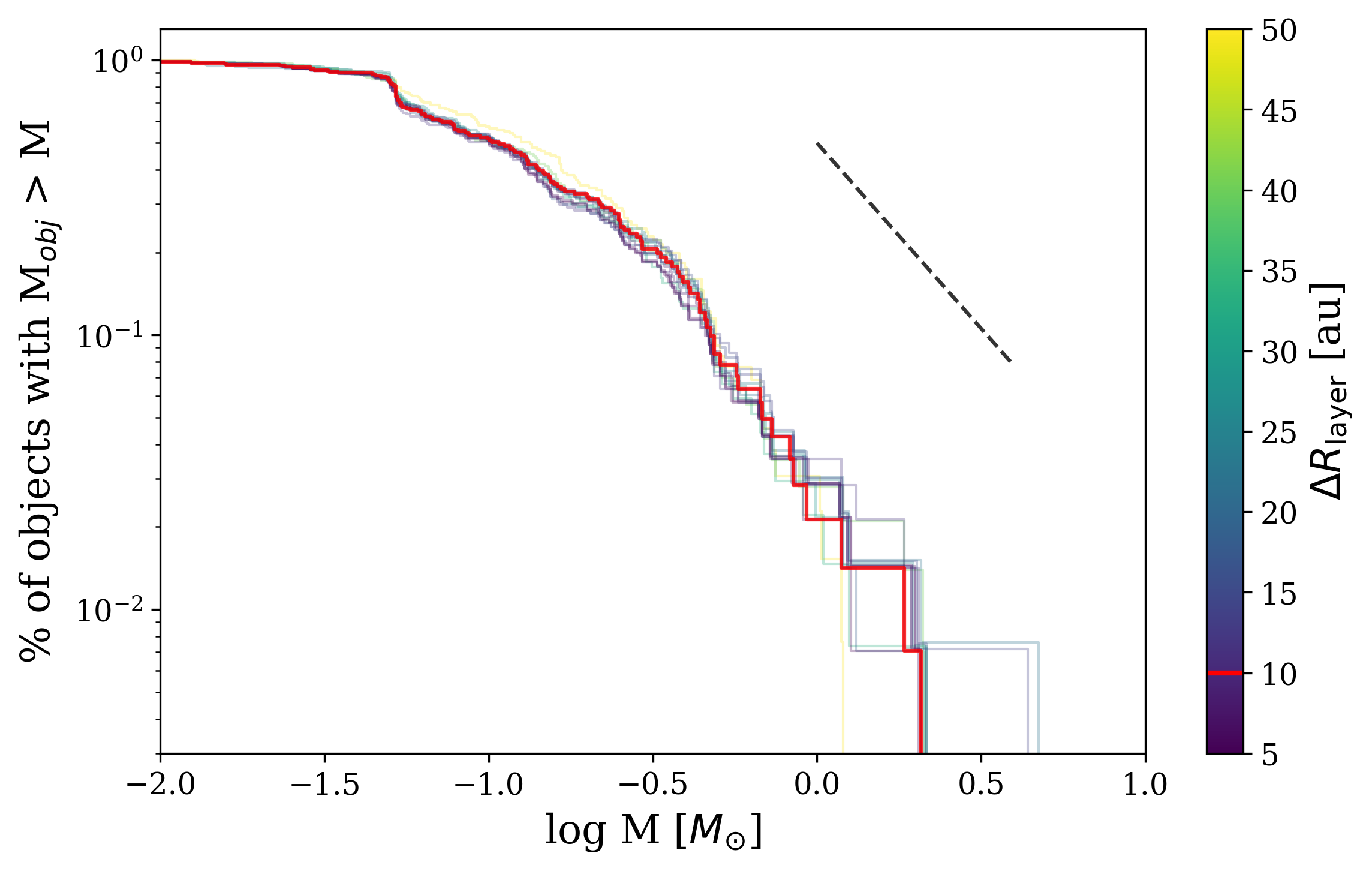}
    \caption{Mass distribution with respect to the convexity (top), elongation (middle), and layer thickness (bottom) parameters. 
    The colored lines correspond to the mass ccdf for various parameter values given by the side colorbars.
    The red lines show the mass ccdf for the reference parameter values. 
    The dashed black lines corresponds to Salpeter slope \(\alpha=-1.35\).}
    \label{fig-benchmark-mass-ccdf-wrt-struc-building}
\end{figure}

The effect of those parameters on the mass distribution of the extracted structures is given in Fig. \ref{fig-benchmark-mass-ccdf-wrt-struc-building}.
The mass distributions are plotted as complementary cumulative distribution functions (ccdf).
No clear trend appears for any of the parameters. 
The mass distributions remain stable with respect to the three tested parameters, with close distribution shapes and maximum mass values.
The extraction method shows a good robustness to the parameters, especially with the purpose of studying mass distributions.
More details are provided in Appendix \ref{appendix-benchmark-struc-building}.

\section{Discussion} \label{section-discussion}

We compared the structures obtained with our virial-based method \textit{vibes} with two extraction algorithms based on density.

\subsection{Density-based extraction algorithms} \label{discussion-subsection-density-algo}

The first extraction algorithm is \textit{hop} \citep{Eisenstein&Hut1998}. 
It assigns each cell above a given threshold \( n_{\rm outer} \) to a peak, by successively linking cells to their densest neighbor, until it reaches a local maximum. 
The cells that are linked to the same peak are grouped together. 
Groups linked by saddle densities higher than a given saddle parameter \( n_{\rm saddle} \) are merged. 
Groups whose density peak is lower than a peak parameter \( n_{\rm peak} \) are discarded. 
Groups whose cell number is lower than a minimum cell number fixed by the user are discarded as well.
We used the implemented ECOGAL wrapper \citep{Colman2024} of the \textit{hop} algorithm. 

The second extraction algorithm is \textit{dendrogram} \citep{Rosolowsky2008}. 
The method is similar to a watershed segmentation, starting from the density peaks and building a hierarchical tree. 
The leaves of the tree (i.e., the last contours without distinct substructures inside) are the structures we keep for this analysis to force focusing on peaks.
As for \textit{hop}, only cells above a given threshold \textit{min\_value} are considered. 
The density step \textit{min\_delta} between isocontours and the minimum number of cells per structure are fixed by the user.
We used the open-source Python package \texttt{astrodendro}\footnote{https://dendrograms.readthedocs.io}.

The extraction algorithms based on density are known to be sensitive to their input parameters \citep{Pineda2009, Smullen2020}.
We tested the effect of the density threshold parameter for both algorithms with variations from \(10^4\) cm\(^{-3}\), density threshold value used in \cite{Offner2025}, to \(10^7\) cm\(^{-3}\), which corresponds to the order of magnitude of the mean density observed in massive dense cores \citep{Louvet2014}.
For \textit{hop}, we kept a fixed peak factor of 2 and a saddle factor of 10, having then \( n_{\rm peak} = 2 ~n_{\rm outer} \) and \( n_{\rm saddle} = 10 ~n_{\rm outer} \) \citep{Colman2024}.
For \textit{dendrogram}, we set the step between density contours \textit{min\_delta} equal to \textit{min\_value}.
For both algorithms, the cell neighborhood was set to 16 neighbors and the minimum number of cell per structure was set to 50, corresponding to the default values in \textit{vibes}.
The contour step parameter \textit{min\_delta} is also critical, since interesting peaks may be missed if it is too large while noisy fluctuations may be kept as real peaks if it is too small. 
We did not study the effect of this parameter. 
The two-dimensional algorithm \textit{getsf} used by many observational core studies will be used for a comparison in a future work.

\subsection{Extraction comparison} \label{discussion-subsection-extraction-comparison}

\begin{figure}[!ht]
    \centering
    \includegraphics[width=9cm]{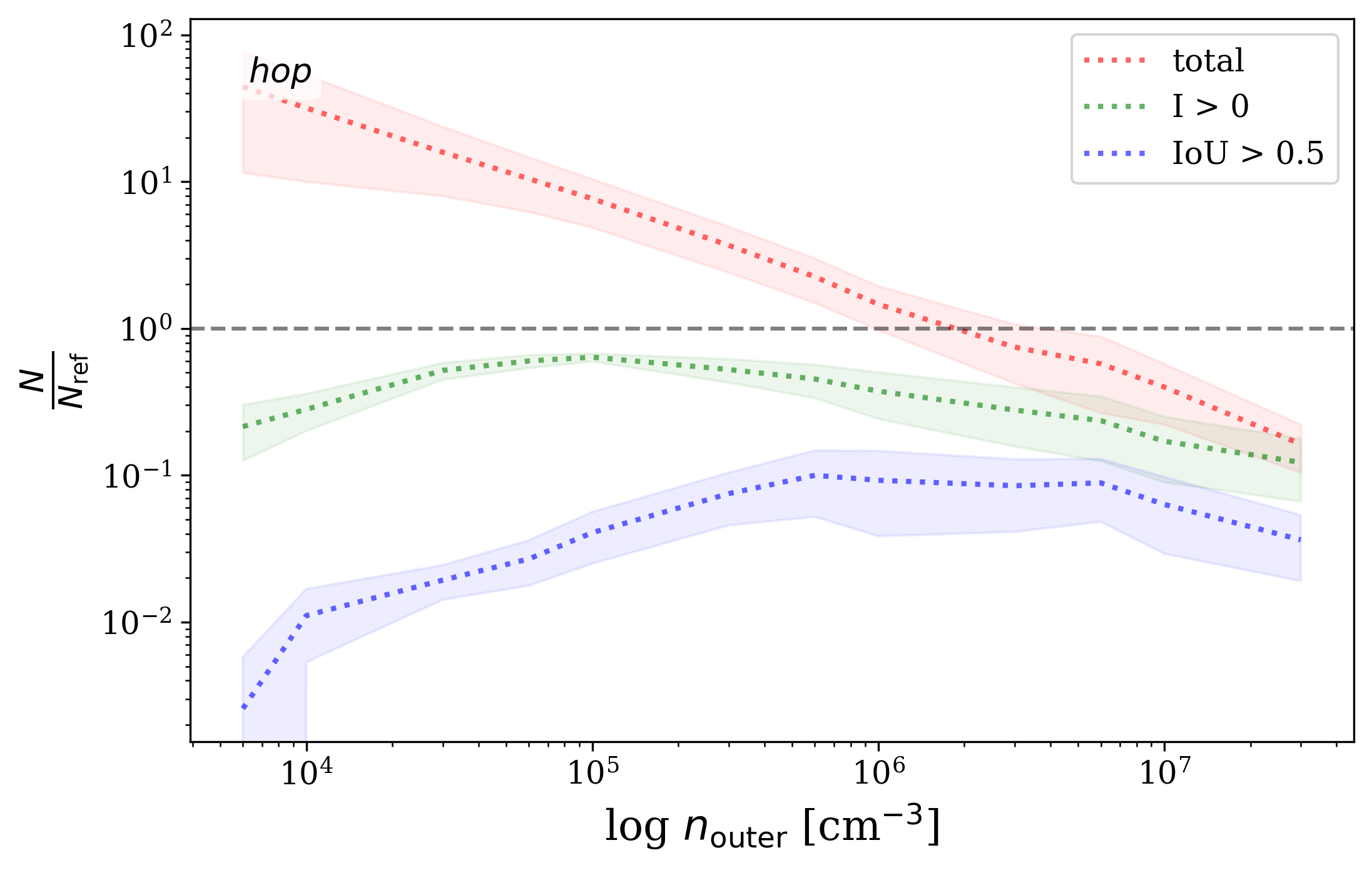}
    \includegraphics[width=9cm]{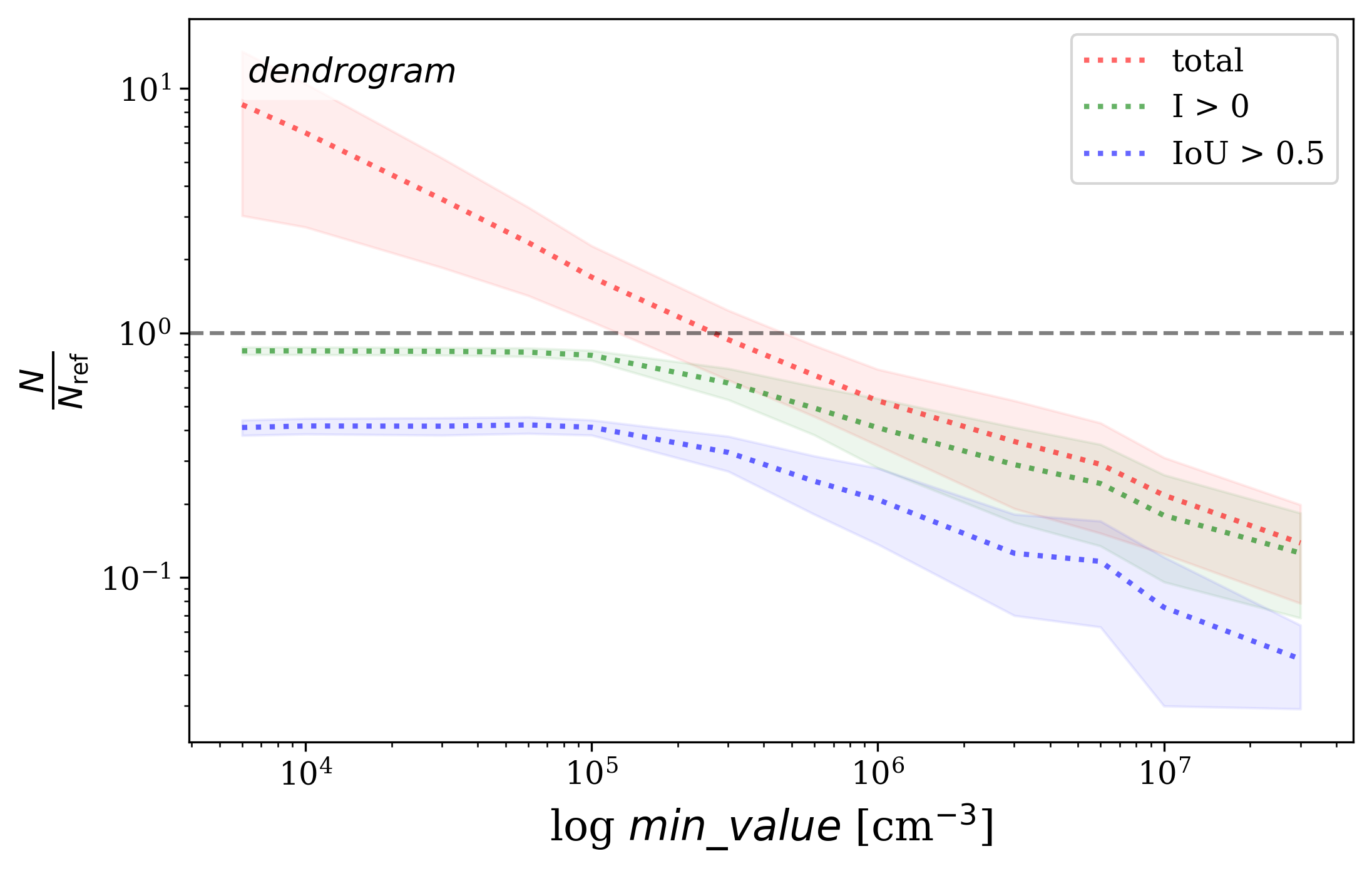}
    \caption{Relative number of extracted objects with respect to the density thresholds, \(n_{\rm outer}\) for \textit{hop} (top) and \textit{min\_value} for \textit{dendrogram} (bottom).
    The dotted red line gives the number of structures extracted at a given threshold value relative to the number of objects from the \textit{vibes} extraction.
    The dashed horizontal gray line shows \(N=N_{\rm ref}\).
    The dotted green and blue lines show the relative number of objects having respectively a nonzero intersection and intersection-over-union higher than 50\% with a \textit{vibes} object associated to the same density peak.
    The colored lines and intervals give the mean and deviation over three snapshots and several \textit{vibes} parameters.}
    \label{fig-discussion-struc-number-comparison}
\end{figure}

The comparison of the number of structures extracted with \textit{vibes}, \textit{hop} and \textit{dendrogram} is given in Fig. \ref{fig-discussion-struc-number-comparison}. 
It is plotted for different values of the threshold parameter, \(n_{\rm outer}\) for \textit{hop} and \textit{min\_value} for \textit{dendrogram}.
The colored intervals give the standard deviation over three snapshots and several \textit{vibes} parameters detailed in Appendix \ref{appendix-density-comparison-complements}.
The number of structures extracted by the density-based algorithms is highly sensitive to the choice of the threshold.
Regarding \textit{hop}, there is a large variation in the number of detected objects with respect to \(n_{outer}\): there is up to fifty times more extracted objects than with \textit{vibes} for \( n_{\rm outer} \) below \( 10^4\) cm\(^{-3} \), while it dramatically drops by two orders of magnitude for \( n_{\rm outer} \) higher than \( 10^7\) cm\(^{-3} \).
The number of detected structure decreases as the threshold parameter increases without convergence, as expected when noise is not a limiting factor. 
The recall of \textit{vibes} objects, given by the number of objects with nonzero intersection, is bad for low \( n_{\rm outer} \) values: almost all of the \textit{vibes} objects are merged into a single \textit{hop} structure and are considered as having a zero intersection with \textit{hop} objects.
The other extracted structures with \textit{hop} correspond to small density fluctuations around the main structure.
The recall then reaches a peak at \( n_{\rm outer} = 10^5 \) cm\(^{-3}\), and finally drops down for higher threshold values: the main structure splits and the number of matching objects between the \textit{vibes} and \textit{hop} increases, until \( n_{\rm outer} \) gets too high and the total number of \textit{hop} objects drops down to zero.
Regarding \textit{dendrogram}, the behavior is qualitatively the same for the total number of extracted objects, with a higher number of structures for low density-threshold values, and a drop to zero when it increases. 
However, the effect of the threshold parameter is weaker. 
There are much less extracted objects than with \textit{hop} for low \textit{min\_value} (up to ten times the \textit{vibes} number, against fifty times for \textit{hop}).
The recall is much better: about 90\% of the \textit{vibes} objects have a nonzero intersection with \textit{dendrogram} objects for low threshold, whereas it was almost zero for \textit{hop}.
In conclusion, many \textit{hop} or \textit{dendrogram} structures extracted at low-density are not gravitationally bound and we need to extract all sources above \(10^5\) cm\(^{-3}\) to be complete.

\begin{figure}[!ht]
    \centering
    \includegraphics[width=9cm]{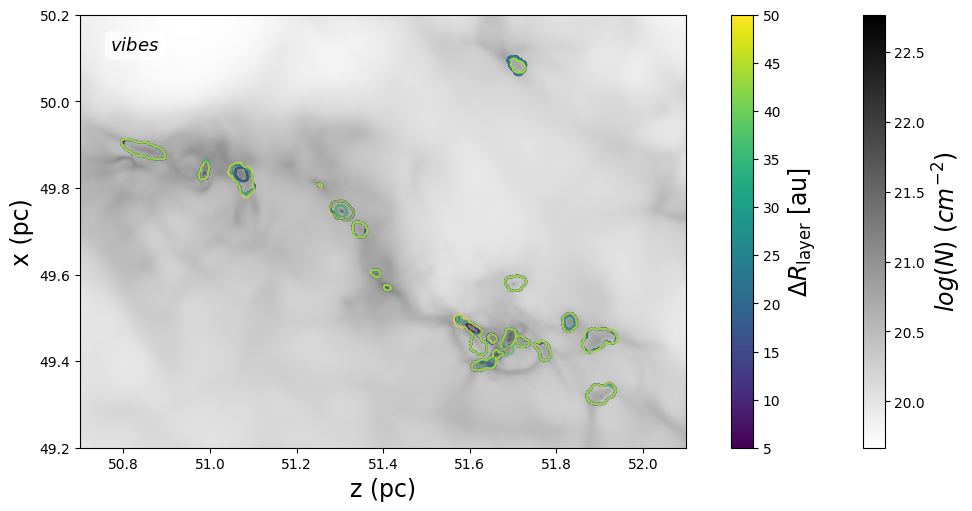}
    \includegraphics[width=9cm]{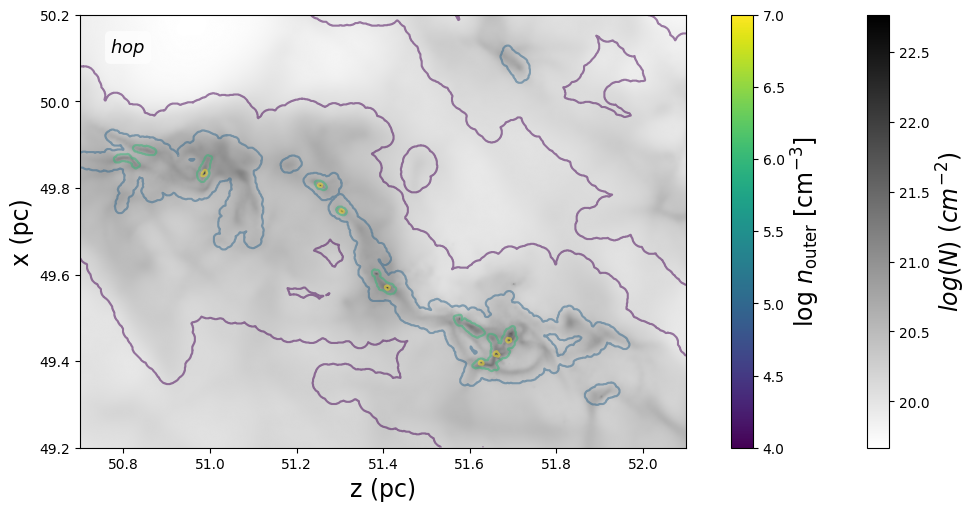}
    \includegraphics[width=9cm]{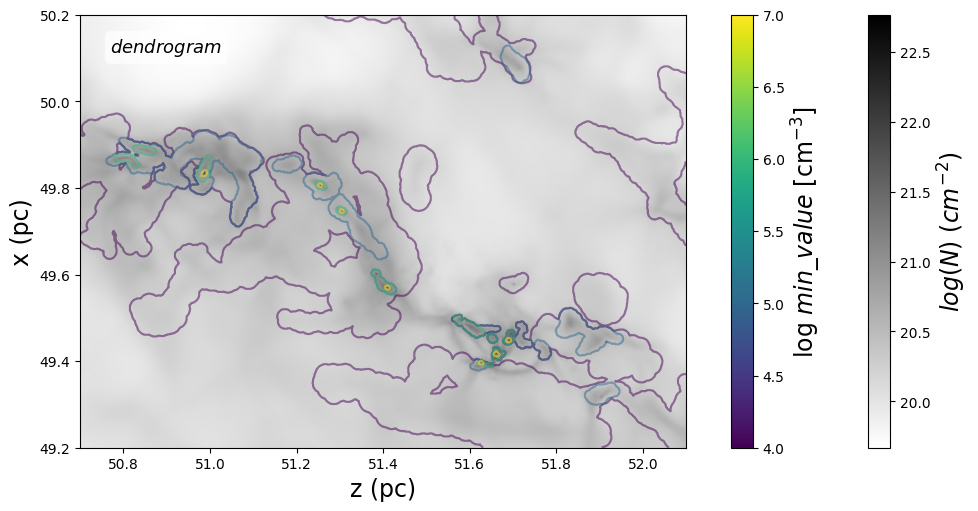}
    \caption{Column density map with the projected extractions obtained with \textit{vibes}, \textit{hop}, and \textit{dendrogram}, with respect to the parameter given by the colorbars.
    The lines are colored with respect to the parameter value corresponding to the extraction.
    The \textit{vibes} extractions (top) are plotted for various \(N_{\rm layer}\), from 5 to 40. 
    The \textit{hop} extractions (middle) are plotted for various \(n_{\rm outer}\), from \(10^4\) to \(10^7\) cm\(^{-3}\). 
    The \textit{dendrogram} extractions (bottom) are plotted for various \(min\_value\), from \(10^4\) to \(10^7\) cm\(^{-3}\).}
    \label{fig-discussion-column-density-map-comparison}
\end{figure}

For the following analysis of the structure properties, we focus on the snapshot at t = 4.93 Myr.
Figure \ref{fig-discussion-column-density-map-comparison} shows column-density maps with the projected extracted structures obtained with the three methods.
It shows the sensitivity of the extractions to the selected parameters.
The density thresholds \(n_{\rm outer}\) for \textit{hop} and \textit{min\_value} for \textit{dendrogram} vary from \(10^4\) to \(10^7\) cm\(^{-3}\).
For \textit{vibes}, the layer size (given by the number of layers, see Sect. \ref{benchmark-subsection-structure-building}) varies from 5 to 40 au.
Contours with different colors correspond to independent extractions.
The sensitivity of \textit{hop} and \textit{dendrogram} on the density threshold is striking, with large structures extracted for low threshold values that are split when the threshold increases.
\textit{Vibes} displays marginal variations with respect to the layer parameter and extracts consistently compact structures near the density peaks.
It converges better and extracts more compact structures at the chosen \(10^5\) cm\(^{-3}\) threshold.
At high density thresholds, the structures extracted with \textit{hop} and \textit{dendrogram} visually tend toward \textit{vibes} structures.
However, for such threshold, many structures are no longer extracted and the completeness is low.
\textit{Vibes} removes the reliance on an arbitrary density threshold and avoids the need to adjust this threshold from subregion to subregion to be able to recover all the structures, a practice that can introduce significant biases in the extraction.

\begin{figure}[!ht]
    \centering
    \includegraphics[width=9cm]{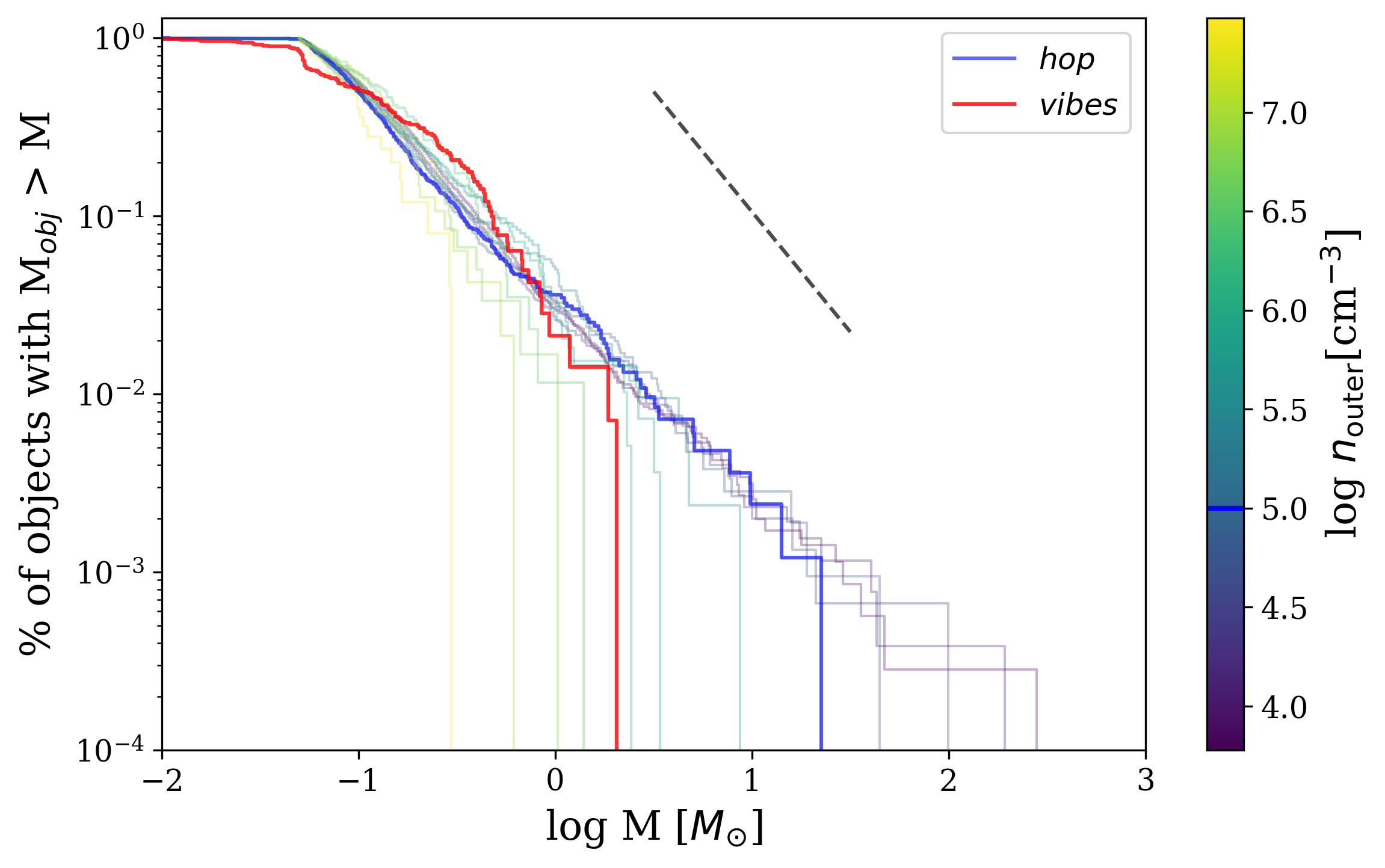}
    \includegraphics[width=9cm]{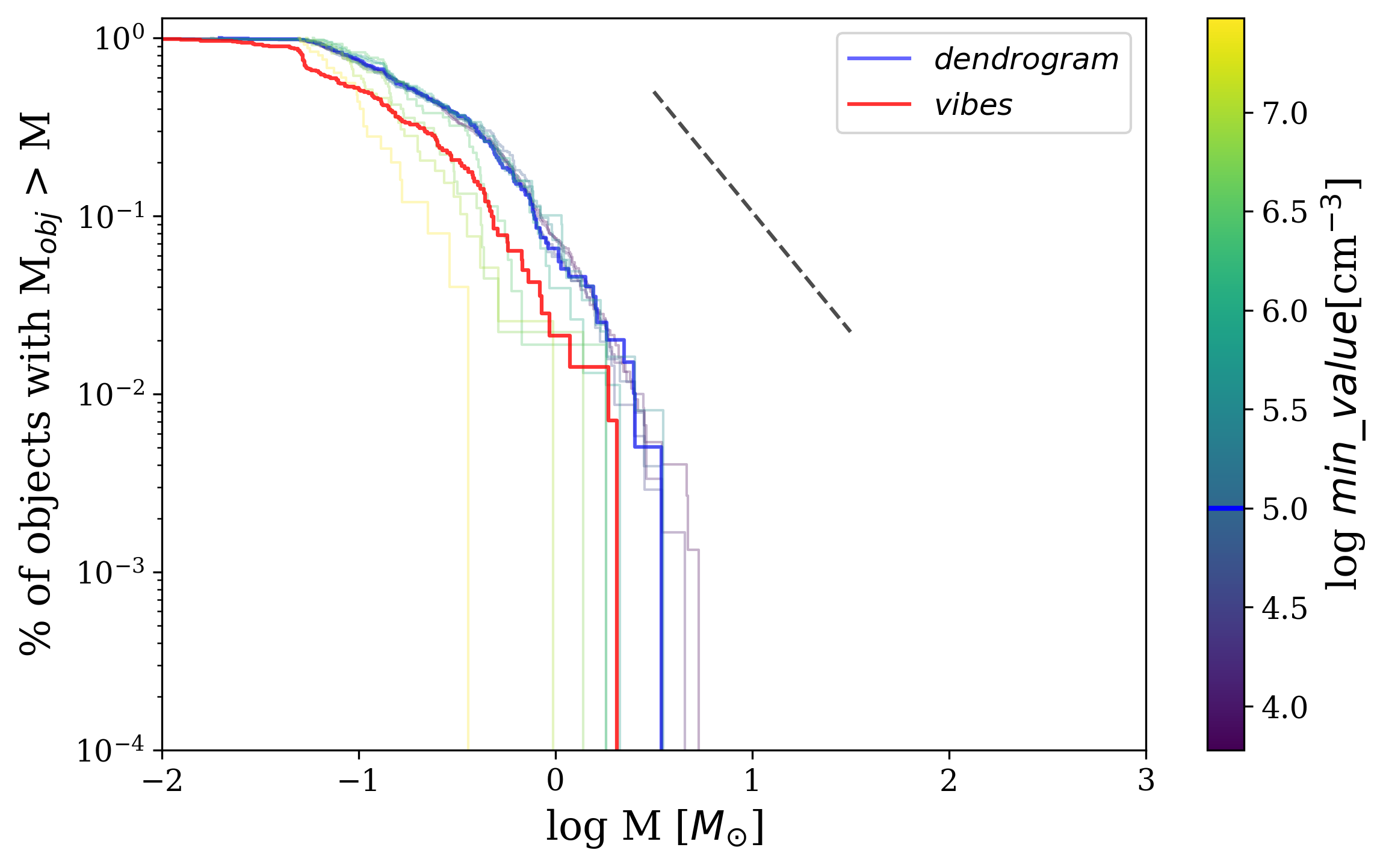}
    \caption{Mass distribution of the structures extracted with \textit{hop} (top) and \textit{dendrogram} (bottom) for different values of the density threshold parameter \(n_{\rm outer}\) and \textit{min\_value}, respectively. 
    The \textit{hop} and \textit{dendrogram} ccdf for the fixed threshold parameter at \(10^5\) cm\(^{-3}\) are given in blue.
    The thin colored lines correspond to the mass ccdf for different values of the algorithm density threshold.
    The \textit{vibes} mass ccdf is given in red. 
    The dashed black line corresponds to Salpeter slope \(\alpha=-1.35\).}
    \label{fig-discussion-ccdf-comparison}
\end{figure}

The comparison of the mass distributions between the \textit{vibes}, \textit{hop}, and \textit{dendrogram} extractions is given in Fig. \ref{fig-discussion-ccdf-comparison}.
This comparison is performed over one single snapshot, with fixed \textit{vibes} parameters.
The shape of the mass distribution is highly affected by the threshold parameter for both density-based algorithms.
The highest masses appear to be very sensitive to the threshold, especially regarding the \textit{hop} extraction: for the minimum tested values of \( n_{\rm outer} \), some objects are extracted having several hundred solar masses.
As observed in Fig. \ref{fig-discussion-column-density-map-comparison}, when the density threshold increases, such big structures are split and the tail of mass distribution gets steeper and steeper.
For \textit{dendrogram}, the effect is less visible but similar: the high-mass part of the distribution shifts toward lower values and the distribution is steeper when the threshold parameter increases.
If such algorithms can be interesting tracers of the density structure of the cloud, those results rise the question of assuming that the extracted structures have mass distributions that can be compared to the IMF.
Details on the sensitivity of the powerlaw fit are given in Appendix \ref{appendix-density-comparison-complements}. 

\begin{figure}[!ht]
    \centering
    \includegraphics[width=9cm]{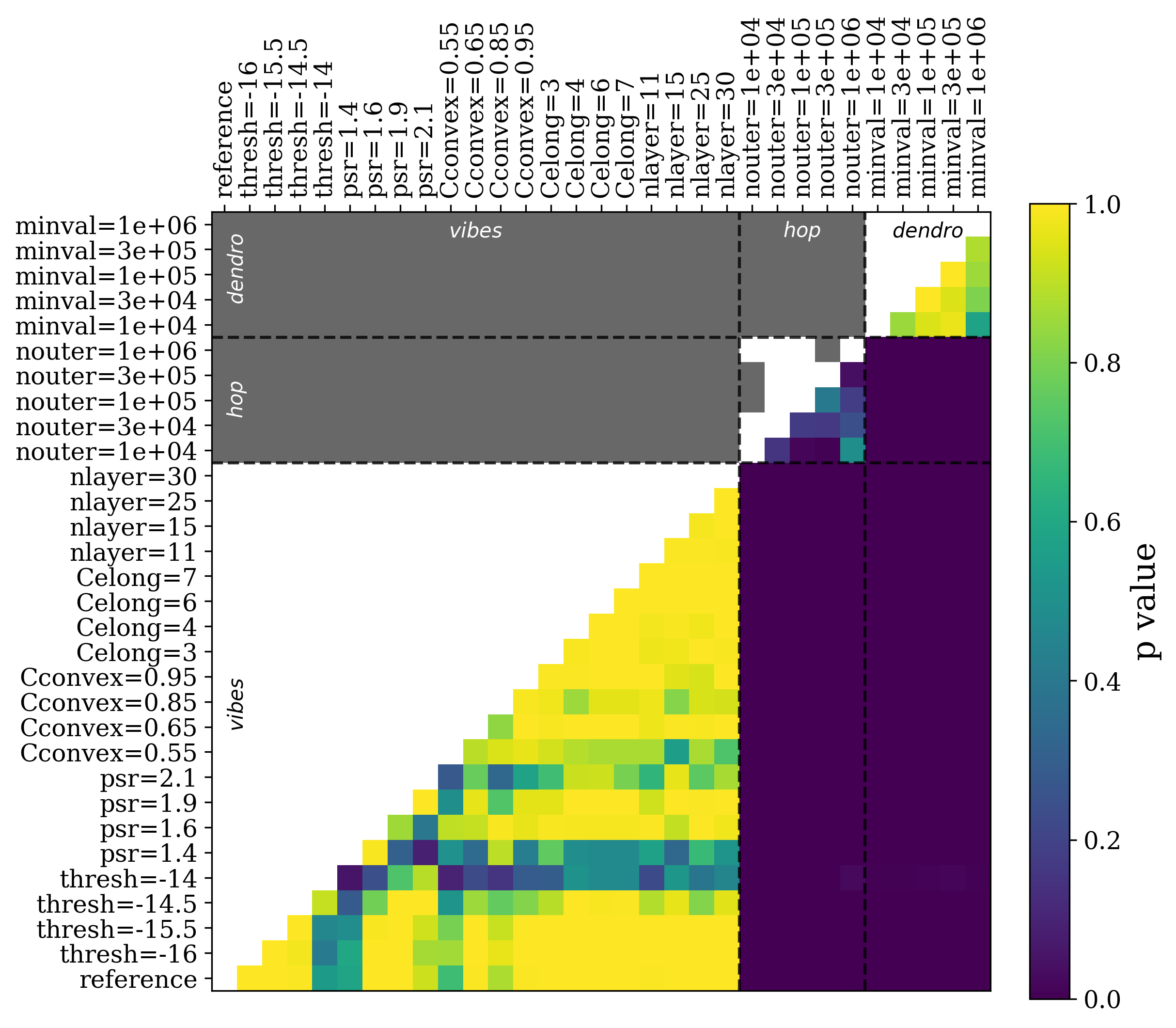}
    \caption{Matrix giving the p values for Kolmogorov-Smirnov tests comparing mass distributions obtained for different parameters.
    The lower triangle shows the p values and the upper triangle indicates whether the null hypothesis (i.e., same underlying mass distribution) is accepted (p > 0.05). 
    Darkened pixels correspond to rejection of the null hypothesis.
    The legend gives the parameter variation for each extraction algorithm (see Table \ref{appendix-table-vibes-params} for \textit{vibes}).
    The dashed black lines split the matrix with respect to the extraction tool: \textit{vibes}, \textit{hop}, and \textit{dendrogram}.}
    \label{fig-appendix-ks-test}
\end{figure}

In order to test the cross-relationship between mass distributions obtained varying the parameters of each algorithm, we performed a series of Kolmogorov-Smirnov tests.
We compared the \textit{hop} and \textit{dendrogram} extractions with different density thresholds to \textit{vibes} extractions with several different parameters (see Table \ref{appendix-table-vibes-params}).
The result of these tests is given in Fig. \ref{fig-appendix-ks-test}.
The incompatibility between the different extraction methods appears clearly, with p values under the 0.05 threshold. 
The \textit{vibes} and \textit{dendrogram} extractions are fully consistent to each other, the mass distributions remaining compatible at 5 \%.
This is not the case for \textit{hop} which has several mass distributions that are not compatible to each other when varying the density threshold.

For the following, the threshold parameters \(n_{\rm outer}\) for \textit{hop} and \textit{min\_value} for \textit{dendrogram} are fixed at \(10^5\) cm\(^{-3}\). 
This value corresponds for both algorithms to the best recall relative to \textit{vibes} structures according to Fig. \ref{fig-discussion-struc-number-comparison}, defining the recall as the highest number of nonzero intersection objects relative to the total number of extracted objects.
This density approximately corresponds to that of low-mass cores \citep{Motte2018b}.

\begin{figure}[!ht]
    \centering
    \includegraphics[width=9cm]{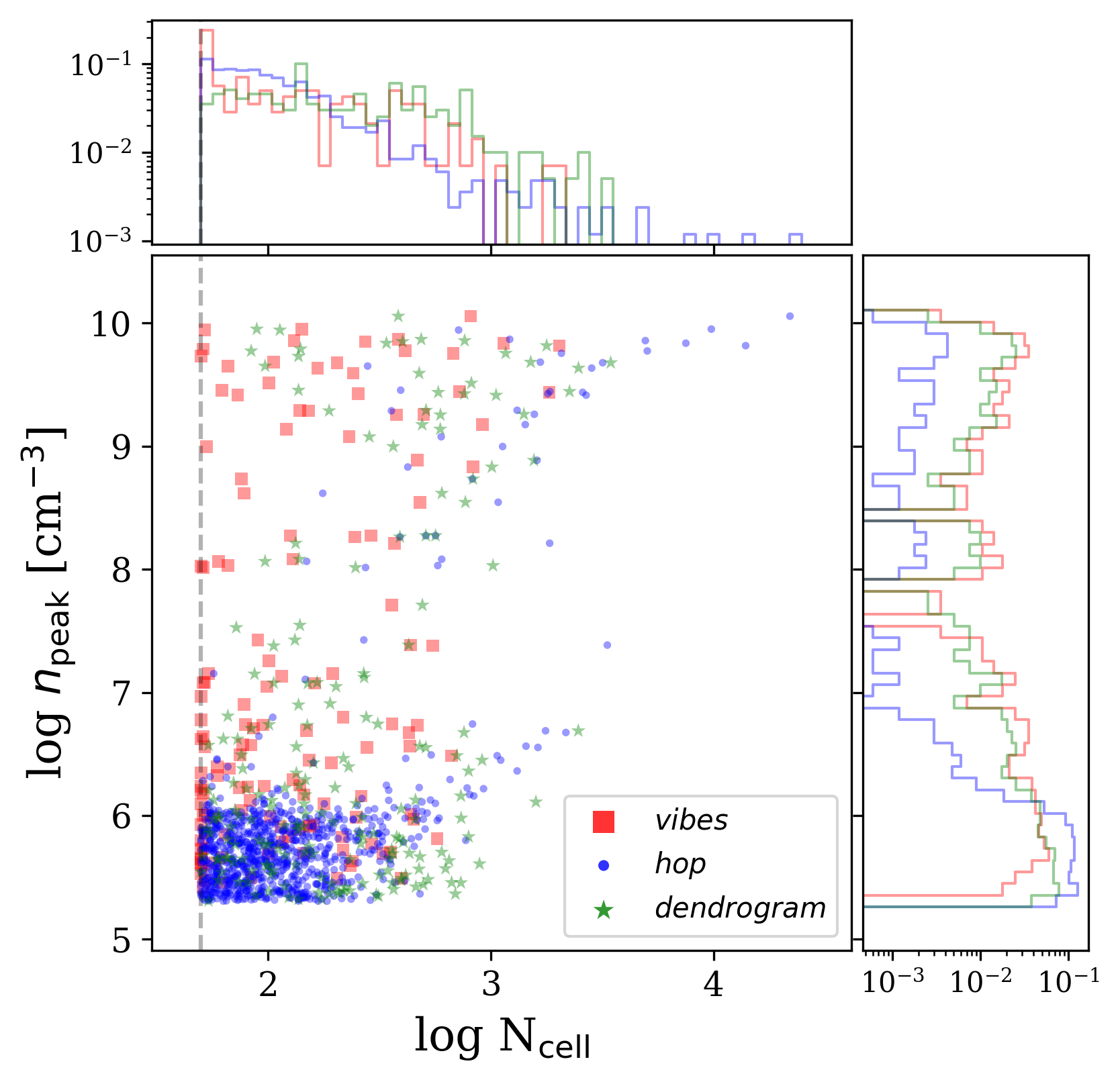}
    \caption{Peak density with respect to the number of cells per structure. 
    The red squares, blue dots, and green stars correspond respectively to \textit{vibes}, \textit{hop}, and \textit{dendrogram} objects. 
    The colors of the side histograms lines correspond to the scatter plot colors.
    The dashed vertical gray line corresponds to the lower limit of 50 cells.}
    \label{fig-discussion-peakdens-ncell-scatter}
\end{figure}

The peak density as a function of the number of cells per structure is given in Fig. \ref{fig-discussion-peakdens-ncell-scatter}. 
\textit{Vibes} structures with similar number of cells have a wide range of peak density: no clear correlation appears between those two quantities.
The distribution reaches a clear maximum at the minimum number of 50 cells (top histogram), which might represent the resolution limit of our method.
Regarding \textit{hop}, the overall distribution of cell numbers per structure is much less uniform.
Most of the extracted objects are located at low peak density and small number of cells. 
The few objects with density peaks higher than \(10^6\) cm\(^{-3}\) are defined with the highest number of cells.
As shown in Fig. \ref{fig-discussion-column-density-map-comparison}, the densest parts of the cloud will be merged into a few very massive objects containing many cells, while the low density regions will be gathered in numerous small structures with low-density peaks (bottom left cluster in Fig. \ref{fig-discussion-peakdens-ncell-scatter}).
The relative number of objects with high peak density is also much lower than for \textit{vibes} (see right histogram in Fig. \ref{fig-discussion-peakdens-ncell-scatter}).
On this plot, the density threshold is set to \(n_{outer} = 10^5\) cm\(^{-3}\): this imbalance between the few big massive high-density objects and the many low-density small objects should be accentuated for lower values of \(n_{\rm outer}\). 
Indeed, the maximum mass in Fig. \ref{fig-discussion-ccdf-comparison} rises up to hundreds of solar masses for the lowest density thresholds.
\textit{Dendrogram} shows an intermediate behavior, with a density peak distribution much closer to \textit{vibes} but still adding a high number of low-density objects.
It appears less sensitive to this minimum number of cells than the two other algorithms.

\begin{figure}[!ht]
    \centering
    \includegraphics[width=9cm]{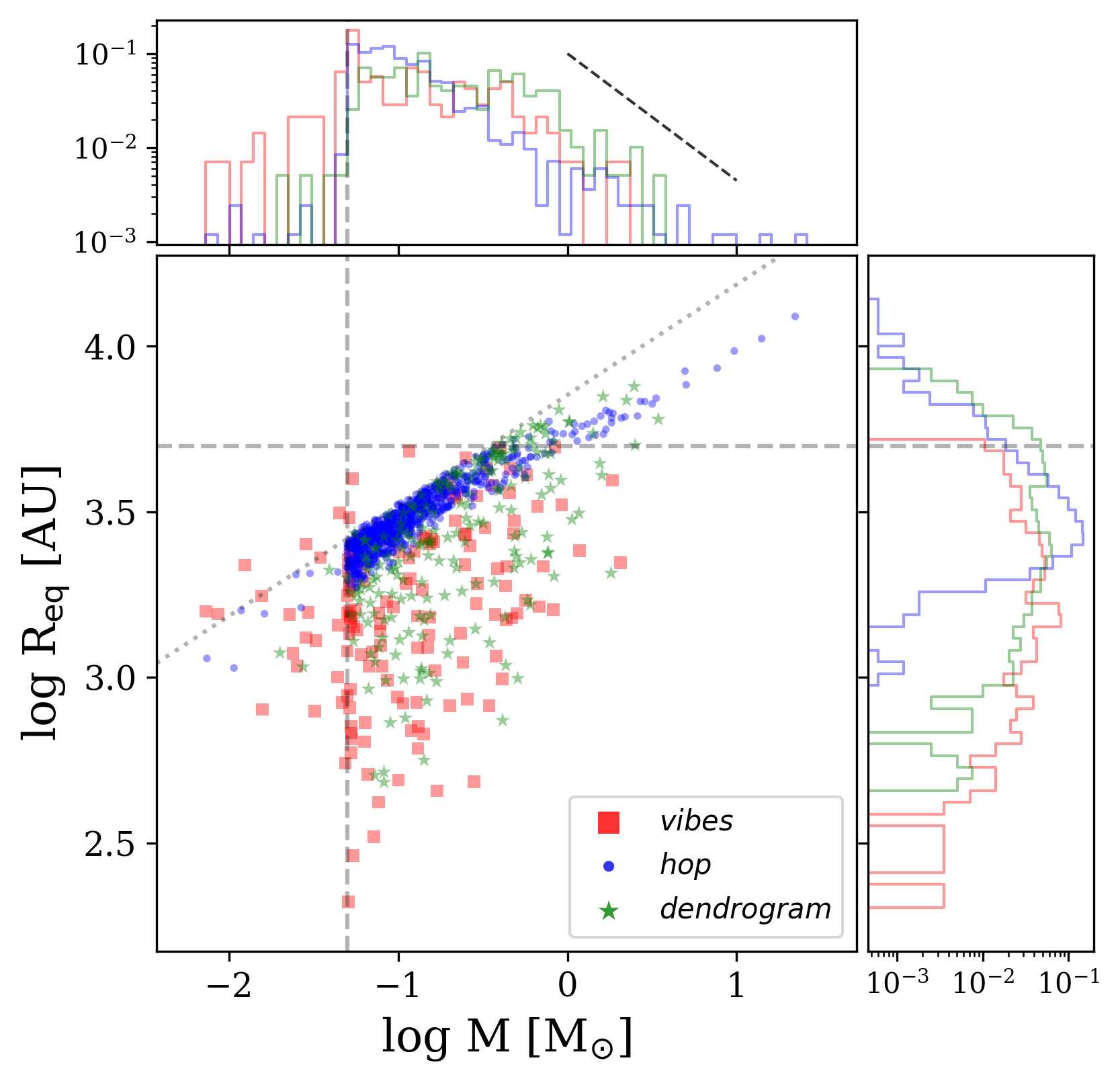}
    \caption{Equivalent radius with respect to the mass of the structures. 
    The red squares, blue dots, and green stars correspond respectively to \textit{vibes}, \textit{hop}, and \textit{dendrogram} objects. 
    The dashed vertical gray line gives to the theoretical lower mass limit of 0.05 \(M_\odot\) (corresponding to 50 cells). 
    The colors of the side histograms lines correspond to the scatter plot colors.
    The dashed horizontal gray line corresponds to the 5000 au maximum size for \textit{vibes} objects. 
    The dashed black line on the top histogram corresponds to Salpeter slope \(\alpha=-1.35\).
    The dotted slope gives \(R_{\rm eq} = \left( \frac{3}{4 \pi n_{\rm outer}} M \right)^{1/3}\) with \(n_{\rm outer} = 10^5\) cm\(^{-3}\).}
    \label{fig-discussion-req-mass-scatter}
\end{figure}

The mass-size diagram is given in Fig. \ref{fig-discussion-req-mass-scatter}. 
The equivalent radius is a direct proxy of the structure volume (Eq. \ref{eq-methods-equivalent-radius}).
Despite the clear maximum in the number of cells distribution observed in Fig. \ref{fig-discussion-peakdens-ncell-scatter} for \textit{vibes} structures, the distribution of physical sizes is rather uniform from 300 to 5000 au in equivalent radius.
This is explained by variations in the local resolution: in highly resolved regions of the simulation, 50 cells cover a much smaller physical volume than in poorly resolved regions.
The distribution of \textit{hop} objects physical size is very different. 
This is a direct consequence of the density-based resolution in the simulation: since Fig. \ref{fig-discussion-peakdens-ncell-scatter} shows small objects in cell number are concentrated in low-density regions (less resolved), the associated physical size is higher than objects with similar number of cells extracted in denser (more resolved) regions with \textit{vibes}.
While \textit{vibes} structures with low number of cells were extracted with very different density peaks (covering regions with different local resolution), all the small \textit{hop} structures are extracted in low density region, at the same local resolution. Thus, they all have the same physical size, given by the distribution maximum (right histogram in Fig. \ref{fig-discussion-req-mass-scatter}), and the same mean density.
On the contrary, the dense and resolved regions will be merged into huge and massive objects. 
This merging of peaks into wide regions that is inherent to \textit{hop} reduces the mean density of such extracted object and make it tend toward the threshold density.
The dependance between the local density and the number of cells for \textit{hop} objects shown in Fig. \ref{fig-discussion-peakdens-ncell-scatter} directly reflects in Fig. \ref{fig-discussion-req-mass-scatter} with a clear correlation between the physical size and the mass.
We note that the structures extracted with \textit{hop} (blue points) closely follow a \(M \propto R_{\rm eq}^3\) relation, where the mean density of all extracted structure is about constant, close to the selected threshold.
Although not necessarily relevant when looking for high-density regions potentially forming stars, the relation highlights a steep underlying density PDF \citep{Ballesteros-Paredes2012}. 
This relation is much less clear for \textit{vibes} and \textit{dendrogram}, whose structures have a mean density that varies significantly.
This density to cell number correlation does not exist for \textit{vibes}, while a physical size to mass correlation still appears.
\textit{Vibes} also appears to be able to cover different ranges of physical sizes, smaller than \textit{hop} in this example, but still higher than two-dimensional extraction algorithms such as \textit{getsf}, which tend to extract only localized peaks \citep{Menshchikov2021}.
\textit{Dendrogram} seems to have an intermediate behavior, with a distribution in physical size that is flatter than with \textit{hop}.
We note that some structures have a lower mass than the theoretical lower mass limit of 0.05 \(M_\odot\), which is directly the minimum number of cells (50) times the mass resolution (\(10^{-3}\) \(M_\odot\)).
This is caused by the stellar feedback implementation in STARFORGE, since particles injected in the simulation by the sinks can have a lower mass than the mass resolution.

\section{Conclusion} \label{conclusion}

We presented \textit{vibes}, a novel method to extract structures from numerical simulations.
The \textit{vibes} extraction tool has been developed from scratch in the format of a python package that is open to the community\footnote{https://gricad-gitlab.univ-grenoble-alpes.fr/chevasim/vibes}.
This method detects overdensities and sets their boundary using an adapted formulation of the virial theorem as a physical criterion. 
We performed different extractions on STARFORGE simulation snapshots, varying internal parameters of the \textit{vibes} algorithm. 
The algorithm sensitivity to those parameters appears to be low.
We compared the \textit{vibes} extraction with two other extraction algorithms based on density, \textit{dendrogram} and \textit{hop}. 
Those algorithms appear highly sensitive to their density threshold parameter, which induces a huge bias in the extractions. 
The choice of the parameters for \textit{hop} and \textit{dendrogram} strongly affects the shape of the CMF, questioning the relevance of a direct comparison to the shape of the IMF.
On the other hand, \textit{vibes} does not require an arbitrary choice of parameters that shape so critically the mass distribution of the structures obtained.
The setting of the structure boundary on a virial criterion rather than a density threshold improves the accuracy and homogeneity of the extracted structures.

Obtaining cores whose boundaries are physically motivated opens a new avenue for the understanding of the origin of the IMF.
Knowing what dominates the energy budget of cores, we will be able to link the general properties of the cloud with what actually triggers the collapse of cores in all types of environments.
This will enable us to understand the key differences between low and high-mass star-formation.
Having cores physically motivated will help tracking them in time in numerical simulations, enabling to assess their lifetime, their fluctuation and how they correlate individually and statistically to the sink particles they release, seeking for the origin of the stellar IMF.
The key question behind is: when the properties of a core (or a sample of cores) are measured in observations, how correlated are these measures with the star(s) the core(s) may form.

\begin{acknowledgements}

The authors would like to thank the anonymous referee for his very useful recommendations.
The authors acknowledge M. Grudi\'c for the STARFORGE data provided, the DAp at CEA Saclay (in particular P. Hennebelle and D. Chapon) for the RAMSES data provided, the CRAL team (especially B. Commerçon) and M. Krumholz, for the very helpful discussions.
The authors acknowledge support from the ANR COSMHIC grant (ANR-20-CE31-0009), and from the Thematic Actions “Physique et Chimie du Milieu Interstellaire” (ATPCMI) and "Physique Stellaire" (ATPS) of CNRS/INSU.
NB acknowledges support from the ANR BRIDGES grant (ANR-23-CE31-0005).
MG acknowledges funding from MICIU/AEI/10.13039/501100011033 and FEDER, EU through project TACOS ( PID2023-146635NA-I00).
The authors also acknowledge the support of the European Research Council (ERC) via the ERC Synergy Grant ECOGAL (grant 855130).

\end{acknowledgements}

\bibliographystyle{aa}
\bibliography{bibfile}

\begin{appendix}

\section{Virial theorem derivation} \label{appendix-virial-theorem-derivation}

The virial theorem provides the acceleration term for the mass distribution of a known structure of volume $\mathcal{V}$ as a sum of various energy contributions.
It is directly derived from the mass conservation equation: 
\begin{equation} \label{eq-appendix-mass-conservation}
    \frac{\partial \rho}{\partial t} = - \nabla \cdot (\rho \mathbf{v})
\end{equation}
, and the momentum conservation equation:
\begin{equation} \label{eq-appendix-momentum-conservation}
    \frac{\partial \rho \mathbf{v}}{\partial t} = - \nabla \cdot \left( \mathbf{\Pi} - \mathbf{T_M} \right) - \rho \nabla \phi
\end{equation}
, where \(\rho\) is the mass density, \(\mathbf{v}\) is the velocity, \(\phi\) the gravitational potential. 
$\Pi$ is the fluid pressure tensor:
\begin{equation} \label{eq-appendix-PI-tensor}
    \mathbf{\Pi} = \rho \mathbf{v} \otimes \mathbf{v} + P \mathbf{I}
\end{equation}
, and $T_M$ is the Maxwell stress tensor:
\begin{equation} \label{eq-appendix-TM-tensor}
    \mathbf{T_M} = \frac{1}{4 \pi} \left( \mathbf{B} \otimes \mathbf{B} - \frac{B^2}{2} \mathbf{I} \right)
\end{equation}
, where \(\mathbf{B}\) is the magnetic field, \(P\) the thermal pressure, and \(\mathbf{I}\) is the identity tensor.

The momentum of inertia of a structure is defined by:
\begin{equation} \label{eq-appendix-momentum-inertia}
    I = \int_\mathcal{V} \rho r^2 ~dV
\end{equation}

In the Eulerian frame, the structure is assumed to be time-independent i.e., the volume $\mathcal{V}$ and the position vectors $\mathbf{r}$ are not functions of time.
The time derivative of the momentum of inertia is given by:
\begin{equation} \label{eq-appendix-momentum-inertia-dot-1}
    \dot{I} = \int_\mathcal{V} \frac{\partial \rho}{\partial t} r^2 ~dV
\end{equation}
Injecting the mass conservation equation (Eq. \ref{eq-appendix-mass-conservation}), Eq. \ref{eq-appendix-momentum-inertia-dot-1} becomes:
\begin{equation} \label{eq-appendix-momentum-inertia-dot-2}
    \dot{I} = - \int_\mathcal{V} \nabla \cdot (\rho \mathbf{v}) r^2 ~dV
\end{equation}
Bringing the $r^2$ term inside the divergence:
\begin{equation} \label{eq-appendix-IPP}
    - \nabla \cdot (\rho \mathbf{v}) r^2 = - \nabla \cdot (\rho \mathbf{v} r^2) + 2 \rho \mathbf{v} \cdot\mathbf{r}
\end{equation}
, and using the divergence theorem:
\begin{equation} \label{eq-appendix-divergence-theorem}
    \int_\mathcal{V} \nabla \cdot \mathbf{a} ~dV = \int_\mathcal{S} \mathbf{a} \cdot \mathbf{dS}
\end{equation}
, Eq. \ref{eq-appendix-momentum-inertia-dot-2} becomes:
\begin{equation} \label{eq-appendix-momentum-inertia-dot-3}
    \dot{I} = - \int_\mathcal{S} ( \rho \mathbf{v} r^2 ) \cdot \mathbf{dS} + 2 \int_\mathcal{V} \rho \mathbf{v} \cdot \mathbf{r} ~dV
\end{equation}
With the same hypothesis of time-independent volume, the second derivative of the momentum of inertia is given by:
\begin{equation} \label{eq-appendix-momentum-inertia-ddot-1}
    \ddot{I} = - \int_\mathcal{S} \left( \frac{\partial \rho \mathbf{v}}{\partial t} r^2 \right) \cdot \mathbf{dS} + 2 \int_\mathcal{V} \left( \frac{\partial \rho \mathbf{v}}{\partial t} \right) \cdot \mathbf{r} ~dV
\end{equation}
Injecting the momentum conservation equation (Eq. \ref{eq-appendix-momentum-conservation}) and dividing by $\frac{1}{2}$ for convenience, Eq. \ref{eq-appendix-momentum-inertia-ddot-1} becomes:
\begin{equation} \label{eq-appendix-momentum-inertia-ddot-2}
    \frac{1}{2} \ddot{I} = - \int_\mathcal{V} \left[ \nabla \cdot (\mathbf{\Pi} - \mathbf{T_M}) - \rho \nabla \phi \right] \cdot \mathbf{r} ~dV - \frac{1}{2} \int_\mathcal{S} \left( \frac{\partial \rho \mathbf{v}}{\partial t} r^2 \right) \cdot \mathbf{dS}
\end{equation}
The tensor terms can be expressed as a sum of a volume and a surface contribution thanks to the divergence theorem:
\begin{align} \label{eq-appendix-divergence-theorem-tensor}
    \int_\mathcal{V} ( \nabla \cdot \mathbf{\Pi} ) \cdot \mathbf{r} ~dV 
    &= \int_\mathcal{V} ( \partial_i \Pi_{ij} ) r_j ~dV \nonumber \\
    &= \int_\mathcal{V} ( \partial_i ( \Pi_{ij} r_j ) - \Pi_{ij} (\partial_i r_j ) ) ~dV \nonumber \\
    &= \int_\mathcal{V} ( \partial_i ( \Pi_{ij} r_j ) - \Pi_{ij} \delta_{ij} ) ~dV \nonumber \\
    &= \int_\mathcal{V} ( \nabla \cdot ( \mathbf{\Pi} \cdot \mathbf{r} ) - Tr\mathbf{\Pi} ) ~dV \nonumber \\
    &= \int_\mathcal{S} ( \mathbf{\Pi} \cdot \mathbf{r} ) \cdot \mathbf{dS} - \int_\mathcal{V} Tr\mathbf{\Pi} ~dV
\end{align}
Thus, the virial theorem can be written as:
\begin{equation} \label{eq-appendix-virial-theorem}
    \frac{1}{2} \ddot{I} = W + T + K + M + \dot{\Phi}
\end{equation}

In our calculations, we compute separately all the following terms:
\begin{align} \label{eq-appendix-virial-theorem-terms-bis}
    &W = - \int_\mathcal{V} \rho \nabla \phi \cdot \mathbf{r} ~dV \nonumber\\
    &T = \underbrace{ \left( \int_\mathcal{V} 3 P ~dV \right) }_{T_V} + \underbrace{ \left( - \int_\mathcal{S} P ~\mathbf{r} \cdot \mathbf{dS} \right) }_{T_S} \nonumber\\
    &K = \underbrace{ \left( \int_\mathcal{V} \rho (v')^2 ~dV \right) }_{K_V} + \underbrace{ \left( - \int_\mathcal{S} \left( \rho \mathbf{v} \otimes \mathbf{v} \cdot \mathbf{r} \right) \cdot \mathbf{dS} \right) }_{K_S} \nonumber\\
    &M = \underbrace{ \left( \int_\mathcal{V} \frac{B^2}{2 \mu_0} ~dV \right) }_{M_V} + \underbrace{ \left( \int_\mathcal{S} \left( \mathbf{T_M} \cdot \mathbf{r} \right) \cdot \mathbf{dS} \right) }_{M_S} \nonumber\\
    &\dot{\Phi} = - \frac{1}{2} \frac{d}{dt} \int_\mathcal{S} r^2 \rho \mathbf{v} \cdot \mathbf{dS} 
\end{align}
The volume and surface contributions are denoted as \(X_V\) and \(X_S\), respectively (\(X\) being \(T\), \(K\) or \(M\)) and are computed separately.
Since we are using single simulation snapshots, no information about the time derivative is available here.
With the same approximation of volume and position being constant with time, \(\dot{\Phi}\) can be written as given by:
\begin{equation} \label{eq-appendix-dotphi-2}
    \dot{\phi} = - \frac{1}{2} \int_\mathcal{S} r^2 \frac{\partial \rho \mathbf{v}}{\partial t} \cdot \mathbf{dS}
\end{equation}
The last term \(\dot{\Phi}\) is then estimated injecting the momentum conservation equation (Eq. \ref{eq-appendix-momentum-conservation}) in Eq. \ref{eq-appendix-dotphi-2}, leading to the expression given by:
\begin{equation} \label{eq-appendix-dotphi-3}
    \dot{\phi} = \frac{1}{2} \int_\mathcal{S} r^2 \left( \nabla \cdot \left( \mathbf{\Pi} - \mathbf{T_M} \right) + \rho \nabla \phi \right) \cdot \mathbf{dS}
\end{equation}

\section{Gradient estimation} \label{appendix-gradient-calculation}

All the derivatives are estimated with the method used in GIZMO (\cite{Hopkins2015}).
For a scalar field \( f \), a vector field \( \mathbf{f} \) or a tensor field \( \mathbf{F} \), the derivative are given by:
\begin{equation} \label{eq-methods-derivative-scalar-field}
    (\mathbf{\nabla} f)_i^\alpha \equiv \sum_{j=1}^{32} \sum_{\beta=1}^3 \left( f_j - f_i \right) \mathbf{B}_i^{\alpha \beta} \left( \mathbf{x}_j - \mathbf{x}_i \right)^{\beta} w_{ij} + \mathcal{O}(h_i^2)
\end{equation}
\begin{equation} \label{eq-methods-derivative-vector-field}
    (\mathbf{\nabla} \cdot \mathbf{f})_i \equiv \sum_{j=1}^{32} \sum_{\alpha=1}^3 \sum_{\beta=1}^3 \left( \mathbf{f}_j - \mathbf{f}_i \right)^{\alpha} \mathbf{B}_i^{\alpha \beta} \left( \mathbf{x}_j - \mathbf{x}_i \right)^{\beta} w_{ij} + \mathcal{O}(h_i^2)
\end{equation}
\begin{equation} \label{eq-methods-derivative-tensor-field}
    (\mathbf{\nabla} \cdot \mathbf{F})_i^{\mu} \equiv \sum_{j=1}^{32} \sum_{\alpha=1}^3 \sum_{\beta=1}^3 \left( \mathbf{F}_j - \mathbf{F}_i \right)^{\mu \alpha} \mathbf{B}_i^{\alpha \beta} \left( \mathbf{x}_j - \mathbf{x}_i \right)^{\beta} w_{ij} + \mathcal{O}(h_i^2)
\end{equation}
, with \( \mathbf{B}_i = \mathbf{E}_i^{-1} \), where \( \mathbf{E}_i^{\alpha \beta} = \sum_j \left( \mathbf{x}_j - \mathbf{x}_i \right)^{\beta} \left( \mathbf{x}_j - \mathbf{x}_i \right)^{\beta} w_{ij} \) is a weighted distance matrix.
The normalized weights \( w_{ij} \) are given by the cubic spline kernel function \( W\left( \left| \mathbf{x} - \mathbf{x}_i \right|, h_i \right) \) (see Eq. \ref{eq-methods-distance-weights}).
\begin{equation} \label{eq-methods-distance-weights}
    w_{ij} = \frac{ W\left( \left| \mathbf{x}_i - \mathbf{x}_j \right|, h_i \right) }{ \sum_k{ W\left( \left| \mathbf{x}_i - \mathbf{x}_k \right|, h_i \right)} }
\end{equation}
The cubic spline kernel \citep{Monaghan&Lattanzio1985} is given by:
\begin{equation} \label{eq-cubic-spline-kernel}
    W(q_{ij}, h_i) = \frac{1}{\pi  h_i^3}
    \begin{cases}
        1 + \frac{3}{4} q_{ij}^2 (q_{ij}-2)   & \text{if} ~ 0 \leq q_{ij} < 1 \\
        \frac{1}{4} (2-q_{ij})^3              & \text{if} ~ 1 \leq q_{ij} < 2 \\
        0                           & \text{if} ~ 1 \leq q_{ij}
    \end{cases}
\end{equation}
, where \( h_i \) is the kernel size and \( q_{ij} = \frac{ \left| \mathbf{x_i} - \mathbf{x}_j \right| }{h_i} \).

For Lagrangian codes, the smoothing length \(h_i\) is directly computed by the code itself. For Eulerian codes, we manage not-regular grids such as AMR by building an artificial smoothing length for each cell solving:
\[ \frac{4}{3} \pi h_i^3 n(h_i) = N_{neigh} \]
, where \(N_{neigh}\) is a fixed parameter set to 32 and \( n(h_i) = \sum_j{W ( \left| \mathbf{r}_j - \mathbf{r}_i \right|, h_i} ) \).
This enables to apply the same derivative estimation process for any kind of code.

\section{Computational test} \label{appendix-test-virial-theorem}

\begin{figure}[!ht]
    \centering
    \includegraphics[width=9cm]{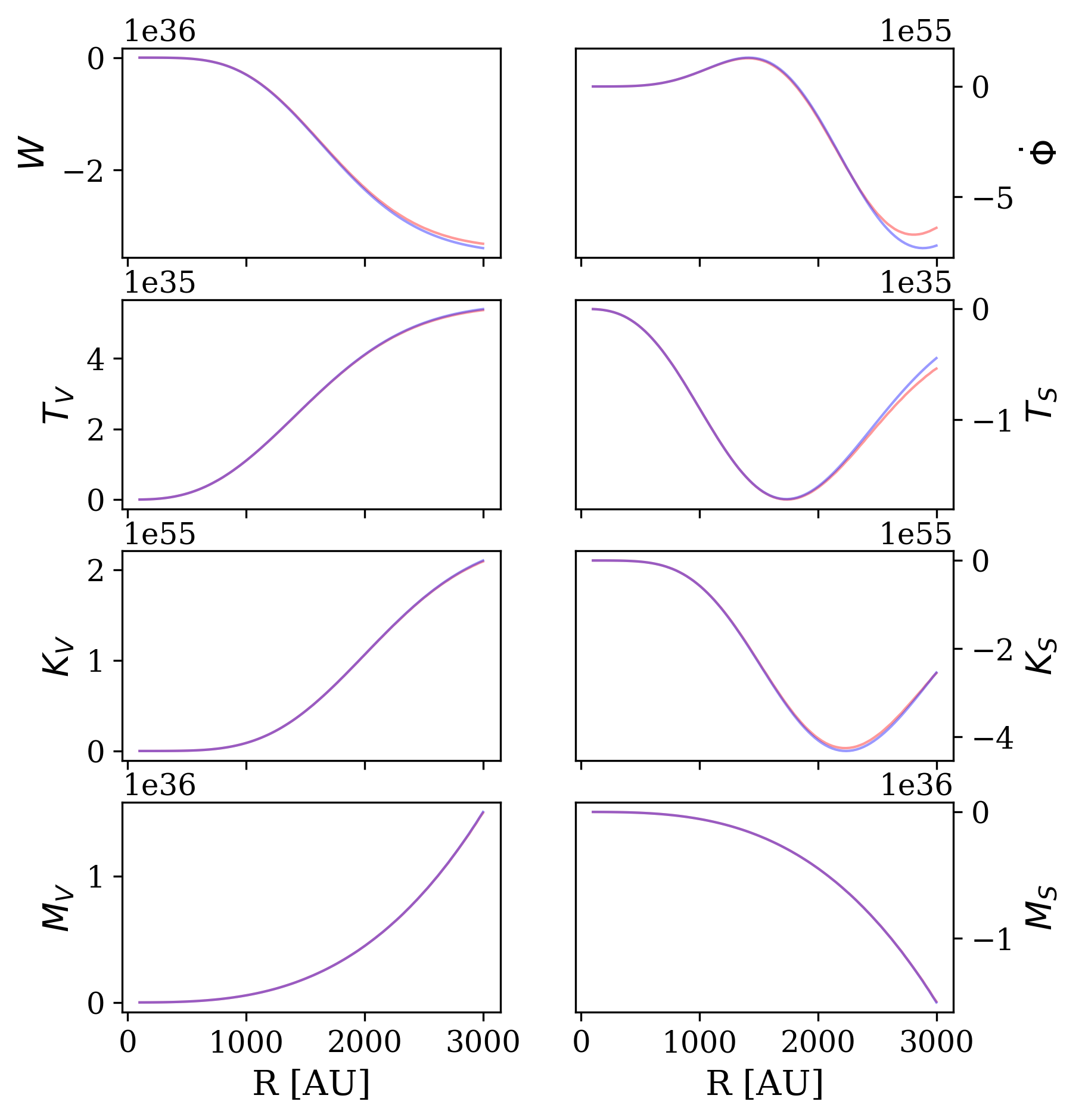}
    \caption{Comparaison between the numerical and analytical solutions for each term of the virial theorem. 
    The red lines give the solution computed by \textit{vibes}. 
    The blue lines give the analytical solution.}
    \label{fig-appendix-test-virial-computation}
\end{figure}

We run a test of our computation of the virial terms, on the following basic configuration:
\begin{itemize}
    \item a box of size 10 kAU, with coordinates normalized between 0 and 1, with center set to (0.5, 0.5, 0.5).
    \item a gaussian density profile \(\rho(r) = \rho_0 ~\exp \left(-\frac{r^2}{2\sigma_0^2}\right)\), with deviation \(\sigma = 0.1\) in box size and maximum density \(\rho_0 = 1e-13\) kg.m\(^{-3}\).
    \item \( 10^6 \) particles, randomly drawn from a gaussian distribution with same deviation than the density profile, so the local number of particle is proportionnal to the physical density.
    \item a constant temperature T = 10 K.
    \item an isothermal equation of state \( P = \rho c_s^2 = \rho \frac{k_B T}{\mu m_H}\), with \(\mu = 2.35\).
    \item a velocity field being the sum of a radial component and a rotational component around the z axis \( \mathbf{v} = c_{rad} \mathbf{r} + c_{rot} \mathbf{\hat{z}} \times \mathbf{r} \), with \(c_{rad} = - 0.01\) s\(^{-1}\) and \(c_{rot} = 0.01\) s\(^{-1}\).
    \item a constant magnetic field along the z axis \( \mathbf{B} = B ~\mathbf{\hat{z}} \) with \(B=10^{-7}\) T.
\end{itemize}
We compute the virial terms for a perfect ball, centered on (0.5, 0.5, 0.5), with radius from 0.01 to 0.3 in box size (from 100 to 3000 AU).
The comparison between the analytical solution and each component of the virial theorem (see Eq. \ref{eq-appendix-virial-theorem-terms-bis}) is given in Fig. \ref{fig-appendix-test-virial-computation}.
We note that the deviation increases for larger radii, where the density of points decreases.

\section{Flag classification} \label{appendix-flags}

A summary of the flag classification is given in Table \ref{appendix-table-flags}.
\begin{table}[!ht]
    \caption{Flag classification.}
    \label{appendix-table-flags}
    \centering
    \begin{tabular}{l c c r}
        \hline
        Flag & Point & \(\Delta R\) & \(\Delta R\) [AU] \\
        \hline
        \hline
        1 & Energy minimum & \( 10 \Delta ~R_{\rm base} \) & 2000\\
        2 & Energy minimum & \( 5 \Delta ~R_{\rm base} \) & 1000\\
        3 & Energy minimum & \( 3 \Delta ~R_{\rm base} \) & 600\\
        4 & Energy minimum & \( 2 \Delta ~R_{\rm base} \) & 400\\
        5 & Energy minimum & \( \Delta ~R_{\rm base} \) & 200\\
        6 & Derivative zero & / & /\\
        7 & Derivative extremum & \( 10 \Delta R_{\rm base} \) & 2000\\
        8 & Derivative extremum & \( 5 \Delta R_{\rm base} \) & 1000\\
        9 & Derivative extremum & \( 3 \Delta R_{\rm base} \) & 600\\
        10 & Derivative extremum & \( 2 \Delta R_{\rm base} \) & 400\\
        11 & Derivative extremum & \( \Delta R_{\rm base} \) & 200\\
        12 & Second derivative zero & / & /\\
        \hline
    \end{tabular}
\end{table}
In the search for inflection points, all the derivative extrema are not considered.
Only the derivative maxima when the energy is negative and the minima when the energy is positive are kept. 
Those correspond to "flat" inflection points rather than maximum slopes.

\section{Benchmark complements} \label{appendix-benchmark-complements}

\subsection{Peak sorting parameters} \label{appendix-benchmark-peak-sorting}

As in Sect. \ref{benchmark-subsection-structure-building}, we studied the effect of the two parameters defining the peak sorting process, the peak threshold and peak-to-saddle ratio parameters (see Sect. \ref{methods-subsection-peak-sorting}), by running several extractions with different values on three different snapshots. 
The peak threshold parameter sets the minimum value for peaks used as seeds for the extraction process. 
The default value is set to \( \rho_{\rm threshold} = 10^{-15} \) kg.m\(^{-3}\).
The peak-to-saddle ratio sets the minimum contrast for the peaks relative to their background to be considered as significant enough and used as seeds for the extraction process. 
The default value is set to psr = 1.75.

\begin{figure}[!ht]
    \centering
    \includegraphics[width=9cm]{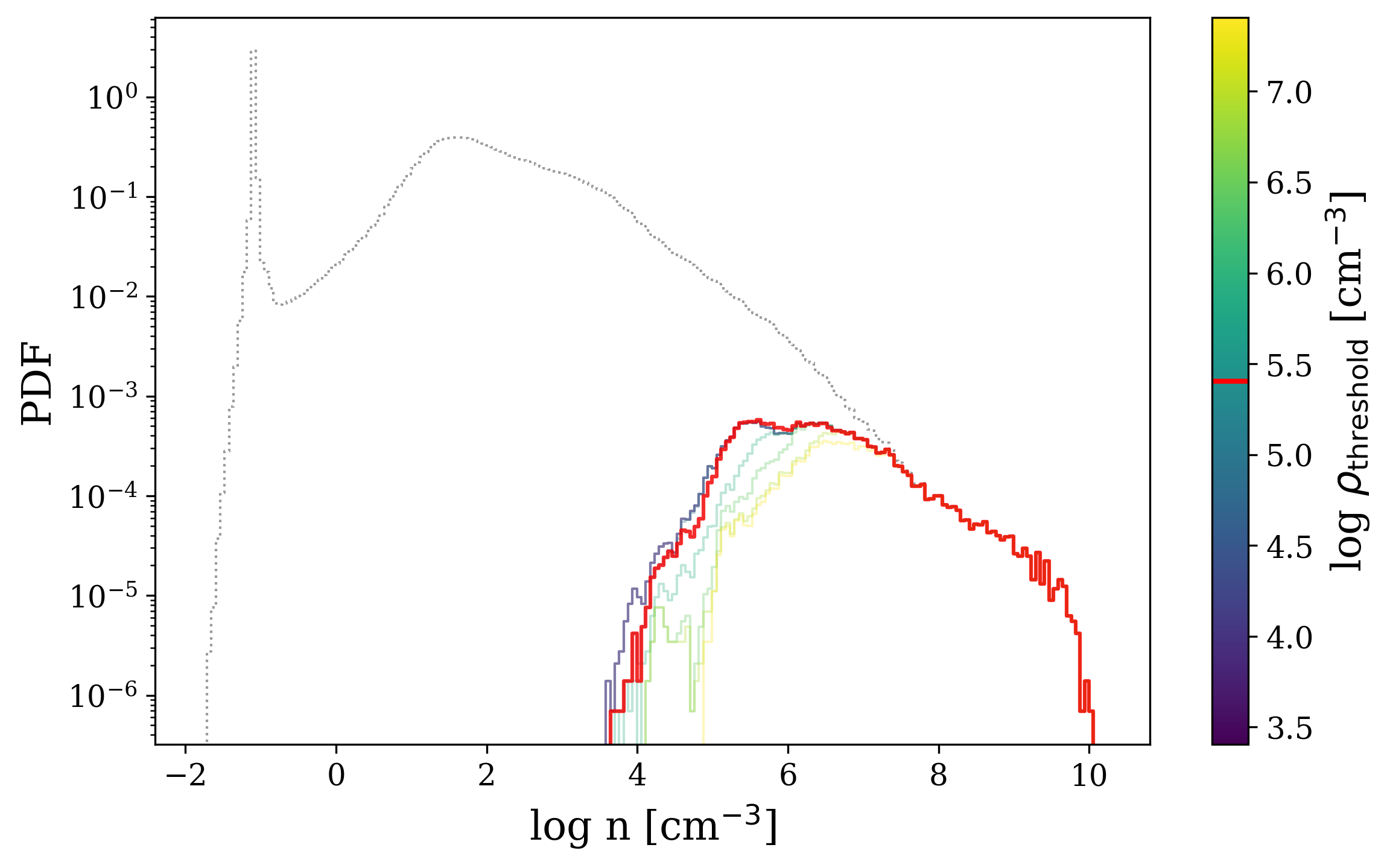}
    \includegraphics[width=9cm]{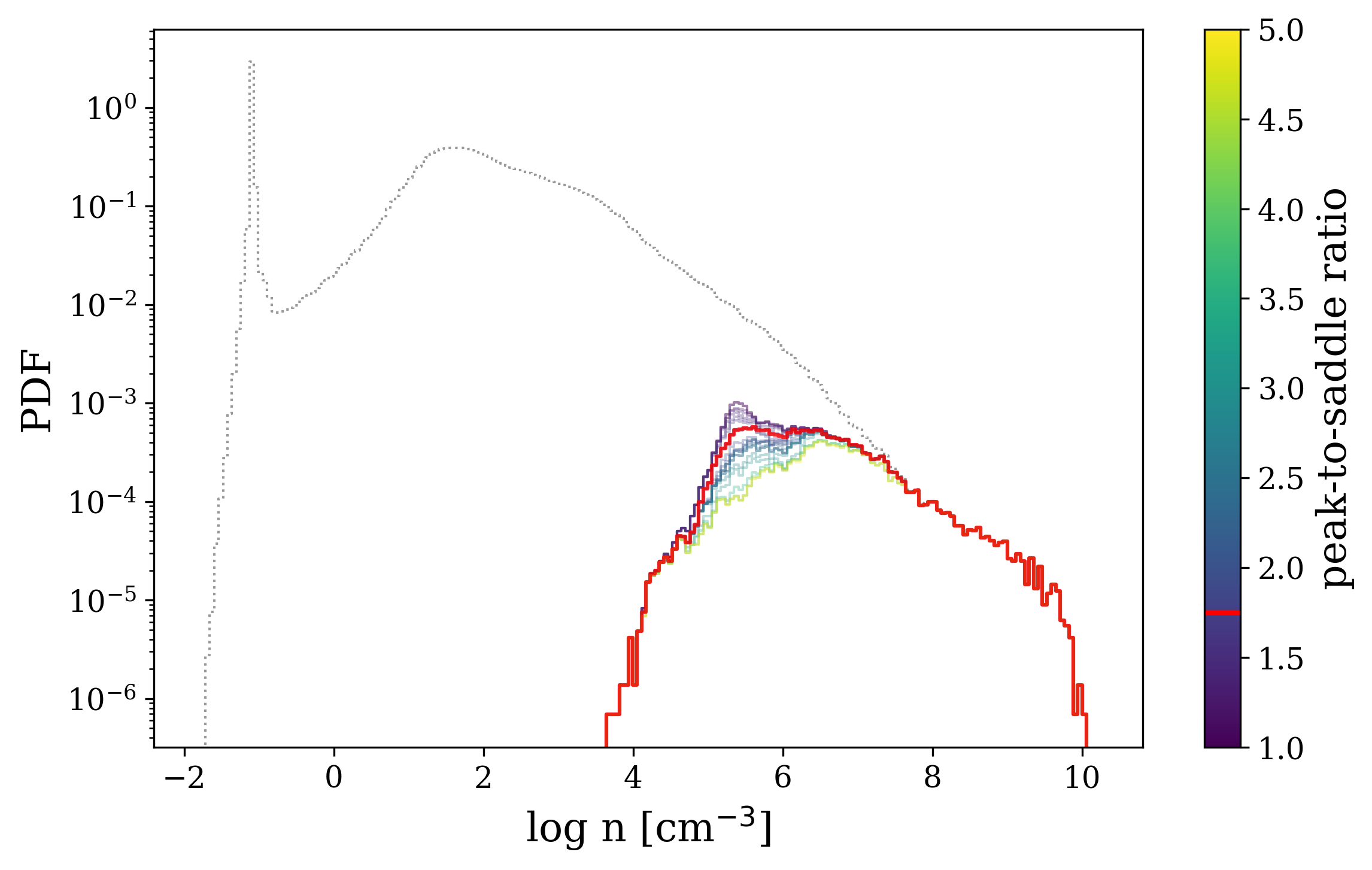}
    \caption{Density pdf with respect to the peak threshold (top) and peak-to-saddle ratio (bottom) parameters. 
    The dashed gray line correspond to the total density pdf. 
    The red line corresponds to the reference extraction. 
    The light colored lines correspond to the extractions with different parameter values.}
    \label{fig-appendix-benchmark-density-pdf-wrt-peak-sorting}
\end{figure}

The comparison of the probability density function (pdf) of the density is given in Fig. \ref{fig-appendix-benchmark-density-pdf-wrt-peak-sorting}. 
Since the high-density structures are not affected by the peak sorting parameters, because the associated peaks are always detected and kept after sorting, the high-density part of the pdf is not expected to vary with the parameter values.
We observe the cut of the lower-density parts for high peak threshold values, corresponding to the structures remove from the reference extractions observed in Fig. \ref{fig-appendix-benchmark-struc-nb-wrt-peak-sorting}. For thresholds lower than the reference value, the extracted density pdf remains almost the same.
The difference is even less significant when varying the peak-to-saddle ratio.

The comparison of the mass distributions for different values of the peak sorting parameters is given in Fig. \ref{fig-discussion-ccdf-comparison}.
\begin{figure}[!ht]
    \centering
    \includegraphics[width=9cm]{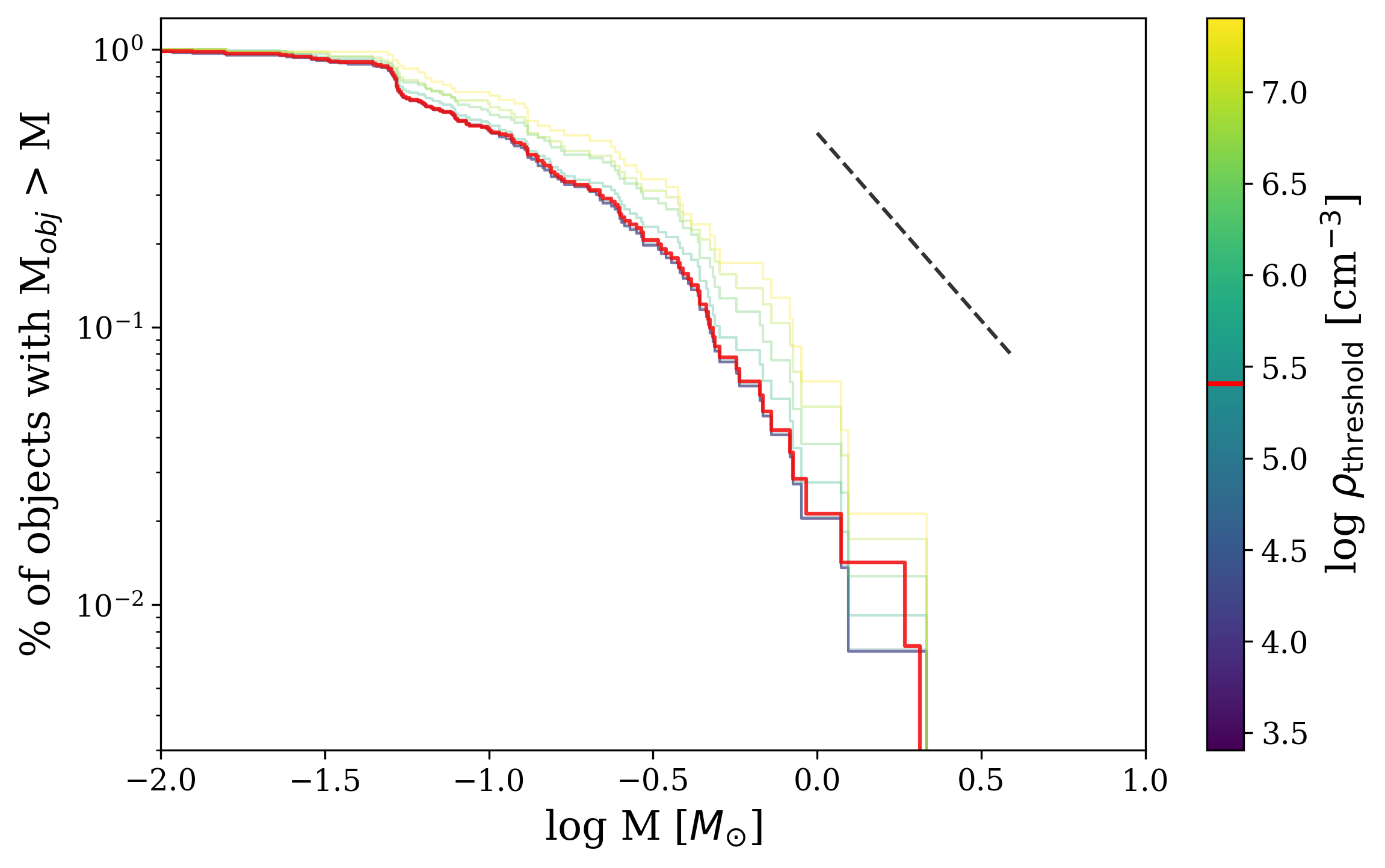}
    \includegraphics[width=9cm]{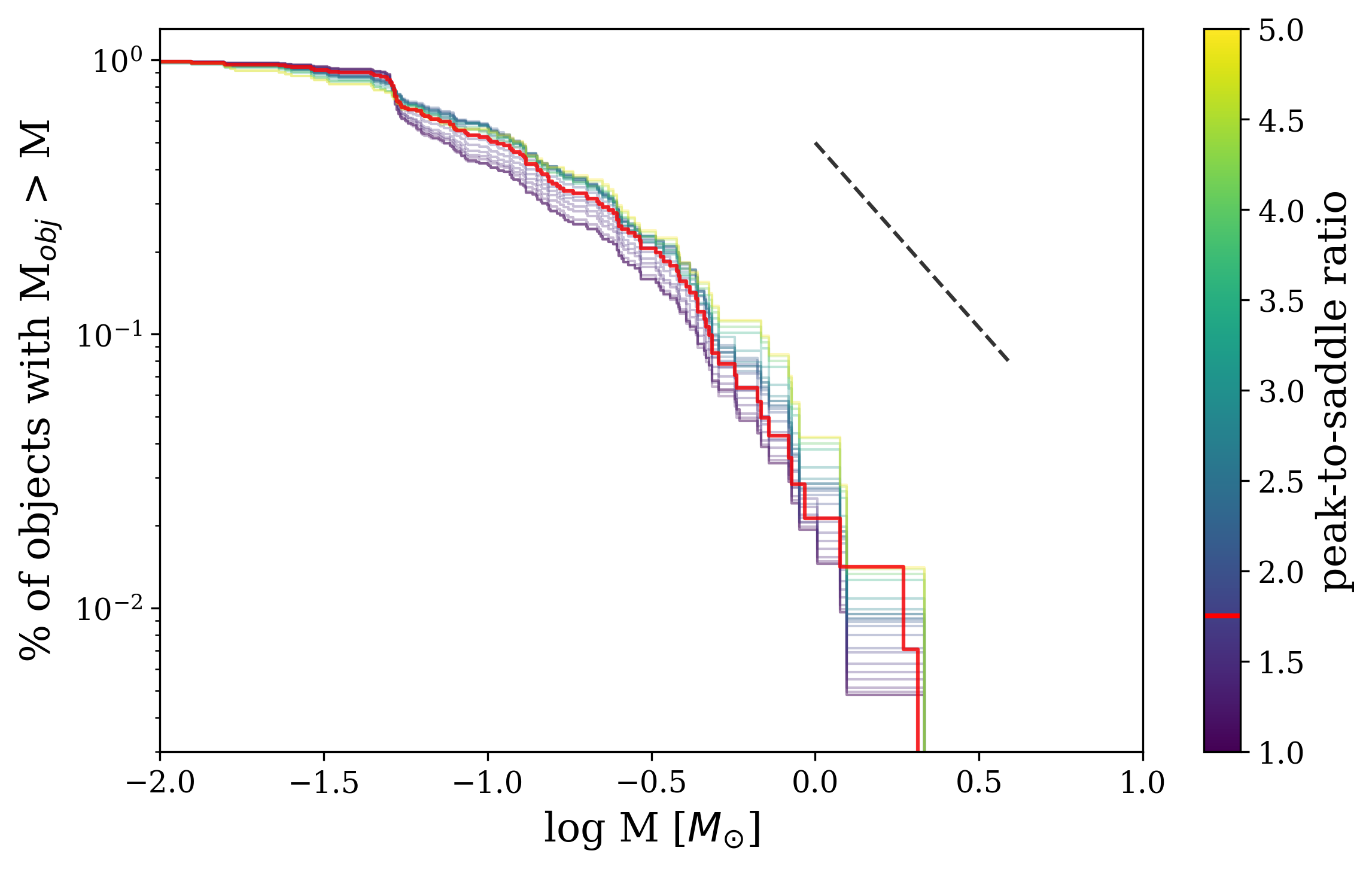}
    \caption{Mass distribution with respect to the peak threshold (top) and peak-to-saddle ratio (bottom). 
    The red line shows the mass ccdf for the reference parameter value. 
    The colored lines correspond to the mass ccdf for the different parameter values given by the side colorbars.}
    \label{fig-appendix-benchmark-mass-ccdf-wrt-peak-sorting}
\end{figure}
The extracted structures being the same in high-density regions, the high-mass part of the distribution is not supposed to be affected by the parameter values. 
Removing the low-mass structures for high peak tresholds slightly shifts the distribution up, but does not significantly affect its shape.
The same effect is observed when varying the peak-to-saddle ratio: the shape remains roughly the same, and is just shifted vertically depending on the fraction of low-mass objects removed by the choice of the parameter value.

\begin{figure}[!ht]
    \centering
    \includegraphics[width=9cm]{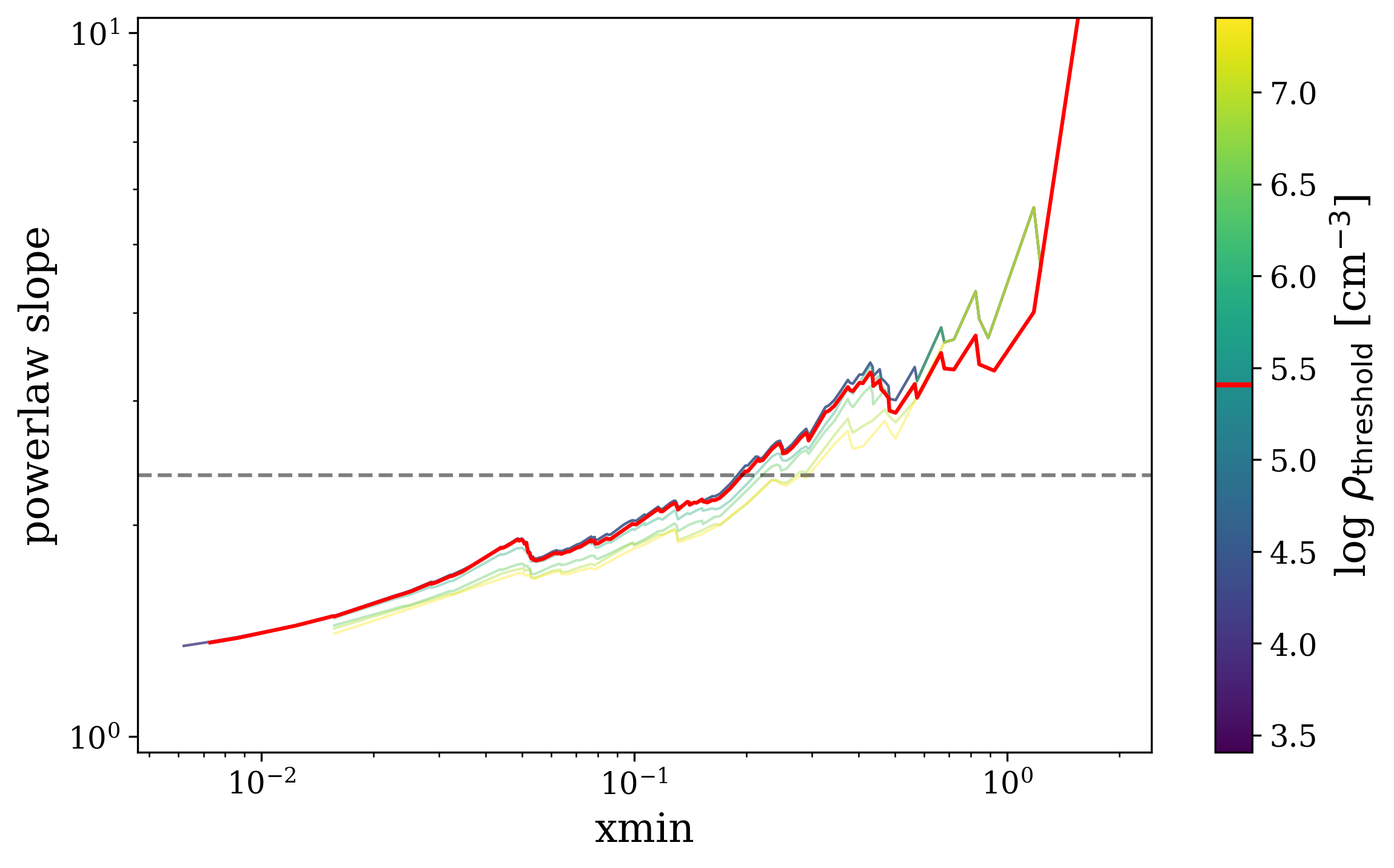}
    \includegraphics[width=9cm]{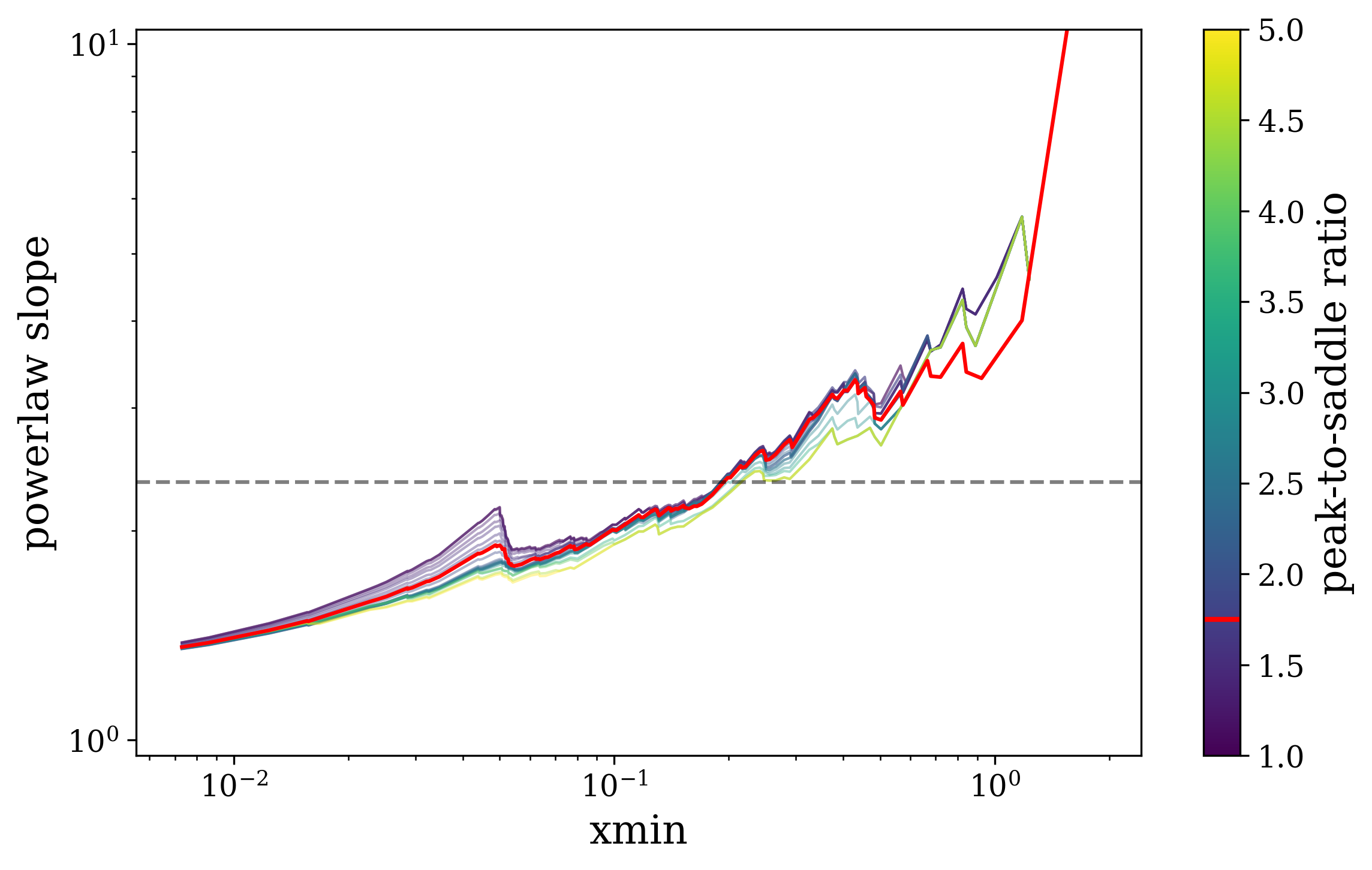}
    \caption{Powerlaw slope as a function of the fit lower limit \(xmin\), for different values of the peak threshold (top) and peak-to-saddle ratio (bottom). 
    The dotted black line corresponds to the Salpeter slope. 
    The colored lines correspond to the powerlaw fits for the different parameter values given by the side colorbars.}
    \label{fig-appendix-benchmark-alpha-mle-wrt-peak-sorting}
\end{figure}

The powerlaw slope estimated for the mass distributions between the \textit{vibes}, \textit{hop} and \textit{dendrogram} extractions is given in Fig. \ref{fig-appendix-benchmark-alpha-mle-wrt-peak-sorting}.
The powerlaw fit has been performed with a maximum likelihood estimation \citep[MLE,][]{Clauset2009} using the \texttt{powerlaw} package \citep{Alstott_2014}. 
The slope is fitted with a MLE considering the values higher than a lower limit \(xmin\).
For both parameters, no significant difference appears when varying the parameters on the selected intervals.

\subsection{Relation between the elongation constraint and the aspect ratio} \label{appendix-benchmark-elongation-aspect-ratio}

In the assumption of a cylindrical filament, of diameter \(d_f\) and length \(L_f\), the structure volume can be written:
\begin{equation}
    V = \frac{4 \pi}{3} R_{\rm eq}^3 = \pi \frac{d_f^2}{4} L_f
\end{equation}
In this configuration, the maximum distance to the center in the definition of the elongation constraint in Eq. \ref{eq-methods-elongation-constraint} is:
\begin{equation}
    d_{n0, max} = \frac{\sqrt{d_f^2+L_f^2}}{2} < C_{\rm elongation} ~R_{\rm eq}
\end{equation}
Combining those two equations, we obtain:
\begin{equation}
    \left(\frac{d_f}{L_f}\right)^2 > \frac{2}{3 C_{\rm elongation}^3} \left(1 + \left(\frac{d_f}{L_f}\right)^2\right)^{3/2}
\end{equation}
Assuming \(\frac{d_f}{L_f} \ll 1\), the aspect ratio \(\frac{L_f}{d_f}\) can be approximately constrained by the Eq. \ref{eq-benchmark-aspect-ratio-constrain}.

\subsection{Structure building parameters} \label{appendix-benchmark-struc-building}

\begin{figure}[!ht]
    \centering
    \includegraphics[width=9cm]{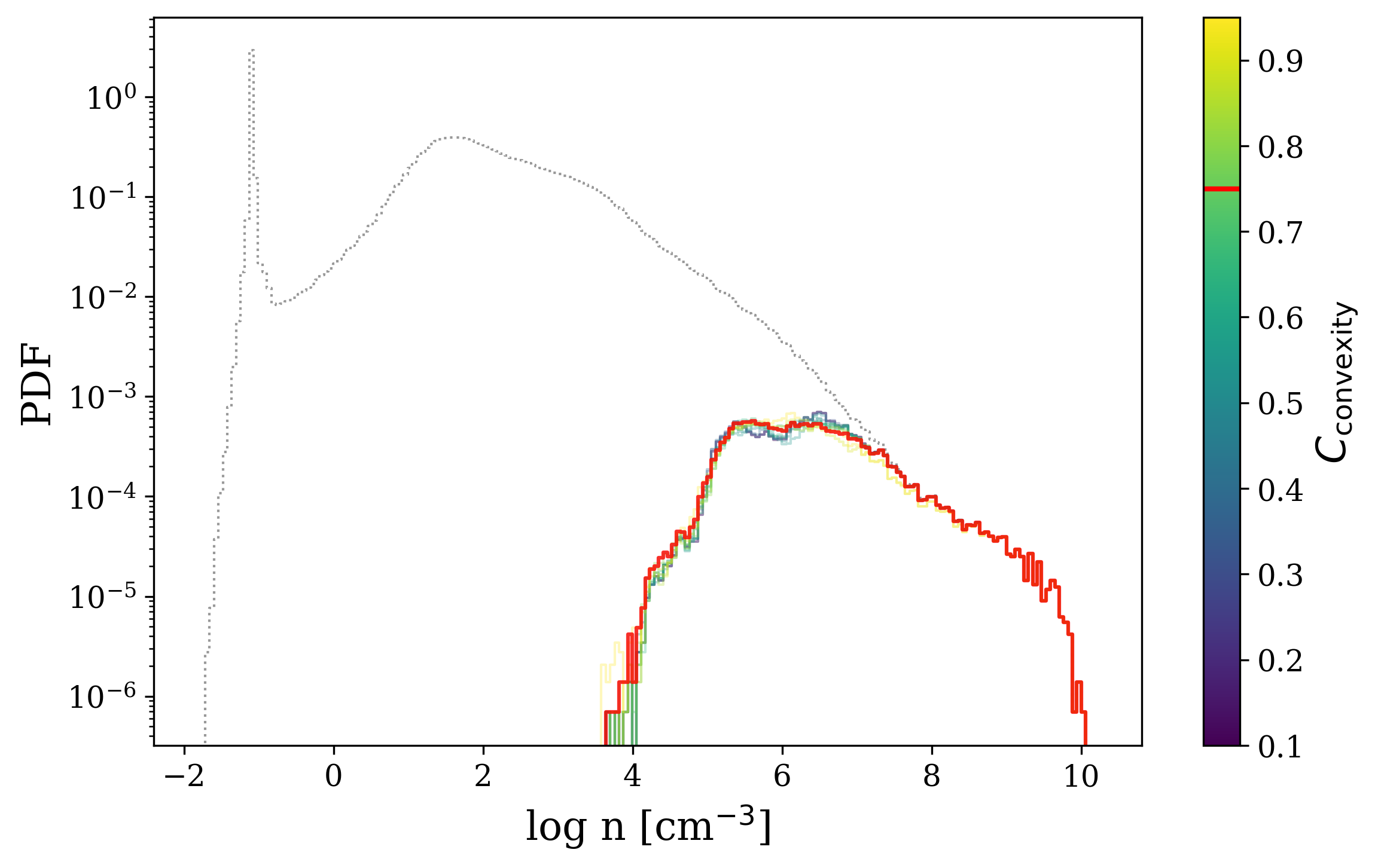}
    \includegraphics[width=9cm]{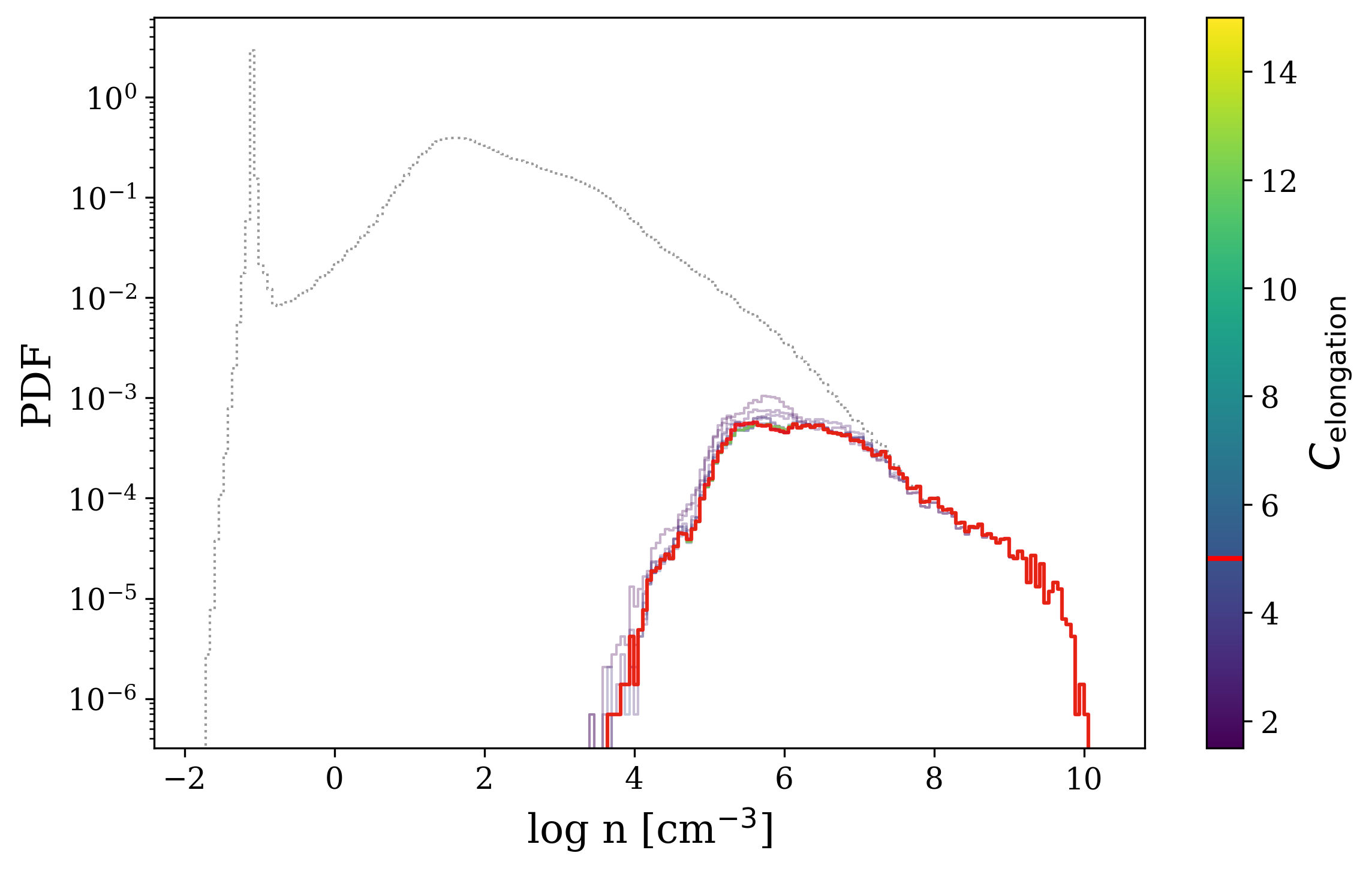}
    \includegraphics[width=9cm]{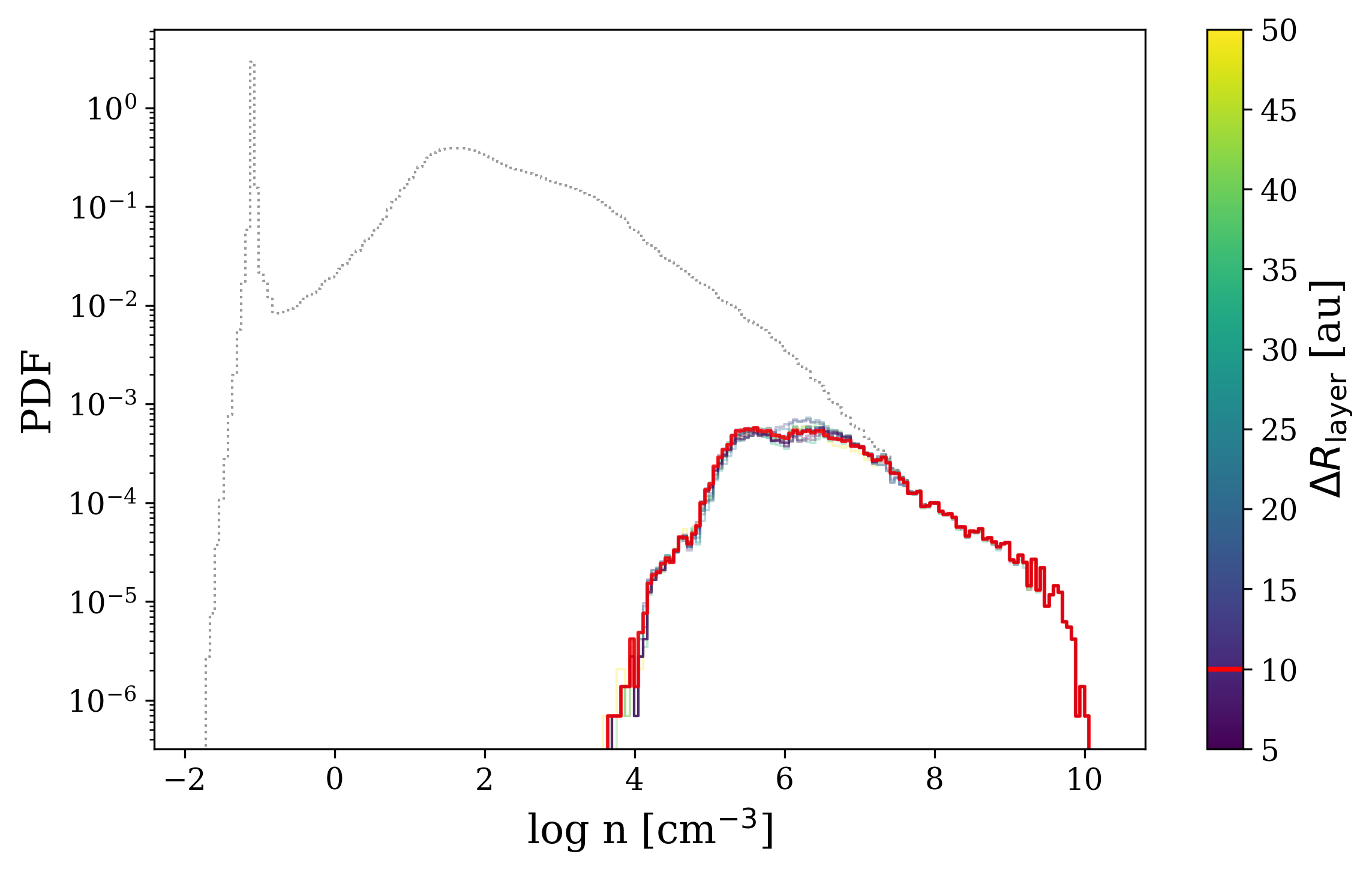}
    \caption{Density pdf with respect to the convexity (top), elongation (middle), and layer number (bottom) parameters. 
    The dashed gray line correspond to the total density pdf. 
    The red line corresponds to the reference extraction. 
    The light colored lines correspond to the extractions with different parameter values.}
    \label{fig-appendix-benchmark-density-pdf-wrt-struc-building}
\end{figure}

In Sect. \ref{benchmark-subsection-structure-building}, we studied the effect of the structure building parameters on the extraction. 
The effect of the structure building parameters on the density coverage of the extraction is given in Fig. \ref{fig-appendix-benchmark-density-pdf-wrt-struc-building}.
For all of them, the density pdf coverage is poorly affected by the variation of the parameter values.

\begin{figure}[!ht]
    \centering
    \includegraphics[width=9cm]{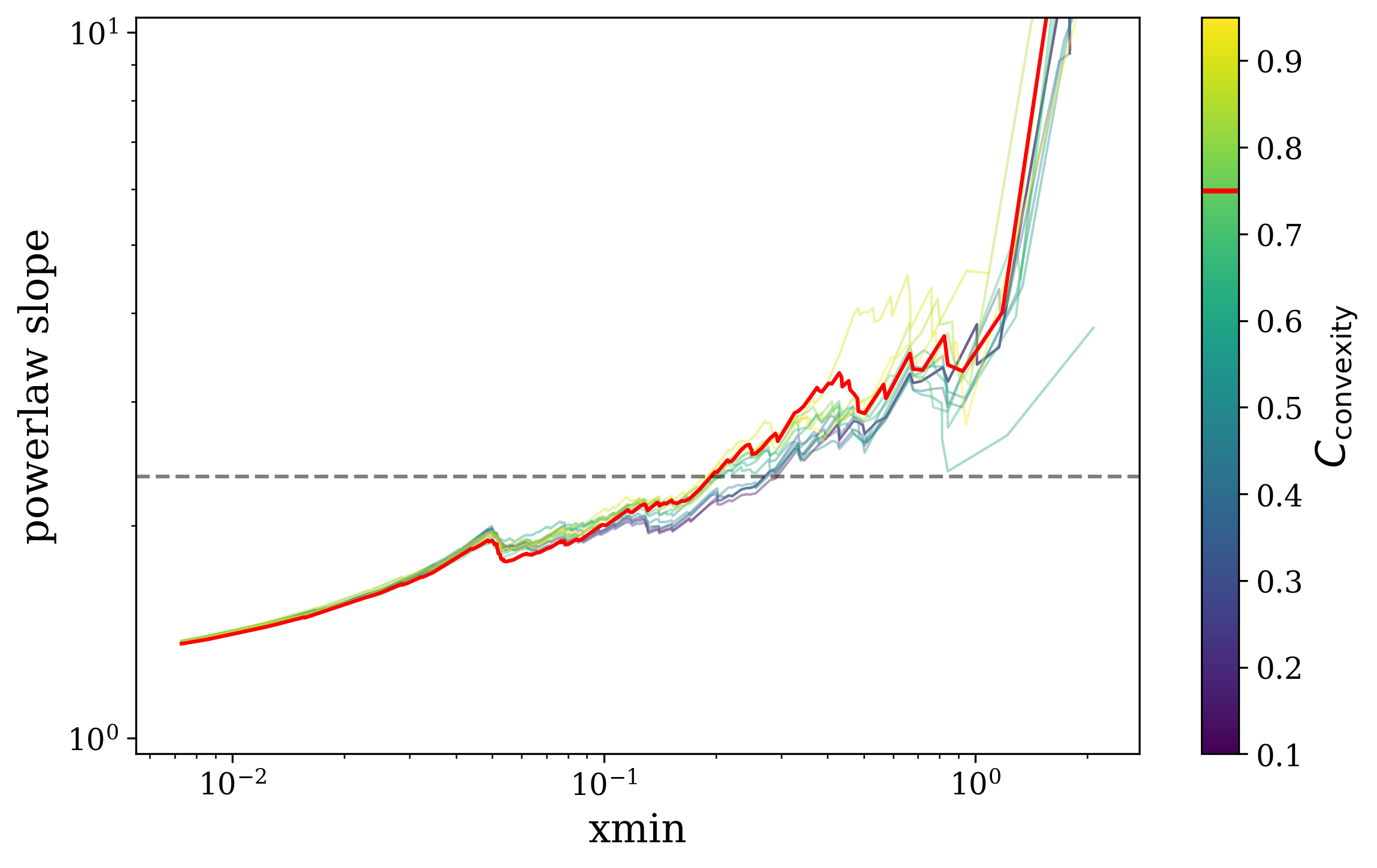}
    \includegraphics[width=9cm]{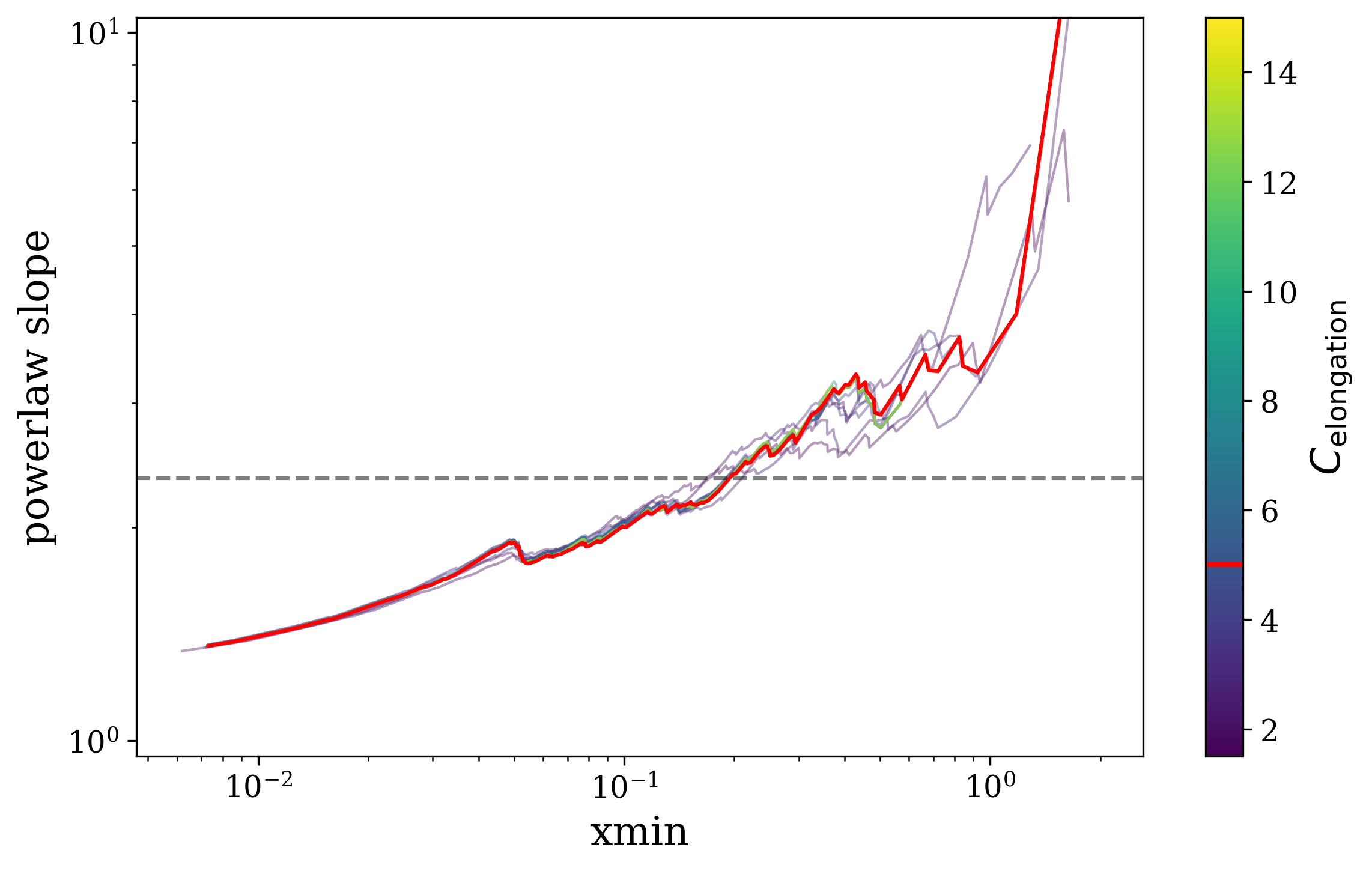}
    \includegraphics[width=9cm]{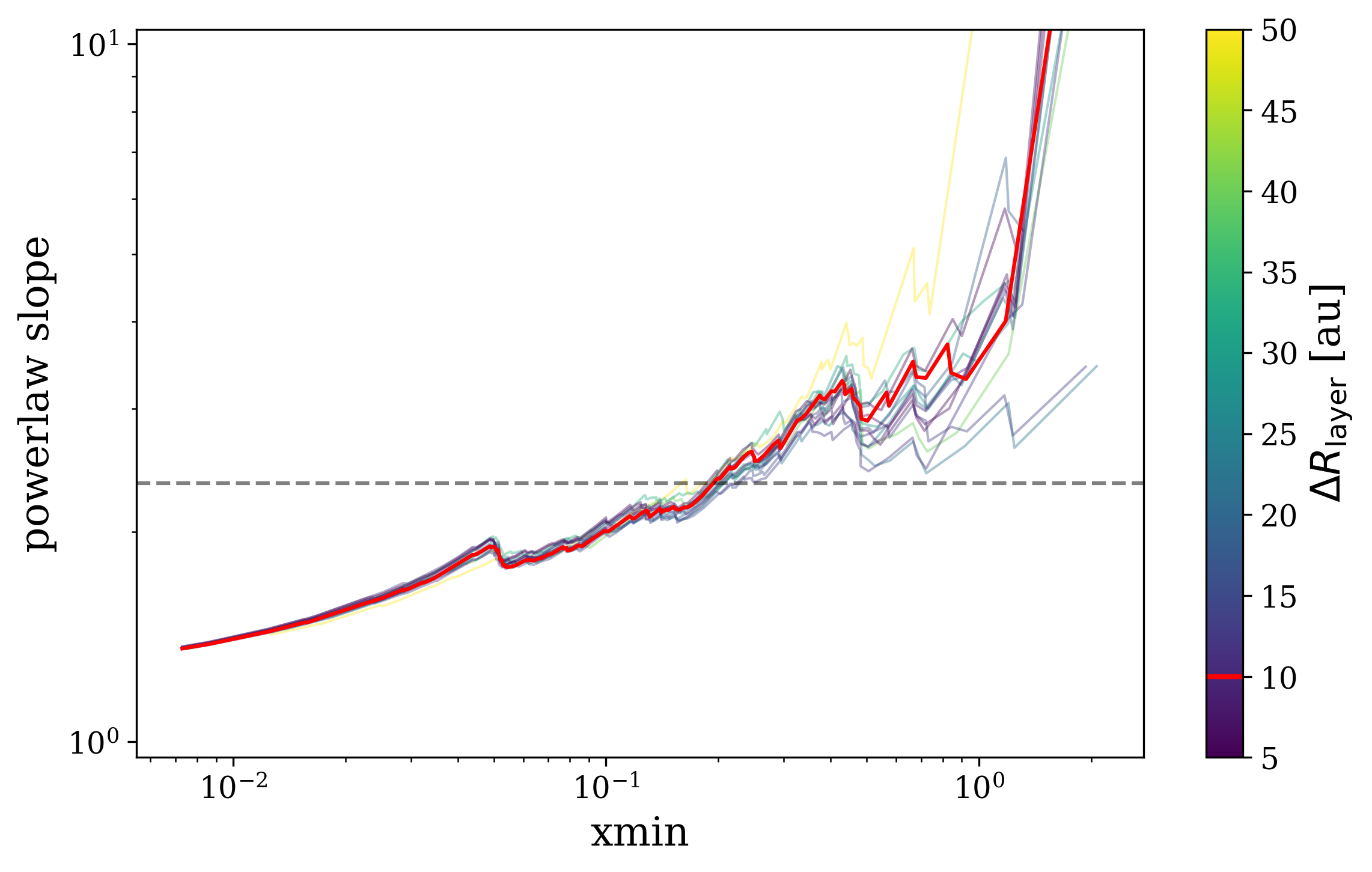}
    \caption{Powerlaw slope as a function of the fit lower limit \(xmin\), for different values of the convexity (top), elongation (middle), and layer number (bottom) parameters. 
    The dotted black line corresponds to the Salpeter slope. 
    The colored lines correspond to the powerlaw fits for the different parameter values given by the side colorbars.}
    \label{fig-appendix-benchmark-alpha-mle-wrt-struc-building}
\end{figure}

The powerlaw slope estimated for the mass distributions is given in Fig. \ref{fig-appendix-benchmark-alpha-mle-wrt-struc-building} as a function of the convexity, elongation and layer number parameters.
For each parameter, the estimated slope is roughly the same for xmin below 0.1.
Above this value, significant differences appear.
The mass maximum decreases for high values of convexity, having for consequence a steeper distribution tail and so a curve shifted to the top here.

\begin{figure}[!ht]
    \centering
    \includegraphics[width=9cm]{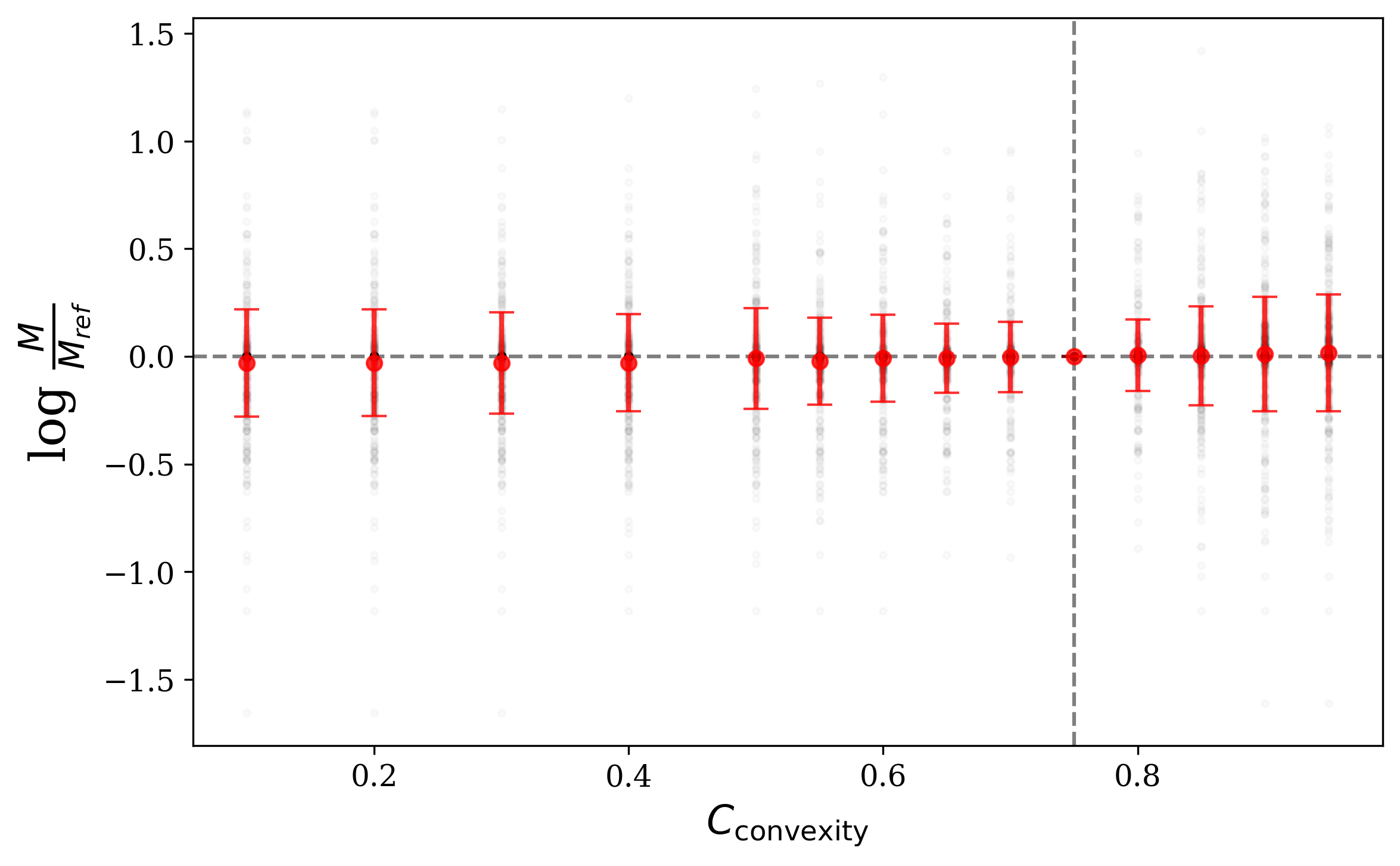}
    \includegraphics[width=9cm]{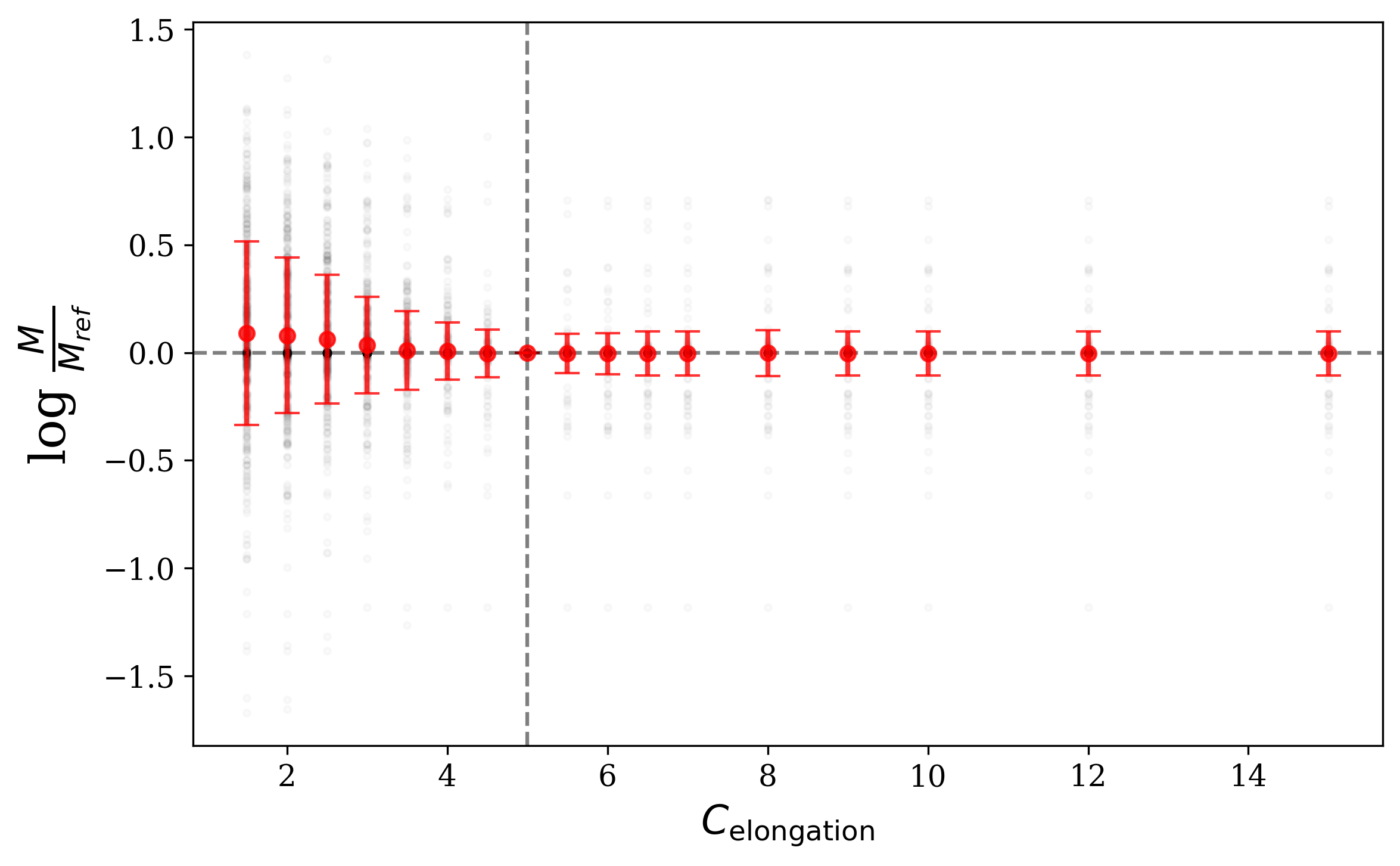}
    \includegraphics[width=9cm]{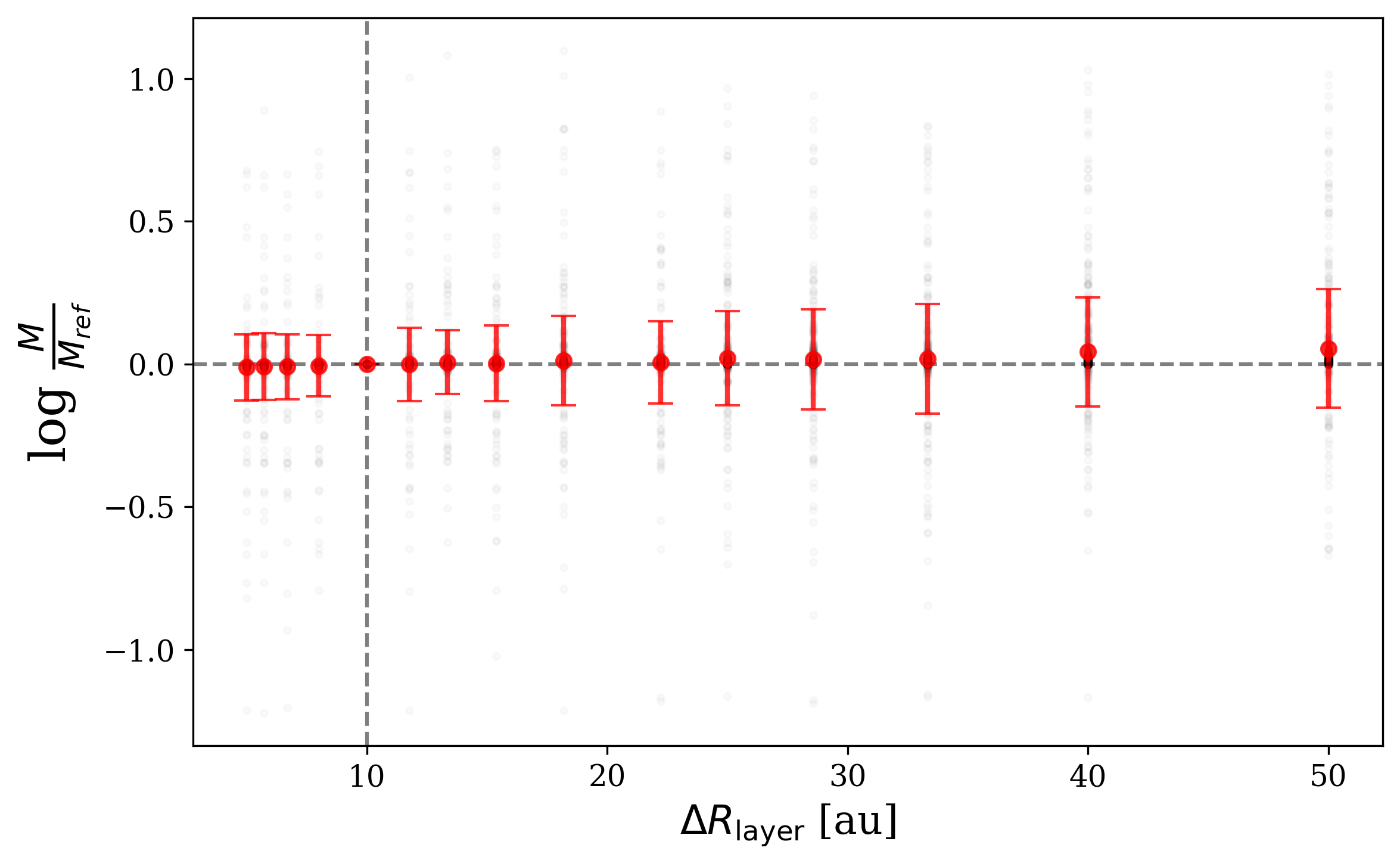}
    \caption{Ratio of the mass of the extracted structures over the mass of the reference structures, with respect to the shape constraint parameters. 
    The low opacity black points correspond to all the individual objects having a nonzero intersection with the reference extraction, over three snapshots. 
    The red points and errorbars correspond to the mean values with standard deviations.}
    \label{fig-appendix-benchmark-mass-ratio-wrt-struc-building}
\end{figure}

We aimed to check the degree of similarity of the extracted objects with the reference extraction.
The masses of the extracted structures relative to the masses of the reference structures are given in Fig. \ref{fig-appendix-benchmark-mass-ratio-wrt-struc-building} with respect to the structure building parameters. 
The reference structures correspond to the structures associated to the same density peaks in the reference extraction.
In this figure, only the structures having a nonzero intersection with an object of the reference extraction are plotted.
The trend is the same for the structure similarity as for the structure recall (i.e., the number of recovered objects from the reference extraction, green and blue dots on Fig. \ref{fig-benchmark-struc-nb-wrt-struc-building}). 
The ratio of the individual masses over the reference remains very close to 1, with a standard deviation lower than 0.25 in log (equivalent to a factor lower than 2) when varying the convexity constraint. 
For the elongation, the similarity is even better for \( C_{\rm elongation} \geq 4 \) but is significantly degraded for lower values, with a standard deviation reaching up to almost 0.5 in log for \( C_{\rm elongation} = 1.5 \), corresponding to a factor higher than 3.
This tends to show that there is a shift value around 4 for the elongation above which the effet of the constraint is limited.
No global trend appears with the evolution of the extraction with respect to convexity, but a convergence is observed for an elongation parameter higher than the reference value \( C_{\rm elongation} \geq 5 \).
In the case of the layer number, no clear trend can be observed, but the structure masses remain close to the reference masses within a deviation lower than 0.25 in log.

\section{Comparison with density-based extraction} \label{appendix-density-comparison-complements}

We performed a comparison of the extractions with \textit{hop} and \textit{dendrogram} to the extraction with \textit{vibes}.
While varying the density thresholds of the density-based alogorithms, we also varied the \textit{vibes} parameters as detailed in Table \ref{appendix-table-vibes-params}.
\begin{table}[!ht]
    \caption{Variations of the \textit{vibes} parameters in the comparison with \textit{hop} and \textit{dendrogram}.}
    \label{appendix-table-vibes-params}
    \centering
    \begin{tabular}{l c r}
        \hline
        Parameter & Tested values & Reference \\
        \hline
        \hline
        log \(\rho_{\rm threshold}\) [kg.m\(^{-3}\)] & [-16, -15.5, -14.5, -14] & -15 \\
        peak-to-saddle ratio & [1.4, 1.6, 1.9, 2.1] & 1.75 \\
        \(C_{\rm elongation}\) & [3, 4, 6, 7] & 5 \\
        \(C_{\rm convexity}\) & [0.55, 0.65, 0.85, 0.95] & 0.75 \\
        \(n_{\rm layer}\) & [11, 15, 25, 30] & 20 \\
        \hline
    \end{tabular}
\end{table}
Those parameters give the standard deviation interval in Fig. \ref{fig-discussion-struc-number-comparison}, and are also used for the Kolmogorov-Smirnov tests detailed in Fig. \ref{fig-appendix-ks-test}.

\begin{figure}[!ht]
    \centering
    \includegraphics[width=9cm]{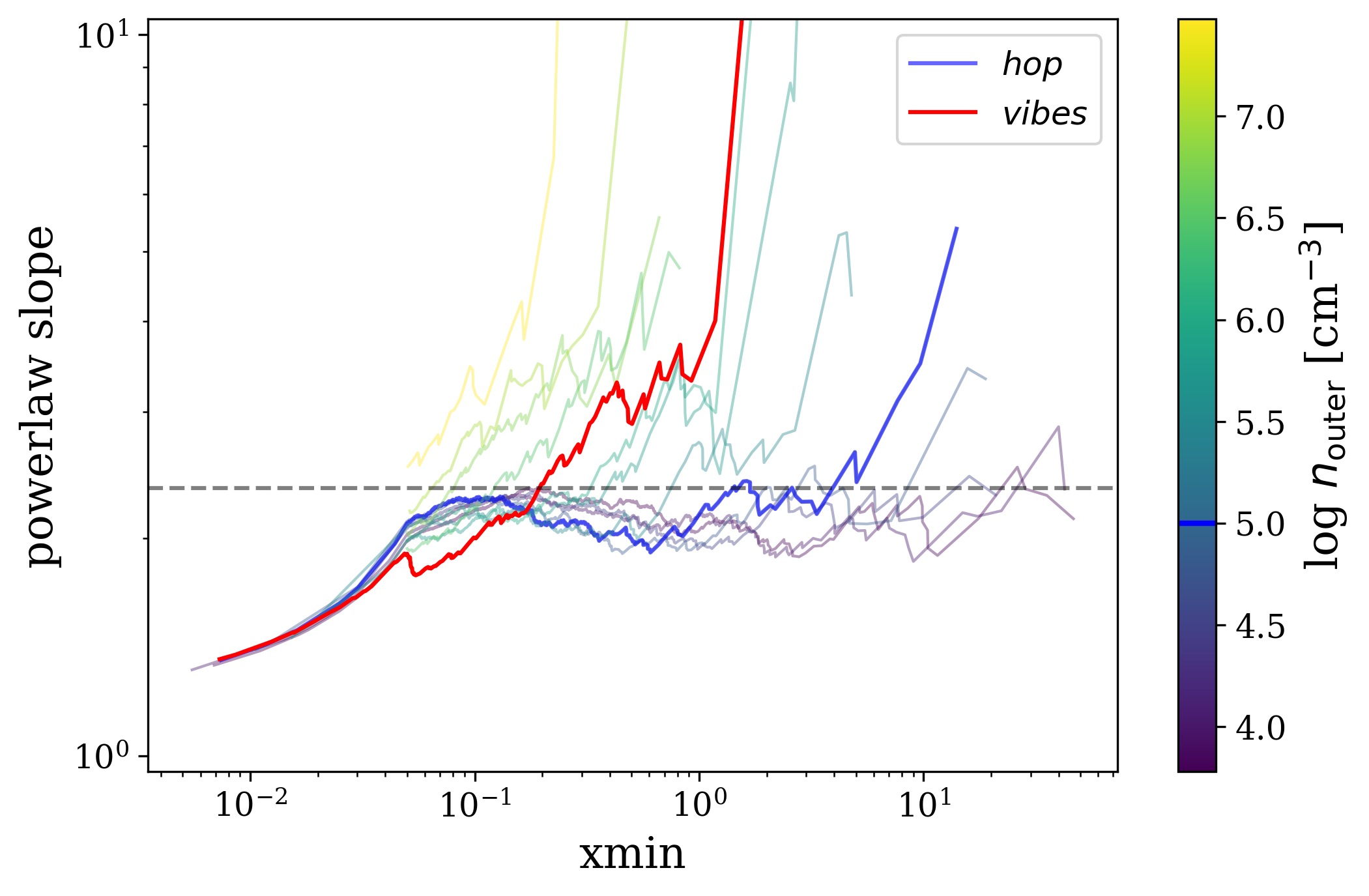}
    \includegraphics[width=9cm]{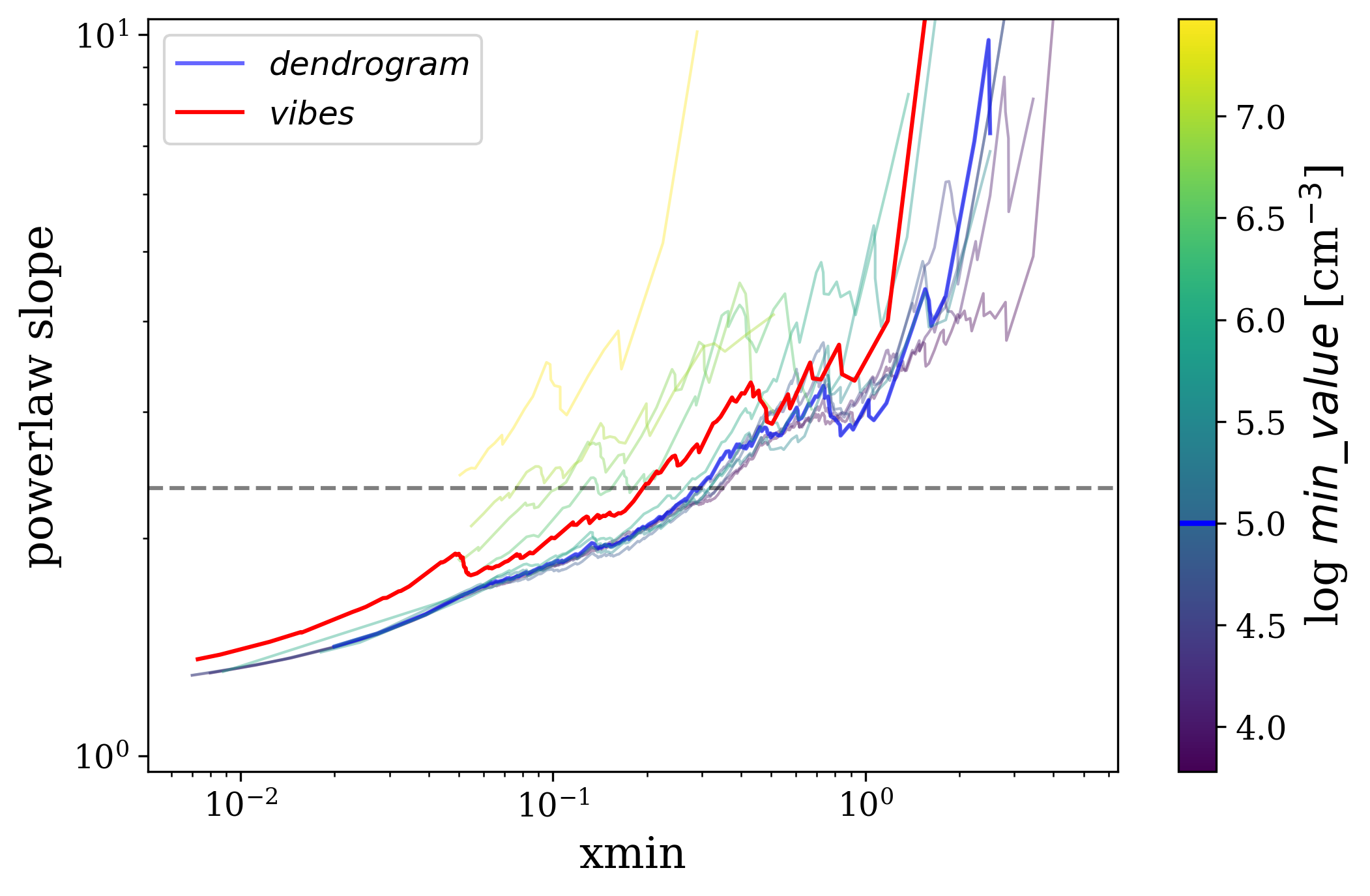}
    \caption{Powerlaw exponent as a function of the threshold parameters, \(n_{\rm outer}\) for \textit{hop} (top) and \(min\_value\) for \textit{dendrogram} (bottom).
    The dotted black line corresponds to the Salpeter slope. 
    The colored lines correspond to the powerlaw fits for the different parameter values given by the side colorbars.}
    \label{fig-appendix-density-extraction-alpha-mle}
\end{figure}

The powerlaw slope estimated for the mass distributions obtained with \textit{hop} and \textit{dendrogram} is given in Fig. \ref{fig-appendix-density-extraction-alpha-mle}.
For \textit{hop}, the estimated powerlaw slope is very sensitive to the threshold parameter.
As discussed in Sect. \ref{discussion-subsection-extraction-comparison}, low values of \(n_{\rm outer}\) induce huge and very massive structures, that are split when the threshold increases. 
Those big objects flatten the mass distribution for low \(n_{\rm outer}\), and the high-mass tail get steeper and steeper as they split while \(n_{\rm outer}\) increases.
The behavior of \textit{dendrogram} is roughly the same, yet less pronounced.

\begin{figure}[!ht]
    \centering
    \includegraphics[width=9cm]{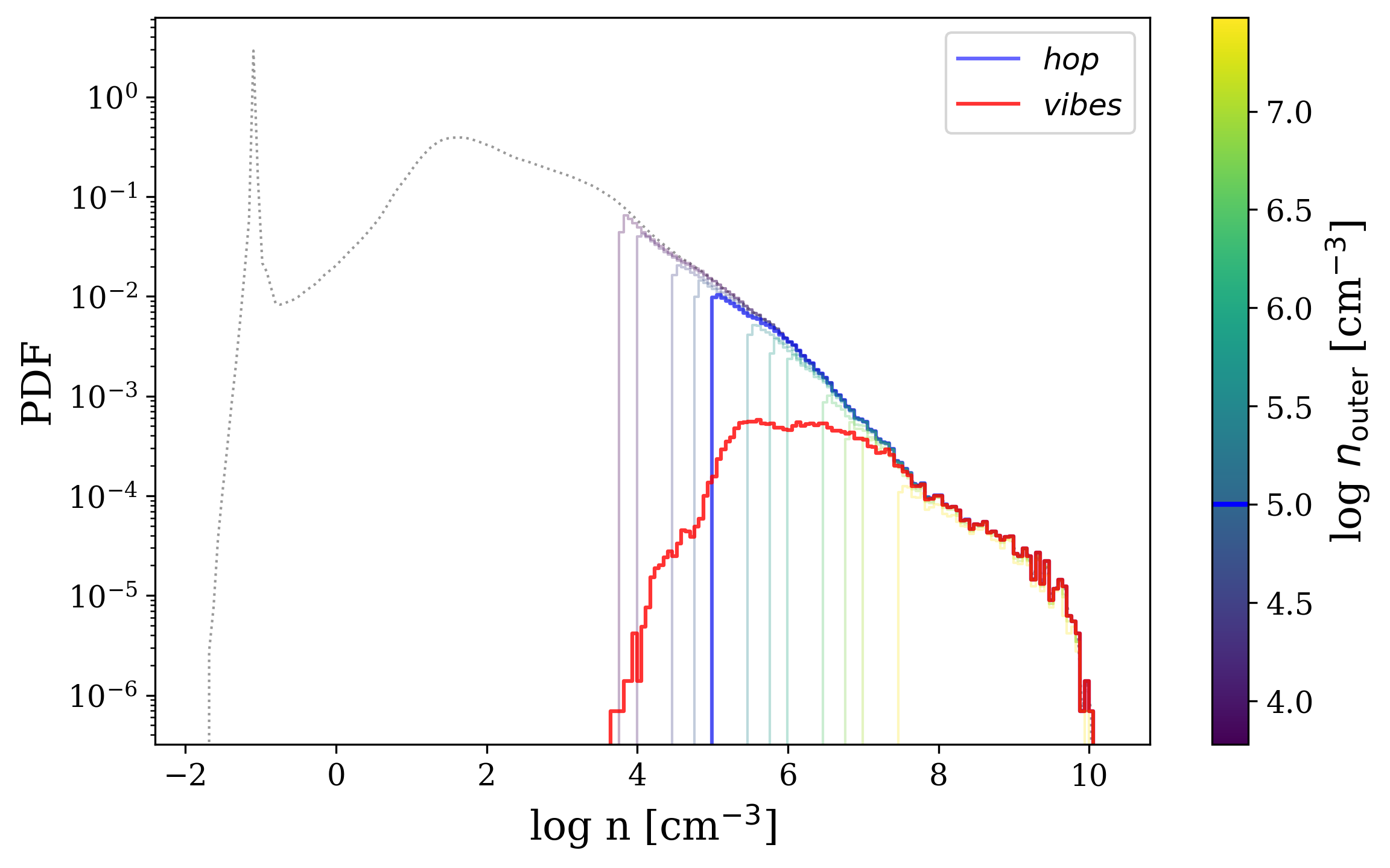}
    \includegraphics[width=9cm]{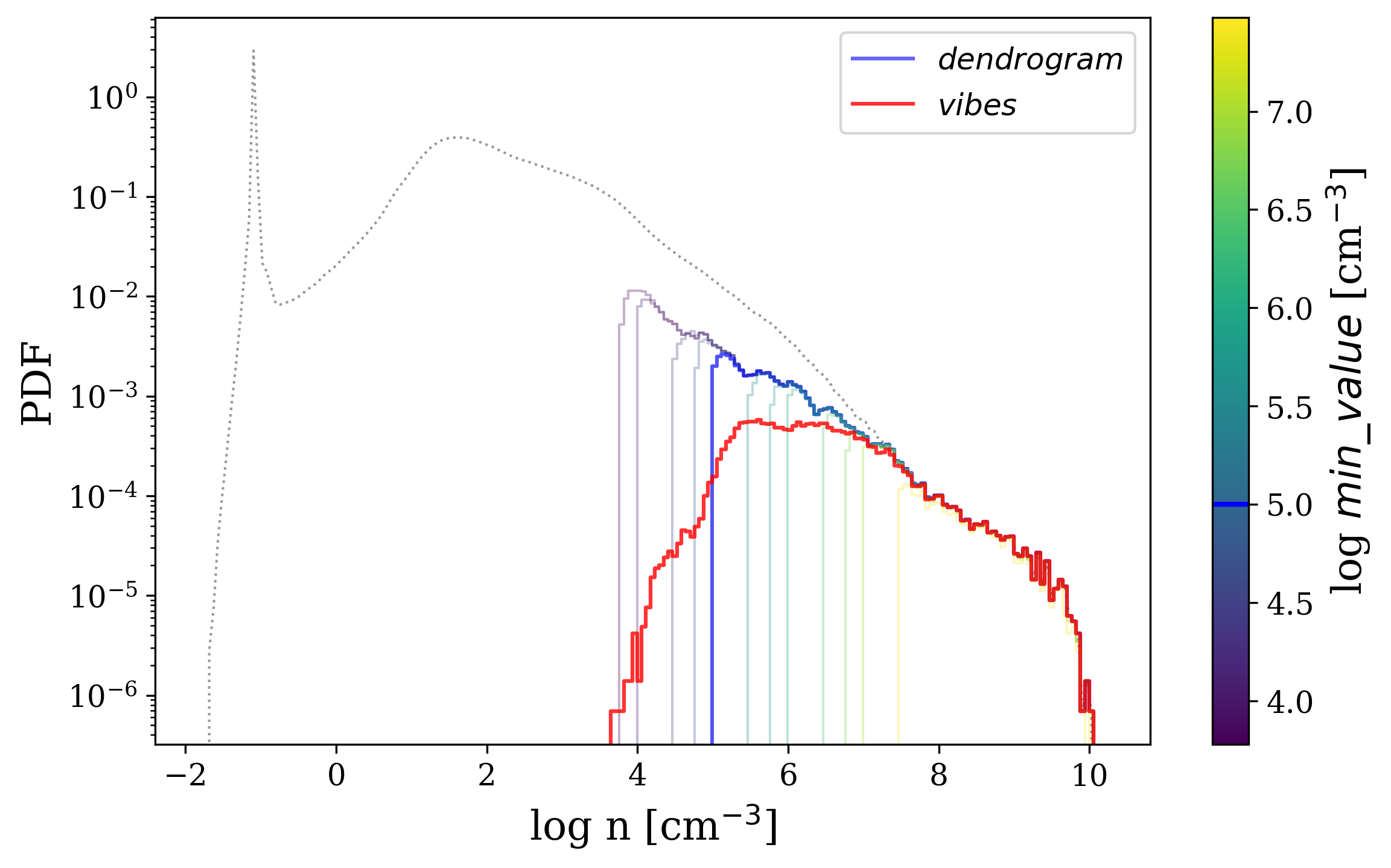}
    \caption{Density pdf of the extractions with \textit{hop} (top) and \textit{dendrogram} (bottom) for different values of the corresponding threshold parameter \(n_{\rm outer}\) and \textit{min\_value}, respectively. 
    The pdf fraction associated to \textit{vibes} structures is given in red. 
    The pdf fraction associated to \textit{hop} and \textit{dendrogram} structures for the fixed threshold parameter at \(10^5\) cm\(^{-3}\) are given in blue.
    The total density pdf is given in gray.}
    \label{fig-discussion-density-pdf-comparison}
\end{figure}

The comparison of the probability density function (pdf) of the density is given in Fig. \ref{fig-discussion-density-pdf-comparison}.
As expected, all the extraction methods cover well the densest parts of the snaphsot, however behaving differently. 
The \textit{hop} extraction shows a sharp transition at the threshold value. 
Almost all the cells above this threshold are associated to a structure in the end. 
This makes the algorithm extremely sensitive to this threshold.
It might be a problem when aiming to define accurate mass reservoirs in the sense of star formation. 
Indeed, the mass of the analyzed structures might be artificially and significantly affected by the way the extraction works.
For \textit{dendrogram}, the same effect is observed but with a significant difference: all the pdf is not covered, meaning all the cells above the threshold are not necessarily assigned to an object.
Part of the mass above the threshold falls between dendrogram contours, and is considered part of the background.
If this selection process might miss part of interesting mass, it makes sense when trying to extract mass reservoirs as peaks above their background, while the background itself is fully set by the density threshold in the case of \textit{hop}.
The \textit{vibes} extraction shows the same behavior, letting significant intermediate-density parts of the simulation (around \(10^5\) cm\(^{-3}\)) outside of the extracted objects, with no sharp transition caused by a density threshold.

\end{appendix}

\end{document}